\documentclass[9pt,journal]{IEEEtran}

\usepackage{amsthm}
\usepackage{amsfonts}
\usepackage{amsmath}
\usepackage{amssymb}
\usepackage{makeidx}
\usepackage{graphicx}
\usepackage{color}
\usepackage{cite}
\usepackage{url}
\usepackage{caption}
\usepackage{subcaption}

\usepackage[ruled,vlined]{algorithm2e}
\usepackage{cite} 	
\usepackage{url}

\usepackage[rgb,letterpaper]{xcolor}
\usepackage{algorithmic}
\usepackage[letterpaper]{psfrag}
\usepackage{tikz}
\usepackage[process=auto,crop=pdfcrop]{pstool}
\usepackage{footnote}

\providecommand{\com[2]}{\begin{tt}[#1: #2]\end{tt}}

\def\S{\mathcal{S}}

\ifCLASSINFOpdf
\else
\fi
\usepackage{pifont}
\usepackage{xcolor,colortbl}
\usepackage{flushend}
\usepackage[normalem]{ulem} 
\usepackage{multicol}
\usepackage{xcolor}

\providecommand{\sortnoop}[1]{}
\usepackage{multibib}
\newcites{ref}{References}
\usepackage{bibentry}  
\nobibliography*       

\begin{document}
\title{Cyber-Physical Systems Security: \\ a Systematic Mapping Study 
\author{Yuriy {Zacchia Lun}$^1$, Alessandro {D'Innocenzo}$^2$, Ivano Malavolta$^3$, 
Maria Domenica {Di Benedetto}$^2$}
\thanks{$^1$Gran Sasso Science Institute, L'Aquila, Italy.}
\thanks{$^2$Center of Excellence DEWS, Univ. of L'Aquila, Italy.}
\thanks{$^3$Vrije Universiteit Amsterdam, The Netherlands.}
\thanks{
The research leading to these results has received funding from the Italian 
Government under Cipe resolution n.135 (Dec. 21, 2012), project \emph{INnovating 
City Planning through Information and Communication Technologies} (INCIPICT). 
}
}
\maketitle
\begin{abstract}

\textit{Context:}
Cyber-physical systems (CPS) are integrations of computation, networking, and 
physical processes. Due to the tight cyber-physical coupling and to the potentially 
disrupting consequences of failures, security is one of the primary concerns for 
this type of systems. CPS security is attracting several research efforts from 
different and independent areas (e.g., secure control, intrusion detection in 
SCADA systems, etc.), each of them with specific peculiarities, features, and 
capabilities, resulting in a considerably variegated and complex scientific body 
of knowledge on the topic.

\noindent \textit{Objective:}
In this study we aim at identifying, classifying, and analyzing existing research 
on CPS security in order to better understand how security is actually addressed 
when dealing with cyber-physical systems. Based on this analysis of the state of 
the art, we also aim at  identifying the implications for future 
research on CPS security.

\noindent \textit{Method:}
In order to achieve this, we designed and conducted a systematic mapping study to 
identify, classify, and compare relevant studies proposing a method or technique 
for cyber-physical systems security. A comparison framework for classifying methods 
or techniques for CPS security has been empirically defined; identified relevant 
studies have been classified on the basis of publication trends, their characteristics 
and focus, and their validation strategies.

\noindent \textit{Results:} We selected a total of 118 primary studies as a result 
of the systematic mapping process. From the collected data we can observe that 
(i) even if solutions for CPS security has emerged only recently, in the last years 
they are gaining a sharply increasing scientific interest across heterogeneous 
publication venues; (ii) the bulk of the works on security for cyber-physical 
systems is focused on power grids, and the approaches considering attacks on 
sensors and their protection completely dominate the scene; regardless of application
field and considered system components, all the works on CPS security deal with 
attacks, in order to either implement or to counteract them, and putting together all 
this studies gives us the possibility to categorize the existing (cyber-physical)
attack models; it comes as surprise that very few papers consider communication 
aspects or imperfections and attempt to provide non-trivial mathematical models 
of the communication; (iii) most advanced and realistic validation methods 
have been exploited in the power networks application domain, but even there 
a benchmark is still missing. 

\noindent \textit{Conclusion:}
The systematic map of research on CPS security provided here is based on, for
instance, application fields, various system components, related algorithms and
models, attacks characteristics and defense strategies. This work presents a
powerful comparison framework for existing and future research on this hot
topic, important for both industry and academia.

\end{abstract} 

\def\cprime{$'$} \newcommand{\noopsort}[1]{} \newcommand{\singleletter}[1]{#1}

\section{Introduction}\label{sec:intro}

Cyber-physical systems (CPS) are integrations of computation, networking, and physical processes \citeref{LeeSeshia10_IntroductionToEmbeddedSystemsCyberPhysicalSystemsApproach,Cardenas:2008:RCS:1496671.1496677}. The key characteristic of cyber-physical systems is their seamless integration of both hardware and software resources for computational, communication and control purposes, all of them co-designed with the physical engineered components \citeref{5512708}.

The economic and societal potential of cyber-physical systems is astonishing, and major investments are being made worldwide to develop the technology \citeref{Lee07_ComputingFoundationsPracticeForCyberPhysicalSystems}. For instance, the December 2010 report of the U.S. President's Council of Advisors on Science and Technology \citeref{PCAST:DesigningDigitalFuture} called for continued investment in CPS research because of its scientific and technological importance as well as its potential impact on grand challenges in a number of sectors critical to U.S. security and competitiveness, including aerospace, automotive, chemical production, civil infrastructure, energy, healthcare, manufacturing, materials and transportation. Also, the anticipated funding to research and education projects on CPS amounts to approximately \$32,000,000 each year \citeref{NSF:CPS}, and the European Union has a similar vision on the importance of research on CPS with fundings focusing on this area. 

Applications of CPS arguably have the potential to dwarf the 20-th century IT revolution \citeref{DBLP:conf/isorc/Lee08a,DBLP:conf/dac/Lee10}. Among the many applications of CPS we can find high confidence medical devices and systems, assisted living, traffic control and safety, advanced automotive systems, process control, energy conservation, environmental control, avionics, instrumentation, critical infrastructure control (electric power, water resources, and communications systems for example), distributed robotics (telepresence, telemedicine), defense, manufacturing, smart structures, etc.

It goes without saying that in this type of systems \textbf{security} is a primary concern and, because of the tight cyber-physical coupling, it is one of the main scientific challenges. Indeed, CPS security is attracting several research efforts from different and independent areas (e.g., secure control, intrusion detection in SCADA systems, etc.), each of them with specific peculiarities, features, and capabilities.

However, if on one side having many research efforts from different and independent areas on CPS security confirms its importance from a scientific point of view, on the other side it is very difficult to have a holistic view on this important domain. Under this perspective, even if the progress of research on cyber-physical systems has started more than ten years ago and the various research communities are very active, \textit{the trends, characteristics, and the validation strategies of existing research on CPS security are still unclear}. With this work we aim at filling this gap.

\textbf{Goal of this work} is to identify, classify, and analyze existing research on cyber-physical systems security in order to better understand how security is actually addressed when dealing with cyber-physical systems. 

In order to tackle our goal we apply a well-established methodology from the Medical and Software Engineering research communities called \textbf{systematic mapping} \citeref{petersen2015guidelines,kitchenham2007guidelines} (see Section \ref{sec:sms}), applying it on the peer reviewed papers which propose and validate a method or technique for CPS security enforcing or breaching.
Through our systematic mapping process, we selected 118 primary studies among more than a thousand entries fitting at best three research questions we identified (see Section \ref{sec:rq}). Then, we defined a classification framework composed of more than 40 different parameters for comparing state-of-the-art approaches, and we applied it to the 118 selected studies. Finally, we analyzed and discussed the obtained data for extracting emergent research challenges and implications for future research on CPS security.
The main \textbf{contributions} of this study are:

\begin{itemize}
\item a reusable \textit{comparison framework} for understanding, classifying, and comparing methods or techniques for CPS security;
\item a \textit{systematic review} of current methods or techniques for CPS security, useful for both researchers and practitioners;
\item a discussion of \textit{emerging research challenges and implications} for future research on CPS security.
\end{itemize}

To the best of our knowledge, this paper presents the first systematic investigation into the state of the art of research on CPS security. 
The results of this study provide a complete, comprehensive and replicable picture of the state of the art of research on CPS security, helping researchers and practitioners in finding trends, characteristics, and validation strategies of current research on security-aware cyber-physical (co-)design, intrusion detection, forecast and response, its future potential and applicability.

The \textbf{main findings} produced by our analysis are discussed below:
\noindent \textit{Publication trends}: 
even if the need for methods and techniques for CPS security has emerged only in 2008, in the last years there is an increasing need and scientific interest on methods and techniques for CPS security. Also, CPS security is turning more and more into a mature field, with more foundational and comprehensive studies published in the recent years. Cyber-physical systems security has a very multidisciplinary nature and it has been broadly considered by researchers with different research interests, such as smart grid, automatic control, communications, networked systems, parallel and distributed systems, etc.

\noindent \textit{Characteristics and focus}: the bulk of the works on CPS security is focused on power grids, while somehow surprisingly, we have not found any work on the cyber-physical security of medical CPS, and only a small part of selected papers is within the application field of secure control of (unmanned) ground vehicles and aerial systems, and of heating, ventilation, and air-conditioning in large functional buildings. All the works considered in this mapping study deal with attacks, in order to either implement or to counteract them: putting together all this studies gives us the possibility to categorize the existing (cyber-physical) attack models. The defense strategies are presented in most of the studies, occupying the central spot of the research efforts on CPS security. More than 90\% of the works are concerned with system integrity, threatened by various types of deception attacks. Regarding the considered system components, the approaches considering attacks on sensors and their protection completely dominate the scene; in fact the resilient state estimation under measurement attacks is a very active research topic within the area of cyber-physical security. Somehow unexpectedly, very few papers consider communication aspects or imperfections and attempt to provide non-trivial mathematical models of the communication; the centralized schemes dominate both attack and defense solutions. 

\noindent \textit{Validation strategies}: most advanced and realistic validation methods have been exploited in the power networks application domain, but even there a benchmark is still missing. Even if the repeatability process, capturing how a third party may reproduce the validation results of the method or technique, is recognized as a good scientific practice, we found no studies providing a replication package. So, we put a particular attention on analysis and description of standard test systems and experimental testbeds used by researchers studying various aspects of CPS security.

By presenting and discussing the above mentioned results we are the first to provide an overview of the state of the art of research in CPS security, thus our work can certainly be useful for both researchers (either young or experienced ones) and practitioners in the field of CPS security.
Finally, we use the results of this study for discussing potential implications for future research on CPS security.

\noindent \textbf{Article outline}.
The article is organized as follows. 
In Section \ref{sec:background} we provide background notions for setting the context
of our study by clarifying and discussing 
(i) cyber-physical systems, 
(ii) CPS security,  
(iii) the methodology we followed (i.e., systematic mapping), and
(iv) related work.
Section \ref{sec:method} describes in details our research methodology in designing, conducting, and documenting the study\footnote{Readers principally interested in the results of our study and future research directions may directly jump to subsequent sections and come back to this section after the first read of the paper.}, followed by a discussion of the obtained results in Sections \ref{sec:trends}, \ref{sec:characteristics} and \ref{sec:validation}. We discuss the implications for future research on CPS security in Section \ref{sec:future} and limitations and threats to validity in Section \ref{sec:threats}. Section \ref{sec:conclusions} closes the article.
%
%
\section{Background}\label{sec:background}

\subsection{Cyber-physical systems}\label{sec:cps}
The term cyber-physical systems (CPS) emerged around 2006, when it was coined at the National Science Foundation (NSF) in the United States \citeref{LeeSeshia10_IntroductionToEmbeddedSystemsCyberPhysicalSystemsApproach}, with the
``cyber" part of the name resulting from the term ``cybernetics", introduced as metaphor apt for control systems \citeref{wiener1965cybernetics}. 

\begin{figure}[htbp]
  \begin{subfigure}[b]{\columnwidth}
     \centering
     \includegraphics[scale=0.1]{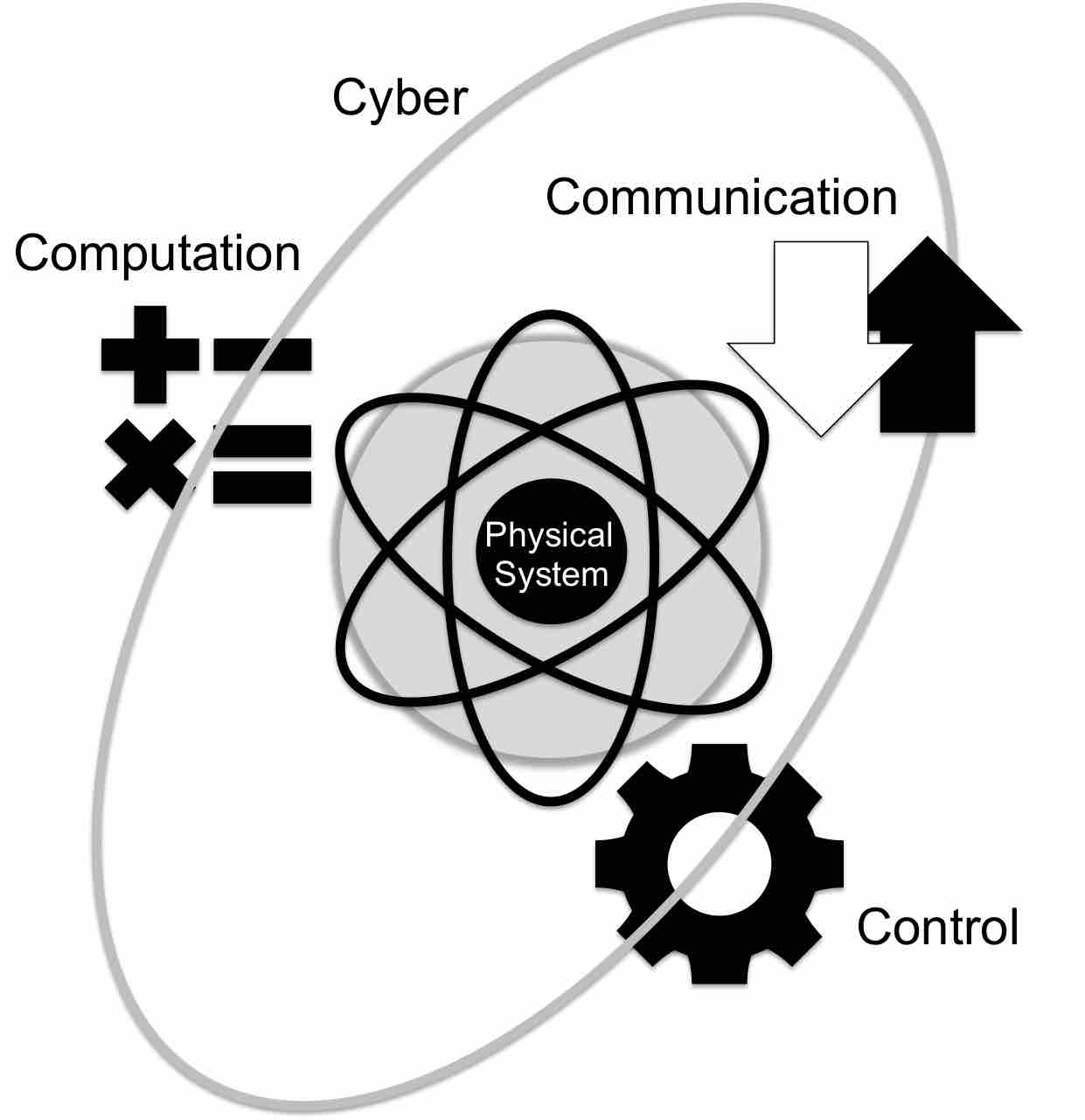}
     \caption{The three main functional components of a CPS}
     \label{fig:cps1}
     \vspace{3mm}
  \end{subfigure}  
  \begin{subfigure}[b]{\columnwidth}
     \centering
     \includegraphics[scale=0.1]{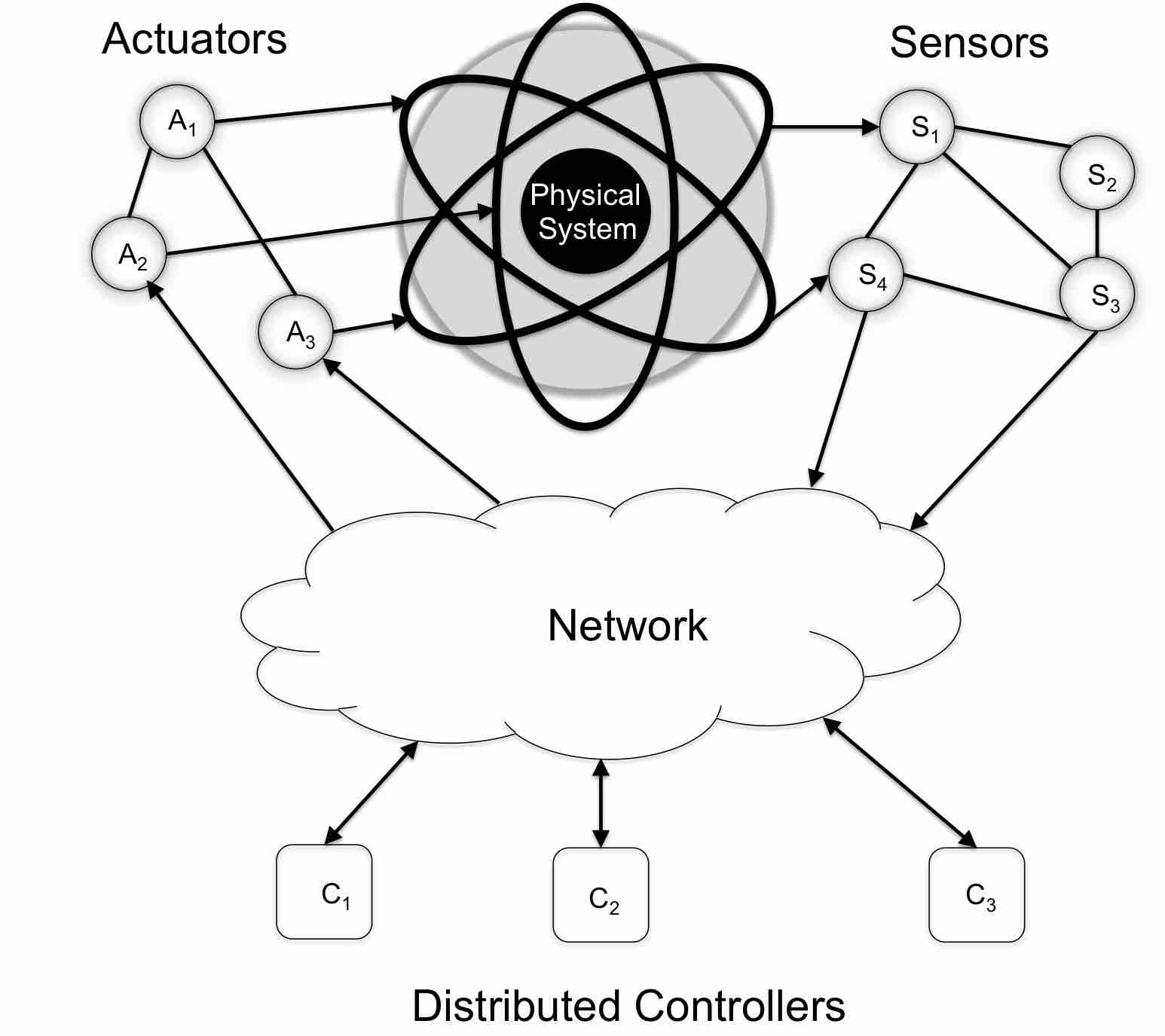}
     \caption{CPS as networked control system~\citeref{4577833}}
     \label{fig:cps2}
  \end{subfigure} 
  \caption{Two main abstractions of cyber-physical systems}
  \label{fig:cps}
\end{figure}

As shown in Figure~\ref{fig:cps}, CPS can be seen as a family of control systems related to the domain of embedded sensor and actuator networks \citeref{Cardenas:2008:RCS:1496671.1496677}, thus close relative of Process Control Systems (PCS) and of Supervisory Control And Data Acquisition (SCADA) systems. However, the {\em seamless integration} of both hardware and software computational, communication and control resources, {\em co-designed} together with physical engineered components \citeref{NistStrategicVisionCPS} is what sets cyber-physical systems discipline apart \citeref{5512708}.

\subsection{Security of CPS}\label{sec:security}
Uncertainty in the environment, security attacks, and errors in physical devices make ensuring overall system security a critical challenge for CPS \citeref{Rajkumar:2010:CSN:1837274.1837461}. Furthermore, a cyber-physical coupling allow sophisticated adversaries to perform attacks threatening also other key attributes of the system, first and foremost safety \citeref{koscher2010experimental, 4531149, 5742014}. This is the reason why, among several crucial requirements of CPS, today many researchers are interested in various (unique) aspects of cyber-physical systems security; for example investigating on combined cyber-physical attack models, reply attacks used to render a pre-defined physical attack to an industrial plant stealthy \citeref{Teixeira:2012:AMS:2185505.2185515}, secure control \citeref{4577833}, anomaly-based intrusion detection \citeref{Denning:1987:IM:22853.22862}, intrusion detection in SCADA systems using multidimensional critical state analysis \citeref{5682374}.

CPS security presents a number of peculiar characteristics that distinguish it from more conventional IT systems security \citeref{NistSP-800-82-Rev2}, \citeref{DHS2009ICS}. For instance, with cyber-physical systems we have real-time requirements, where response is time-critical, modest throughput is acceptable, high delay and/or jitter is not tolerable, and response to human or other emergency interaction is essential. Such systems are often resource-constrained and may not tolerate typical IT security practices. Even the usual definition of security as the combination of three primary security attributes of confidentiality, integrity and availability \citeref{1335465} assumes for the cyber-physical systems a completely new meaning \citeref{4577833}. Given that the estimation and control algorithms used in CPS are designed to satisfy certain {\em operational goals}, such as, closed-loop stability, safety, liveness, or the optimization of a performance function, {\em availability} in CPS can be viewed as the ability to maintain the operational goals by preventing or surviving denial-of-service (DoS) attacks \citeref{Mirkovic:2004:TDA:997150.997156} to the information collected by the sensor networks, the commands given by the controllers, and the physical actions taken by the actuators. Similarly, CPS {\em integrity} aims to maintain the operational goals by preventing, detecting, or surviving deception attacks \citeref{6580348} in the information sent and received by the sensors, the controllers, and the actuators. The intent of {\em confidentiality} in CPS is to prevent an adversary from inferring the {\em state} of the physical system by eavesdropping on the communication channels between the sensors and the controller, and between the controller and the actuator or by means of side channel attacks \citeref{Tiri:2007:SAP:1278480.1278485} on sensors, controllers and actuators.



In the literature there are several approaches addressing the primary security objectives of cyber-physical system availability, integrity and confidentiality. From a high-level point of view, security can be seen as a system-wide concern that takes into account both (i) design for security \citeref{Ravi:2004:SND:996566.996771} and (ii) security mechanisms \citeref{Bishop:2002:ASC:579090}. 

\noindent \textbf{Design for security}. Multiple Independent Levels of Security/Safety (MILS) \citeref{alves2006mils} approach, Defense in Depth \citeref{Bakolas2011184} strategy, and Moving Target Defense \citeref{Casola2014MTD} paradigm, together with classic Saltzer and Schroeder's  considerations \citeref{1451869}, provide relevant design principles. Since cyber-physical systems may be subject to attacks from resourceful adversaries \citeref{Wang:2010:SIC:1953383.1953487}, in the design and analysis of security-aware CPS \citeref{6843720}, it is important to include the trust \citeref{1335465} analysis of the architecture, consider realistic and rational adversary models \citeref{Teixeira:2012:AMS:2185505.2185515}, and employ quantitative security metrics, e.g. \citeref{5482589, 6042046,  sallhammar2007stochastic, 6687271}. To gain confidence in the security and in the correctness of the system design and implementation, formal verification approaches \citeref{6730832, 6386929, wang2014} such as Theorem Proving and/or Model Checking should be applied. For instance, Common Criteria (ISO 15408) standard for Information Technology Security Evaluation requires the use of formal methods for the high Evaluation Assurance Levels (5 to 7). 

\noindent \textbf{Security mechanisms}. A typical-cyber security preventive technical mechanisms \citeref{5590215} related but not specific to CPS include authentication, authorization/access control, accountability, cryptography, and boundary protection. Reactive security mechanisms for cyber-physical systems, a.k.a. intrusion detection \citeref{Mitchell:2014:SID:2597757.2542049}, together with automatic response and recovery, can instead greatly benefit from the particular characteristics of this type of systems, thanks to the possibility to use the models of the physical system \citeref{Cardenas:2008:RCS:1496671.1496677} to reveal anomalies in the behavior.

\subsection{Systematic mapping studies}\label{sec:sms}

A systematic mapping study (or scoping study) is a research methodology particularly intended to provide an \textbf{unbiased, objective and systematic instrument} to answer a set of research questions by finding all of the relevant research outcomes in a specific research area (CPS security in our paper) \citeref{petersen2015guidelines}. 
Research questions of mapping studies are designed to provide an overview of a research area by classifying and counting research contributions in relation to a set of well-defined categories such as publication type, forum, frequency, assumptions made, followed research method, etc.
\citeref{kitchenham2007guidelines,petersen2008systematic}. The mapping process involves searching and analyzing the literature in order to identify, classify, and understand existing research on a specific topic of interest.

In the recent years many researchers are conducting systematic mapping studies on a number of areas and using different guidelines or methods (e.g., on
technical debt \citeref{li2015systematic}, search-base software engineering \citeref{lopez2015systematic}, model-driven engineering for wireless sensor networks \citeref{malavolta2014study}).
In a recent study \citeref{petersen2015guidelines} it emerged that at least ten different guidelines have been proposed for designing the systematic mapping process. We conducted our study by considering the two most commonly accepted and followed guidelines according to \citeref{petersen2015guidelines}, specifically: the ones proposed by Kitchenham and Charters \citeref{kitchenham2007guidelines} and Petersen et al. \citeref{petersen2008systematic}, respectively. Also, we refined our mapping process according to the results of a consolidating update on how to conduct systematic mapping studies proposed by Petersen et al. in 2015 \citeref{petersen2015guidelines}.
Finally, due to the various specificities of existing research on CPS (e.g., the presence of many different definitions of CPS, the intrinsic multidisciplinarity of existing research on CPS, etc.), we found it appropriate to tailor the method and classification schemes proposed in the guidelines according to our topic. The method we followed in our systematic mapping study is detailed in Section \ref{sec:method}.

\subsection{The need for a systematic mapping study on security for CPS}\label{sec:need}
As it was outlined in the introduction, there is a lack of systematic studies on cyber-physical systems security.
In order to ground this claim and establishing the need for performing a mapping study on 
security for cyber-physical systems, we searched a set of electronic data sources (i.e., those listed in Section~\ref{sec:search}), for systematic studies on security-aware cyber-physical co-design, self-protection and related security mechanisms specific to CPS\footnote{Search performed on January 5, 2015.} without any success. None of the retrieved publications was related to any of our research questions detailed in Section \ref{sec:rq}. So, we can claim that our research complements the related works described in Section~\ref{sec:related} to investigate the state-of-research about cyber-physical systems security.

In this systematic mapping study we {\em aim} to identify, classify, and understand existing research on cyber-physical systems security. Those activities will help researchers and practitioners in identifying limitations and gaps of current research \citeref{kitchenham2007guidelines} on security-aware cyber-physical (co-)design, intrusion detection, forecast and response, its future potential, and its potential applicability in the context of real-world projects.

\subsection{Related studies}\label{sec:related}

Cyber-physical systems security within the smart grid domain has been reviewed by Mo, Kim, Brancik, Dickinson, Lee, Perrig and Sinopoli \citeref{6016202} and by Sridhar, Hahn and Govindarasu \citeref{6032699}.

The work from Mo et al. \citeref{6016202} is a good starting point to face the area of CPS security since it gives a broad overview on cyber and system-theoretic approaches to security and shows how a combination of both of them together can provide better security level than traditional methods. The provided example describes defense against replay attack \citeref{5394956} following secure control \citeref{4577833} method. 

The article from Sridhar, Hahn and Govindarasu \citeref{6032699} is more domain-specific. Since power system is functionally divided into generation, transmission, and distribution, the survey considers cyber vulnerabilities and security solutions for each of the underlying fields. Notably, it deals with a wide range of (sophisticated) attacks \citeref{5590115, Liu:2011:FDI:1952982.1952995, 6039809}, some bad data detection techniques \citeref{4112982, bobba2010detecting} and mentions attack resilient control. This work provides also an overview on supporting infrastructure security, with a look on secure communication, device security, security management and awareness, cyber security evaluation, and intrusion tolerance. All in all, the paper identifies the importance of combining both power application security and supporting infrastructure security into the risk assessment process and provides a methodology for impact evaluation. Conclusively, it lists a number of emerging research challenges in risk modeling and mitigation, pointing out the importance of attack resilient control, domain-specific anomaly detection and intrusion tolerance. 

Both of previous surveys \citeref{6016202,6032699} are focused on smart grid domain-specific security. Moreover, based on 
the guidelines for performing systematic literature reviews from Kitchenham and Charters \citeref{kitchenham2007guidelines}, these studies cannot be considered as a systematic literature reviews but as {\em informal literature surveys}.

The intrusion detection techniques for different CPS applications were surveyed by Mitchell and Chen \citeref{Mitchell:2014:SID:2597757.2542049}. 
For each presented intrusion detection system (IDS) design it was analyzed which, if any, distinguishing characteristics of cyber-physical intrusion detection were considered. The unique characteristics of cyber-physical intrusion detection listed in this study are physical process monitoring, closed control loops, attack sophistication and legacy technologies. The conclusion was that there is a lack of IDS techniques that specifically consider most or all distinguishing aspects of CPS. Other notable remark was that behavior-specification-based detection, which formally define legitimate behavior and detects an intrusion when the system departs from this model, has a potential to be the most effective one and deserves more research attention. A similar inference was made by Zhu and Sastry in their survey of SCADA-specific IDS \citeref{zhu2010scada}. Although the works on intrusion detection are relevant for our study, our goal is to give a much broader holistic view on cyber-physical security, and not only on a particular family of mechanisms.
 
\section{Method}\label{sec:method}

\begin{figure*}[!htbp]
	\centering
	\includegraphics[width=1.8\columnwidth]{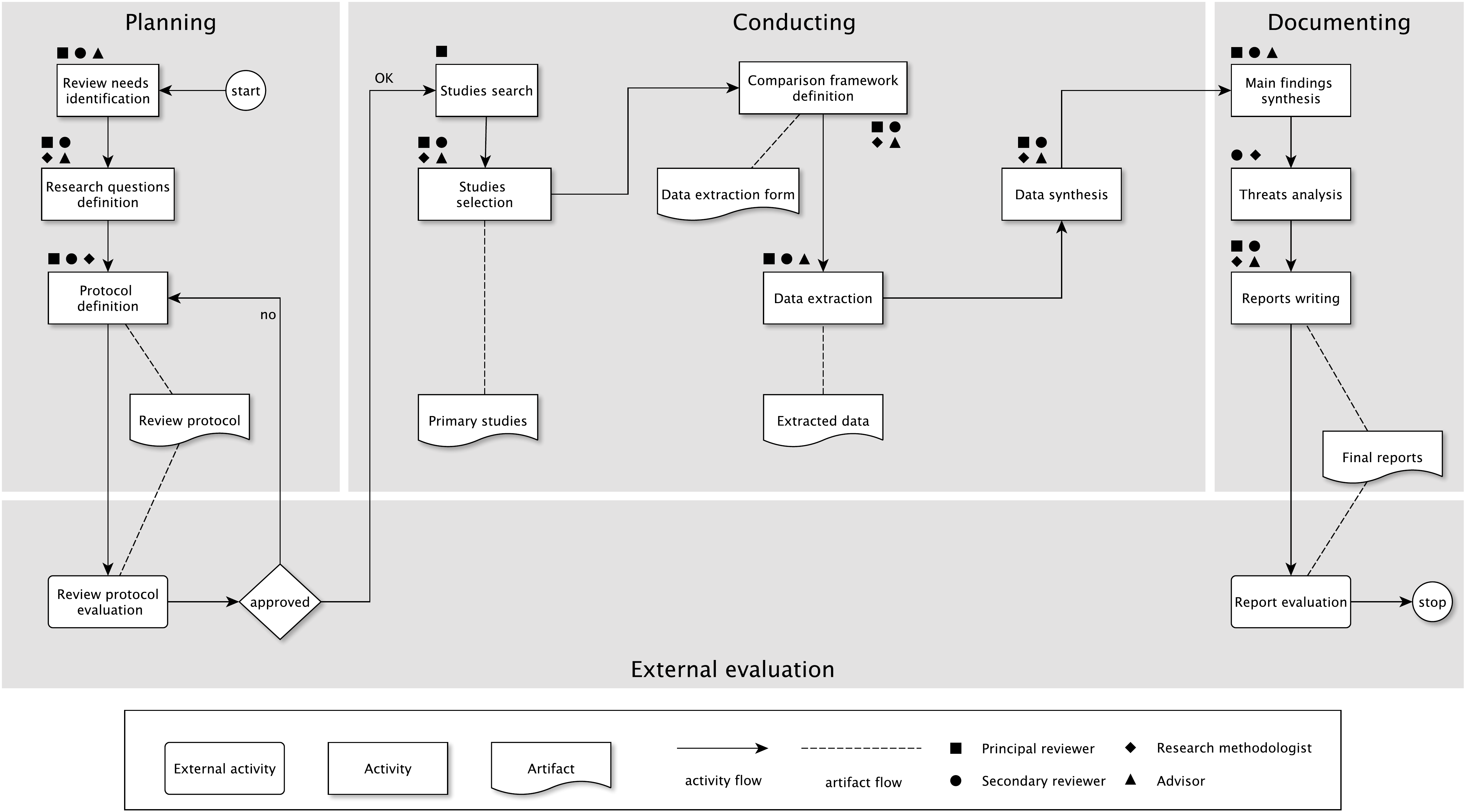}
	\caption{Overview of the whole review process}
	\label{fig:process}
\end{figure*}

Figure~\ref{fig:process} shows the overview of the process we followed for carrying on our study.
The overall process can be divided into three main phases, which are the well-accepted ones for performing a systematic study~\citeref{kitchenham2007guidelines,wohlin2012experimentation}: planning, conducting, and documenting.

Each phase has a number of output artifacts, e.g., the planning phase produces the protocol we followed in our study. In order to mitigate potential threats to validity and possible biases, some of the produced artifacts has been circulated to external experts for independent review. More specifically, we identified two classes of external experts: SLR experts who focused on the overall design of the study and domain experts focusing more on aspects related to security for cyber-physical systems. We contacted and received the feedback of one SLR expert and two domain experts, who reviewed our review protocol and final report independently.

In the following we will go through each phase of the process, highlighting its main activities and produced artifacts.

\subsubsection{Planning}\label{sec:planning}
In addition to establishing the need for performing a mapping study on 
security for cyber-physical systems, in this phase we identified the main research questions (see Section~\ref{sec:rq}), and we produced a well-defined review protocol describing in details the various steps we had to follow in our study. The produced review protocol has been independently evaluated by the previously named SLR- and domain-experts, and it has been refined according to their feedback. The final version of the review protocol is publicly available as part of the replication package of this study\footnote{Replication package of this study: \url{http://cs.gssi.infn.it/CPSSecurity}}.

\subsubsection{Conducting}\label{sec:conducting}
In this phase we set the previously defined protocol into practice. More specifically, 
we performed the following activities:
\begin{itemize}
  \item \textit{Studies search}: we performed a combination of techniques for identifying the comprehensive set of candidate entries on security for cyber-physical systems. Section~\ref{sec:search} will describe in details the search strategy of this research.
  \item \textit{Studies selection}: the candidate entries identified in the previous activity has been filtered in order to obtain the final list of primary studies to be considered in later activities of the protocol.
  The details of this phase are given in Section \ref{sec:search}.
  \item \textit{Comparison framework definition}: in this activity we defined the set of parameters for comparing the primary studies. The main outcome of this activity is the data extraction form, which is a document explaining the possible values and the meaning of each parameter of the comparison framework (see Section~\ref{sec:extraction}). The data extraction form is available as part of the replication package of our study. 
  \item \textit{Data extraction}: In this activity we went into the details of each primary study, and we filled a corresponding data extraction form, as defined in the previous activity. Filled forms has been collected and aggregated in order to be ready to be analyzed during the next activities. More details about this activity will be presented in Section~\ref{sec:extraction}.
  \item \textit{Data synthesis}: this activity focuses on a comprehensive summary and analysis of the data extracted in the previous activity. The main goal of this activity is to elaborate on the extracted data in order to address each research question of our study (see Section~\ref{sec:rq}). This activity involves both quantitative and qualitative analysis of the extracted data. The details about this activity are in Section~\ref{sec:synthesis}.
\end{itemize}

\subsubsection{Documenting}\label{sec:documenting}
This phase is fundamental for reasoning on the obtained findings and for evaluating the quality of the systematic literature review. The main activities performed in this phase are: (i) a thorough elaboration on the data extracted in the previous phase with the main aim at setting the obtained results in their context, (ii) the analysis of possible threats to validity, and (iii) the writing of a set of reports describing the performed mapping study to different audiences. Produced reports have been evaluated by SLR- and domain- experts. This article itself is an example of produced final report.

\subsection{Research questions}\label{sec:rq}
It is fundamental to clearly define the research questions of a systematic literature 
study~\citeref{Brereton2007571}.
Before going into the details of the identified research questions, we formulate the goal of this research by using the Goal-Question-Metric perspectives (i.e., purpose, issue, object, viewpoint~\citeref{gqm}). Table \ref{tab:gqm} shows the result of the above mentioned formulation.

\begin{table}[!htbp]\caption{Goal of this research}
    \vspace*{-1mm}
    \begin{tabular}{p{1.15cm} | p{6.85cm} }
    \textit{Purpose} & Analyze the \\ 
    \textit{Issue} & publication trends, characteristics, and validation strategies \\
    \textit{Object} & of existing methods and techniques for CPS security  \\ 
    \textit{Viewpoint} & from a researcher's point of view. \\[-1mm] 
    \end{tabular}
    \label{tab:gqm}
\end{table}

The goal presented above
can be refined into the following main research questions. For each research question we also provide its primary objective of investigation. The research questions of this study are:

\begin{itemize}
  	\item \textit{RQ1 - What are the publication trends of research studies on cyber-physical systems security?}

			Objective: to classify primary studies in order to assess interest, relevant venues, and contribution types; depending on the number of primary studies, trends can be assessed over the years. 

      \item \textit{RQ2 - What are the characteristics and focus of existing research on cyber-physical systems security?}

      Objective: to analyze and classify all the existing approaches for CPS security with respect to the specific concerns they want to address (e.g., cyber and physical security, secure control, physical-model-based and network-model-based intrusion detection, or any combination of them).
      
      \item \textit{RQ3 - What are the validation strategies of existing approaches for cyber-physical systems security?}
        
        Objective: to analyze and classify all the existing approaches for CPS security with respect to the strategies used for assessing their validity (e.g., controlled experiment, industrial application, prototype-based experiment, test bed, simple examples, correctness by construction, formal proofs). 
\end{itemize}

Answering RQ1 will give a detailed overview about publication trends, venues, and research groups active on the topic.
The classification resulting from our investigation on RQ2 and RQ3 will provide a solid foundation for a thorough comparison of existing and future solutions for cyber-physical systems security.
This contribution is especially useful for researchers willing to further contribute this research area with new approaches to cyber-physical systems security, or willing to better understand or refine existing ones. 

The above listed research questions drove the whole systematic mapping study, with a special
influence on the primary studies search process, the data extraction process, and the data analysis process. 

\subsection{Search strategy}\label{sec:search}

Goal of our search strategy is to detect as much relevant material

\noindent as possible, because leaving relevant results out of a systematic literature study may lead to inaccurate evidence, thus resulting in an internal threat to validity \citeref{4343749}. 

\begin{figure}[!htbp]
	\centering
	\includegraphics[width=\columnwidth]{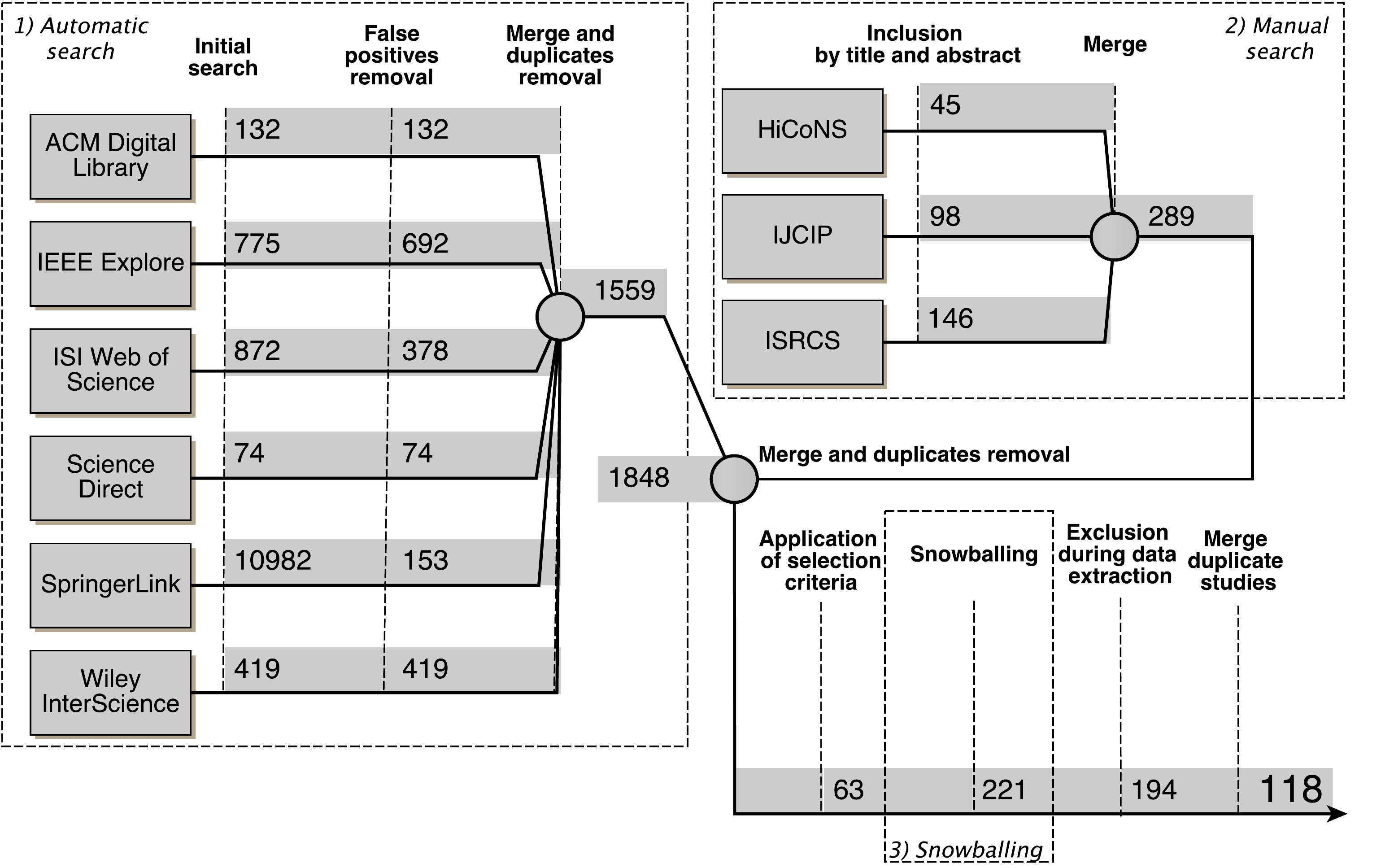}
	\caption{Overview of the search and selection process}
	\label{fig:search}
\end{figure}

Figure \ref{fig:search} shows the details about our search strategy.
In order to achieve maximal coverage, our search strategy consists of three complementary methods: an automatic search, a manual search, and the snowballing. 

\subsubsection{Automatic search}
It refers to the execution of a search string on a set of electronic databases and indexing systems, in the literature it is the dominant method for identifying potentially relevant papers \citeref{Chen:2010:TEU:2227057.2227074}. The applied search string is the following:
\begin{center}
\texttt{((((``cyber physical" OR ``cyber-physical" OR cyberphysical OR ``networked control") AND system*) OR CPS OR NCS) AND (attack* OR secur* OR protect*))}
\end{center}

In the spirit of Zhang, Babar and Tell \citeref{zhang2011identifying}, we established a {\em quasi-gold standard} (QGS) for creating a good search string for the automatic search. 
This procedure requires a manual search in a small number of venues (see Table \ref{manual-search-venues}) and the results of these manual searches have been treated as a QGS by cross-checking the results obtained from the automatic search. 
So, we iteratively defined and modified the search string and conducted automatic searches on the electronic data sources until the quasi-sensitivity was above the established threshold of $80\%$.
When the {\em quasi-sensitivity} became greater than $80\%$, the search performance was considered acceptable and the results from the automated search have been merged with the QGS. The details of the above mentioned process are provided in the replication package of this study.

In this stage it was fundamental to select papers objectively so, following the suggestions from Wohlin et al. \citeref{wohlin2012experimentation}, two researchers assessed a 
random sample of the studies and the inter-researcher agreement has been measured using
the Cohen Kappa statistic~\citeref{cohen1968weighted}.
Each disagreement has been discussed and resolved, with the intervention of the team administrator, if necessary, until the Cohen Kappa statistic reached a result above or equal to $0.80$.

Our automatic search is performed on the six electronic data sources listed in Table \ref{automatic-search-libraries}.
As suggested in \citeref{kitchenham2013systematic}, in order to cover as much relevant literature as possible, we chose six of the largest and most complete scientific databases and indexing systems available in computer science. 
The selection of these electronic databases and indexing systems is guided also by their high accessibility and their ability to export search results to well-defined formats.

\begin{table}[!htbp]\caption{Electronic data sources targeted with search strings}
\begin{center}
   \begin{tabular}{| l | l |} 
   \hline
   {\bf Library} & {\bf Website} \\ \hline \hline   
   ACM Digital Library & \url{http://dl.acm.org} \\ \hline
   IEEE Explore & \url{http://ieeexplore.ieee.org} \\ \hline  
   ISI Web of Science & \url{http://apps.webofknowledge.com} \\ \hline 
   ScienceDirect &  \url{http://www.sciencedirect.com} \\ \hline
   SpringerLink & \url{http://link.springer.com} \\ \hline
   Wiley InterScience & \url{http://onlinelibrary.wiley.com/+} \\ \hline
   \end{tabular}
   \label{automatic-search-libraries}
   \end{center}
\end{table}

Among the results of the automatic searches we removed a set of  \textit{false positives} in order to work on a polished set of potentially relevant studies (see Figure \ref{fig:search}). Examples of false positives include proceedings of conferences or workshops, tables of contents, maps, lists of program committee members, keynotes, tutorial or invited talks, and messages from (co-)chairs. As shown in Figure \ref{fig:search}, our automatic search resulted in 1559 potentially relevant studies.

For the sake of replicability, we provide all the details, data, and results of our automatic search in the \textit{Automatic search} report in the replication package of this study. 

\subsubsection{Manual search} 
By following the quasi-gold standard procedure defined in \citeref{zhang2011identifying}, we (i) identified a subset of important venues for the domain of cyber-physical systems security (they are shown in Table~\ref{manual-search-venues}), and (ii) we performed a {\em manual search} of relevant publications in those venues.
The search have been performed by considering {\em title} and {\em abstract} of each publication and the considered time interval is between December 2008 and November 2014 (since the earliest of above mentioned venues dates back to December 2008).
By referring to Figure \ref{fig:search}, we manually searched and selected 289 potentially relevant studies.

\begin{table}[!htbp]\caption{Selected venues for manual search}
\begin{center}
   \begin{tabular}{| p{6.5cm} | l |} 
   \hline
   {\bf Venue} & {\bf Publisher} \\ \hline \hline
   International Conference on High Confidence Networked Systems (HiCoNS) & ACM \\ \hline
   International Journal of Critical Infrastructure Protection (IJCIP) & Elsevier \\ \hline
   International Symposium on Resilient Control Systems (ISRCS) & IEEE \\ \hline
   \end{tabular}
   \label{manual-search-venues}
\end{center}
\end{table}

The outcomes of the automatic and manual searches have been suitably merged in order to have one single source of information for the subsequent selection and snowballing activities. After merging all the studies and removing duplicates we obtained 1848 potentially relevant studies.
In order to further restrict the number of studies to be considered during the snowballing activity, we applied the selection process depicted in Section \ref{sec:selection} to the current set of studies, thus obtaining 63 potentially relevant studies.
For the sake of replicability, we provide all the details, data, and results of our manual search in the \textit{Manual search} report in the replication package of this study. 

\subsubsection{Snowballing} 
We applied the {\em snowballing} technique for identifying additional sources published in other journals or venues \citeref{Greenhalgh:2005},
which may not have been considered during the automatic and manual searches.
So, as recommended in \citeref{Jalali:2012:SLS:2372251.2372257}, we applied (backward and forward) snowballing on the primary studies selected by the automatic and manual searches. 
More specifically, we considered all the studies selected by the automatic and manual searches and we automatically searched all the papers referring them (i.e., forward snowballing \citeref{Wohlin:2014:GSS:2601248.2601268}); then, we scrutinized also the references of each selected study to identify important studies that might have been missed during the initial search (i.e., backward snowballing \citeref{Wohlin:2014:GSS:2601248.2601268}).
We provide all the details, data, and results of our snowballing activities in  the \textit{Snowballing} report in the replication package of this study.

In all considered search methods we examined {\em title}, {\em keywords} and {\em abstract}.

\subsection{Selection strategy}\label{sec:selection}
As shown in Figure \ref{fig:search}, after the search activity we considered all the collected studies and filtered them according to a set of well-defined inclusion and exclusion criteria.
 In the following we provide the inclusion criteria of our study: 
\begin{itemize}
  \item Studies focusing on security of cyber-physical systems. 
  \item Studies proposing a method or technique for cyber-physical system security enforcing or breaching. 
  \item Studies providing some kind of validation of the proposed method or technique 
        (e.g., via formal analysis, controlled experiment, exploitation in industry, example usage).
\end{itemize}

The exclusion criteria of our study are:
\begin{itemize}
  \item Studies not subject to peer review~\citeref{wohlin2012experimentation} 
        (e.g., journal papers, papers published as part of conference proceedings are considered, whereas white papers are discarded).
  \item Studies written in any language other than English. 
  \item Studies focusing on security method or technique not specific to cyber-physical system (e.g studies focusing on either the physical or cyber part only of the system under consideration).
  \item Studies published before 2006 
        (because the cyber-physical systems discipline has emerged in 2006).
  \item Secondary or tertiarty studies (e.g., systematic literature reviews, surveys, etc.).  
  \item Studies in the form of tutorial papers, short papers, poster papers, editorials, because they do not provide enough information.
\end{itemize}

In this context, a study was selected as a primary study if it satisfied \textit{all} inclusion criteria, and it was  discarded if it  met \textit{any} exclusion criterion.
In order to reduce bias, the selection criteria of this study have been decided during the review protocol definition (thus they have been checked by three external reviewers).

In order to handle studies selection in a cost effective way we used the adaptive reading depth~\citeref{mapping_se}, as the full-text reading of clearly excluded approaches is unnecessary. So, we considered {\em title}, {\em keywords} and {\em abstract} of each potentially relevant study and, if selection decision could not be made, other information (like {\em conclusion} or even {\em full-text}) have been exploited \citeref{zhang2011identifying}. By following the approach proposed in \citeref{ali2014evaluating}, two researchers classified each potentially relevant study either as \textit{relevant}, \textit{uncertain}, or \textit{irrelevant}; any study classified as \textit{irrelevant} has been directly excluded, whereas all the other approaches have been discussed with the help of a third researcher.

When reading a primary study in details for extracting its information, researchers could agree that the currently analysed study was semantically out of the scope of our research, and so it has been excluded (see the \emph{Exclusion during Data Extraction} stage in Figure \ref{fig:search}), resulting in 194 potentially
primary studies. 

As suggested in \citeref{wohlin2012experimentation}, if a primary study was published in more then one paper (e.g., if a conference paper has been extended to a journal version) then we considered only one reference paper as primary study; in those cases we considered all the related papers during the data extraction activity in order to obtain all the necessary data \citeref{kitchenham2007guidelines}.
The final set of primary studies is composed of 118 entries, the detailed list of our primary studies is provided in Appendix \ref{app:primary}.

\subsection{Data extraction}\label{sec:extraction}
Data extraction refers to the recording of all the relevant information from the primary studies required to answer the research questions \citeref{wohlin2012experimentation}.
Before analysing each primary study, we defined a \textit{comparison framework} for classifying research studies on cyber-physical systems security. 

To help the definition of a sound and complete comparison framework, we selected and adapted suitable dimensions and properties found in existing surveys and taxonomies related to CPS security, such as those proposed in \citeref{Yuan2014SSS,1335465,6309293,Yampolskiy:2013:TDC:2461446.2461465}.
In addition, we defined several parameters for classifying methods and techniques for CPS security; we grouped those parameters into three main dimensions: method or technique's {\bf Positioning}, {\bf Characterisation} and {\bf Validation}.

The \textbf{Positioning} dimension characterizes the objectives and intent of existing research on CPS security (the $WHAT$ aspect of each method or technique). For example, this dimension includes the following parameters:
\begin{itemize}
\item {\em CPS application field}, such as power distribution, unmanned aerial systems, etc.;
\item {\em considered security attributes} like availability, integrity, and confidentiality;
\item {\em system components}, including sensors, actuators, network, controllers and plant.
\end{itemize}

The \textbf{Characterization} dimension concerns the classification of studies based on $HOW$ CPS security is addressed in research. It include several parameters, like:
\begin{itemize}
\item {\em theoretical foundations}, such as control theory, compressed sensing, graph theory, computational complexity, etc.;
\item {\em defense strategy}, like detection, mitigation, protection-based prevention, etc.
\end{itemize}

The \textbf{Validation} dimension concerns the strategies researchers apply for providing evidence about the validity of proposed methods or techniques for CPS security. 
Examples of relevant parameters of this dimension are the following:
\begin{itemize}
\item {\em Simulation test systems}, such as IEEE 24-bus reliability test system, Tennessee Eastman challenge, etc.;
\item {\em Repeatability}, to capture how a third party may reproduce the validation results of the method or technique. We considered the repeatability of a study as {\em high} when the authors provide enough details about the steps performed for evaluating or validating the study, the developed or used software, the used or simulated testbed, if any, and any other additional resource; {\em low} otherwise.
\end{itemize}

All the dimensions and parameters of our comparison framework have been encoded in a dedicated \textit{data extraction form}, which can be seen as the implementation of a \textit{comparison framework}.
The final data extraction form is composed of a list of attributes representing the set of data items extracted from the primary studies. 
Our data extraction form has been designed to collect such information from each primary study; it includes both standard information (such as name of reviewer, date of data extraction, title, authors and publication details of the study) \citeref{kitchenham2007guidelines} and the set of parameters to compare the primary studies according to the three dimensions described above (e.g., the used state estimation model, attack model, experimental testbed, etc.). For the sake of brevity we do not provide the description of all the parameters of our data extraction form, we will briefly elaborate on each of them while discussing the results of this study in Sections \ref{sec:trends}, \ref{sec:characteristics} and \ref{sec:validation}; the interested reader can refer to the \textit{Data extraction form} document of our replication package for thorough and extensive discussion of all parameters of our classification framework.

As suggested in \citeref{wohlin2012experimentation}, the data extraction form (and thus also the classification framework) has been independently piloted on a sample of primary studies by two researchers, and iteratively refined accordingly. Then, the data extraction activity has been conducted by two researchers who manually filled a copy of the data extraction form for each primary study; the overall effort to complete this activity can be estimated as 3 man-months with full-time commitment. 

\subsection{Data synthesis}\label{sec:synthesis}
The data synthesis activity involves collating and summarizing the data extracted from the primary studies with the main goal of understanding, analyzing, and classifying 
current research on security for cyber-physical systems \citeref[$\S$~6.5]{kitchenham2007guidelines}. 

We analyzed the extracted data to find trends and collect information about each research question of our study.
Depending on the parameters of the classification framework (see Section~\ref{sec:extraction}), in this research we applied both quantitative and qualitative synthesis methods, separately. When considering qualitative data, we applied the \textit{line of argument} synthesis~\citeref{wohlin2012experimentation}, that is: firstly we analyzed each primary study individually in order to document it and tabulate its main features with respect to each specific parameter of the classification framework defined in Section~\ref{sec:extraction}, then we analyzed the set of studies as a whole, in order to reason on potential patterns and trends.
When both quantitative and qualitative analyses have been performed, we integrated their results in order to explain quantitative results by using qualitative results~\citeref[$\S$~6.5]{kitchenham2007guidelines}. 
In the following sections we present the results of our analysis of the 
extracted data. In total 118 publications have been selected and analyzed as the subjects of
our study. For the sake of clarity we organized the results of the analysis according to our research questions (see Section~\ref{sec:rq}). 


\section{Results - Publication Trends (RQ1)}\label{sec:trends}
In order to assess the publication trends about security for cyber-physical systems we identified
a set of variables focusing on the publication and bibliographic data of each primary study. For each primary study we collected its title, authors, authors' institutions, authors' countries, publication year, publication venue (i.e., journal, conference, workshop, book), as well as  other bibliographic data. In the following we describe the main facts emerging from our analysis.

\subsection{Publication timeline}

\begin{figure}[!htbp]
	\centering
	\includegraphics[width=\columnwidth]{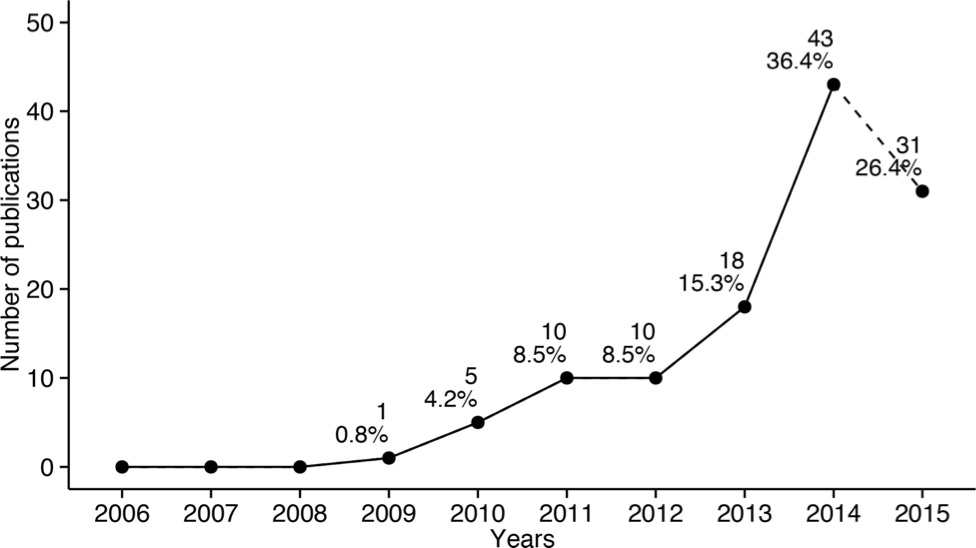}
	\caption{Distribution of primary studies by year (partial data for 2015)}
	\label{fig:years}
\end{figure}

Figure~\ref{fig:years} presents the distribution of the selected publications\footnote{See Section \ref{sec:selection} for details on selection strategy, which, of course, determined the results presented here.} on security for cyber-physical systems over the time period from 2006 to 2015. The first interesting result of our study is the growth of the number of those publications in the last years. Indeed, we can observe that there was a relatively low number of publications on this topic over the time period from 2006 (zero publications) to 2010 (5 publications). Starting from 2011, we see a continuous growing trend over the years, culminating in the 2014 and 2015 years, which together amount for the 61.8\% of the selected studies. 

From the collected data, we can offer the following observations: 
\begin{itemize}
  \item there are no selected studies until 2009; this may be because the main concepts and research interest on cyber-physical systems emerged only around 2006 \citeref{LeeSeshia10_IntroductionToEmbeddedSystemsCyberPhysicalSystemsApproach}, and the need for methods and techniques for CPS security has emerged only recently; 
  \item there is a sharp increase in the number of selected studies between 2012 and 2014; we can trace this observation to the fact that (i) in the last years methods and techniques for CPS security are gaining increasing interest and attention from a scientific point of view and (ii) methods and techniques for CPS security are getting urgently needed to produce industry-ready systems with the required levels of security and reliability;
  \item our study covers the studies published before April 2015; nevertheless, in this year 31 studies have been already published on CPS security, representing the 26.4\% of the whole set of primary studies of our research; this result further confirms the growing attention and need of research on CPS security; we expect that this growing trend will continue;
  \item finally, we can notice that 117 (99.2\%) out of the 118 selected studies were published during the last five years; this can be seen as an indication that CPS security is a relatively new area, which is gaining more and more traction from a scientific point of view; this observation is further strengthened by the fact that the highest slope is between 2013 and 2014, where the number of publications has more than doubled, going from 18 (15.3\%) to 43 (36.4\%).
\end{itemize}

\subsection{Publication venues}

\begin{figure}[!htbp]
	\centering
	\includegraphics[width=.65\columnwidth]{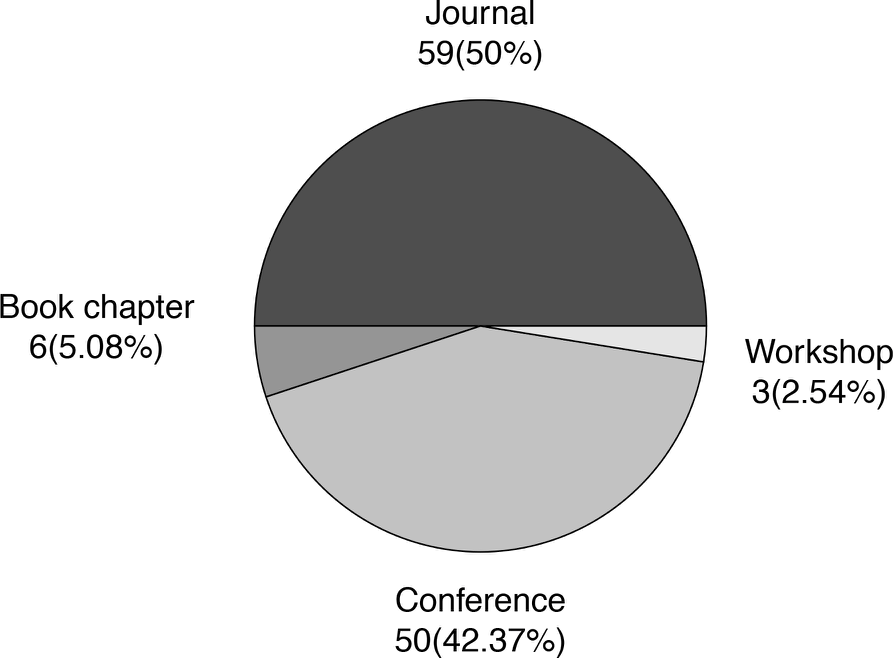}
	\caption{Distribution of primary studies by type of publication}
	\label{fig:pub_types1}
	\end{figure}

In accordance with our selection strategy, we selected publications which have been subject to peer review. Indeed, each primary study was published either as a journal paper, conference paper, workshop paper, or book chapter. Figure \ref{fig:pub_types1} shows the distribution of primary studies over their publication types. The most common publication types are journal and conference, with 59 (50.01\%) and 50 (42.37\%) of the primary studies, respectively. Book chapter and workshop are the least popular publication types, with only 6 (5.08\%) and 3 (2.54\%) studies falling into their categories, respectively. Such a high number of journal and conference papers on CPS security may indicate that CPS security is becoming more and more a mature research theme, despite its relative young age (the first publication on CPS security was in 2009). 

	Moreover, the very low number of workshop papers may be an indication of two facts: on one side researchers on CPS security are valuing more other types of publications (e.g., journal papers), given the high effort and skills required to contribute in this research area; on the other side, it may be an indication that actually the research community on CPS security still does not have a clearly defined identity, and a symptom of this situation may be the lack of a workshop or conference fully dedicated to CPS security. We will detail more on this aspect when analyzing the targeted publication venues (see Table \ref{tab:venues}).
	
\begin{figure}[!htbp]
   \centering
   \includegraphics[width=\columnwidth]{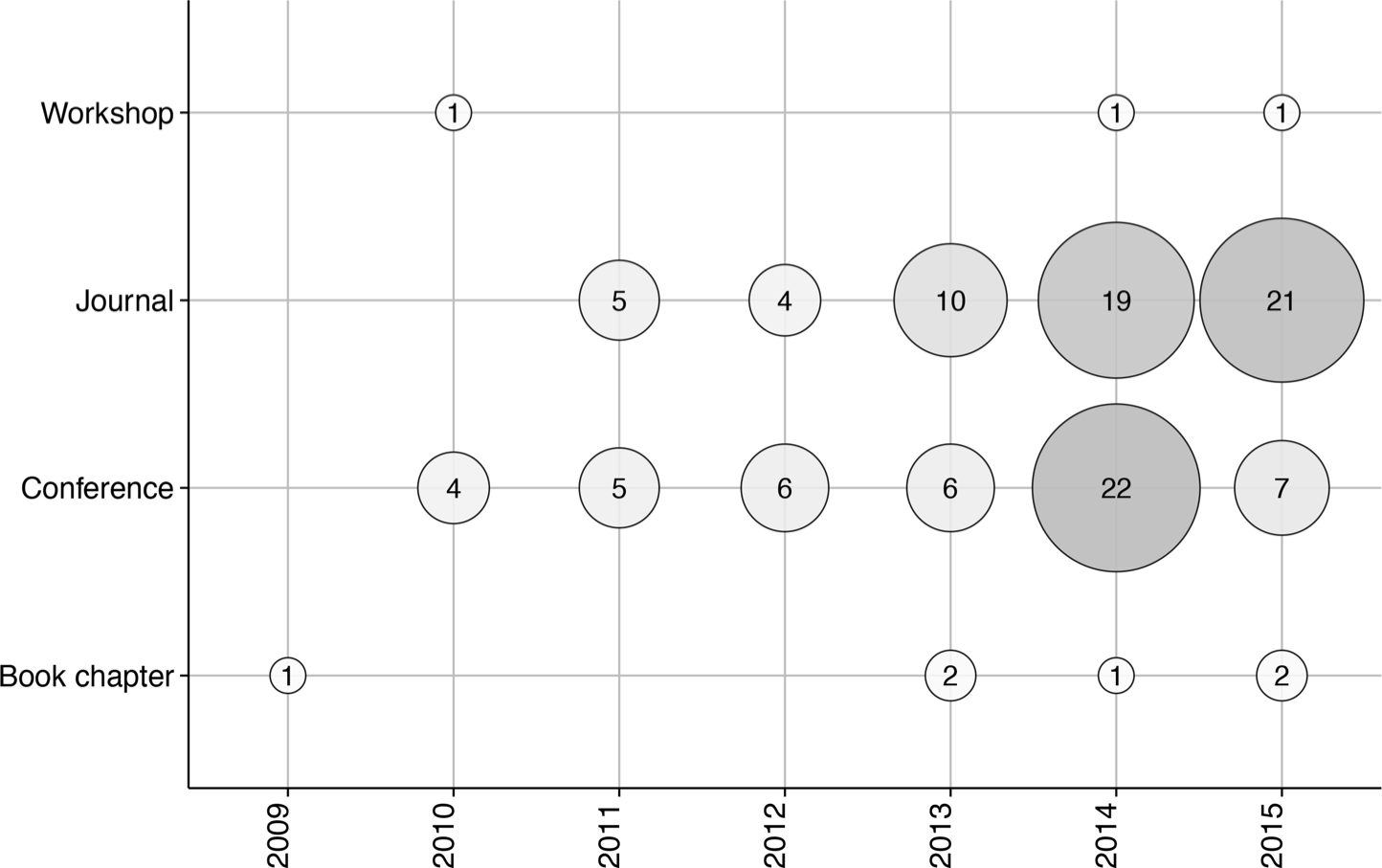}
   \caption{Distribution of primary studies by type of publication and over the years (partial data for 2015)}
   \label{fig:pub_types2}
\end{figure}
	
For what concerns the evolution of publication types of the years, Figure \ref{fig:pub_types2} shows that there is a growing trend in the publications in journals and conference proceedings, as 84 out of 118 studies are journal and conference papers published between 2013 and 2015. Also, almost all book chapters have been published between 2013 and 2015 (5 out of 6 book chapters). Again, this may be a further confirmation that CPS security is turning more and more into a mature field, with more foundational and comprehensive studies published in the recent years.

By looking at the specific targeted publications venues we can notice that research on CPS security is published across a number of venues spanning different research areas, such as automatic control, networked systems, smart grid, security for information systems. Indeed, the 118 selected papers of our study were published at 53 different venues. Table \ref{tab:venues} shows the publication venues with more than one selected study, specifying venue name, type, and number of selected studies.

\begin{savenotes}
\begin{table}[!htbp]\caption{Publication venues with more than one selected study}
\scriptsize
\begin{center}
   \begin{tabular}{| p{4.92cm} | p{1.15cm}  | p{1.3cm}  |} \hline
   {\bf Publication venue} & {\bf Type} & {\bf \#Studies}  \\ \hline \hline
   IEEE Transactions on Smart Grid & Journal & 19 (16.10\%) \\ \hline
   IEEE Conference on Decision and Control (CDC) & Conference & 11 (9.32\%) \\ \hline
   IEEE Transactions on Automatic Control & Journal & 9 (7.62\%) \\ \hline
   American Control Conference (ACC) & Conference & 6 (5.08\%) \\ \hline
   IEEE Journal on Selected Areas in Communications & Journal & 6 (5.08\%) \\ \hline
   IEEE Conference on Smart Grid Communications (SmartGridComm) & Conference & 6 (5.08\%) \\ \hline
   International Conference on High Confidence Networked Systems (HiCoNS) & Conference & 4(3.38\%)\footnote{The HiCoNS conference has been merged into the International Conference on Cyber-Physical Systems (ICCPS) since 2015.} \\ \hline
   IEEE Control Systems & Journal & 3 (2.54\%) \\ \hline
   Global Communications Conference (GLOBECOM) & Conference & 3 (2.54\%) \\ \hline
   IEEE Transactions on Parallel and Distributed Systems & Journal & 3 (2.54\%) \\ \hline
   IEEE Transactions on Power Systems & Journal & 3 (2.54\%) \\ \hline
   Automatica & Journal & 2 (1.69\%) \\ \hline
   ACM Symposium on Information, Computer and Communications Security (ASIACCS) & Conference & 2 (1.69\%) \\ \hline
   Cyber Physical Systems Approach to Smart Electric Power Grid & Book & 2 (1.69\%) \\ \hline
   International Journal of Systems Science & Journal & 2 (1.69\%) \\ \hline \hline
   \textbf{TOTAL} & - & \textbf{81 (68.64\%)} \\ \hline
\end{tabular} \label{tab:venues}
\end{center}
\end{table}
\end{savenotes}

Firstly, the clear winner is the \textit{IEEE Transactions on Smart Grid}, with a total of 19 studies out of 118, representing the 16.10\% of all selected studies; then the \textit{IEEE Conference on Decision and Control (CDC)} and the \textit{IEEE Transactions on Automatic Control} follow with 11 and 9 studies, respectively. Those publication venues can be considered as the de facto leading venues for publishing studies on CPS security. Other publication venues follow until reaching a total number of 81 selected studies, which represent the 68.64\% of all selected studies. From the collected data we can offer the following observations:
\begin{itemize}
  \item the most targeted venues are heterogeneous and pertain to different research areas, such as smart grid, automatic control, communications, networked systems, parallel and distributed systems, etc.; this is a clear indication of the very multidisciplinary nature of cyber-physical systems, even in a specific sub-area like CPS security; this finding indicates also that CPS security has been broadly considered by researchers with different research interests;
  \item according to two well-acknowledged international rankings the most targeted venues for CPS security are all top-level and very reputable in their research area. Indeed, all journals are ranked in the first quartile according to the SCImago Journal Rank (SJR) indicator \citeref{GonzalezPereira2010379}, and all conferences are ranked either as A or B according to the computer science conference rankings (CORE)~\citeref{core} (depending on data availability); 
  \item interestingly, there is a whole book in the set of most targeted venues and it is the sole publication venue specifically targeted to research on CPS. The book is titled \textit{Cyber Physical Systems Approach to Smart Electric Power Grid} \citeref{khaitan2015cyber} and it has been published in 2015. It aims at presenting the recent advances in the field of modeling, simulation, control, security and reliability of CPS in power grids; this book can be a useful reading for current and future researchers in the area of CPS security, with a special emphasis on power grids. 
\end{itemize}

\subsection{Research institutions}


Research on CPS security is pursued in different research institutions worldwide,
with a high degree of collaboration across institutions.
Indeed, our study reveals that 127 unique research institutions have been involved in at least one
selected study, and that in average 1.79 research institutions were involved for each selected study.

\begin{figure}[!htbp]
	\centering
	\includegraphics[width=\columnwidth]{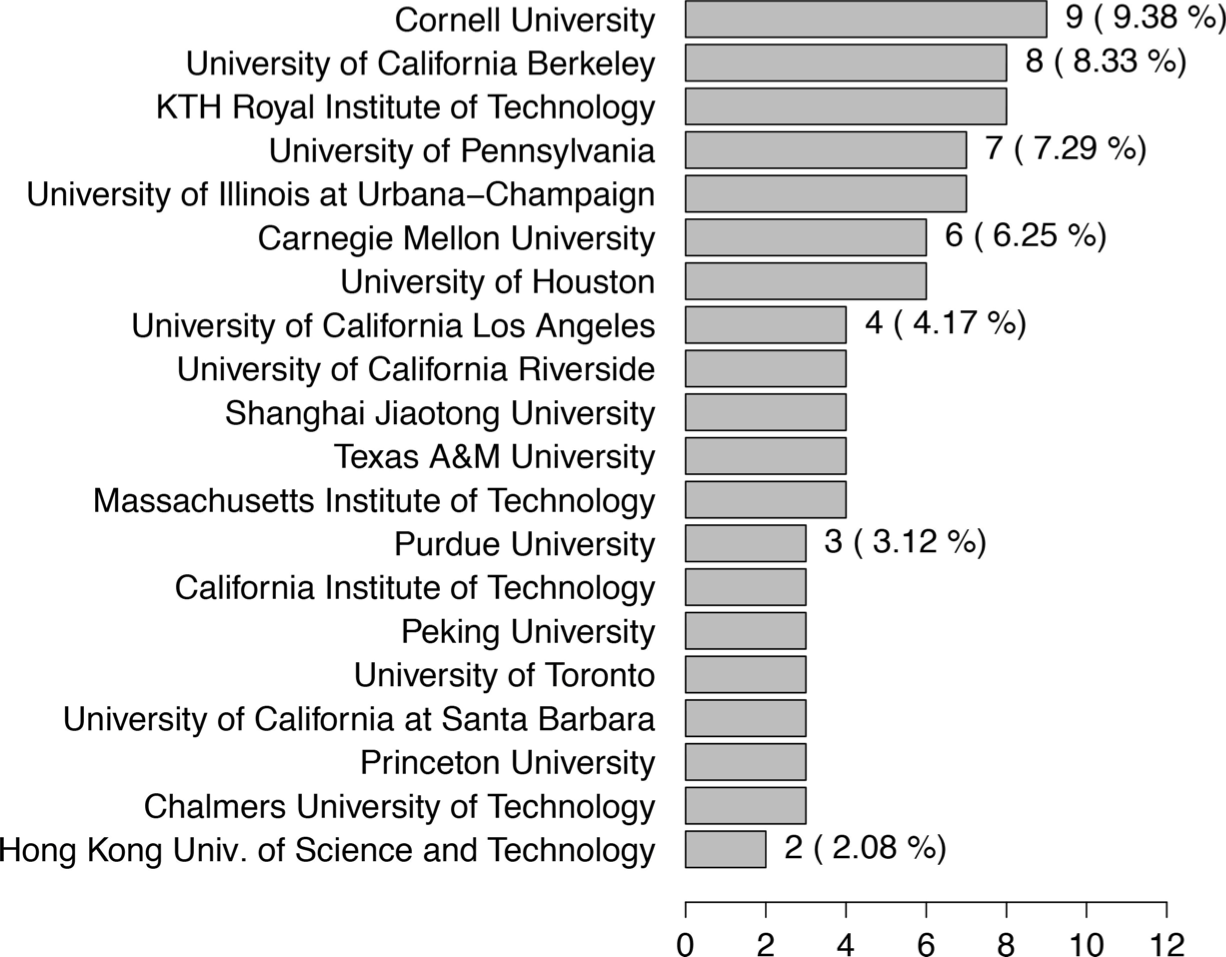}
	\caption{Distribution of primary studies by institution (top 20)}
	\label{fig:institutions}
\end{figure}

Figure \ref{fig:institutions} focuses on the top 20 research institutions involved in at least one selected publication. Our study reveals that the three most active research institutions on CPS security are: the \textit{Cornell University} (USA), the \textit{University of California Berkeley} (USA), and the \textit{KTH Royal Institute of Technology} (Sweden), with 9 (9.38\%), 8 (8.33\%), and 8 (8.33\%) publications, respectively. In addition to the results and discussion of this study, interested and future researchers on CPS security can use the list of institutions as a reference for identifying relevant literature on the topic.
For a complete overview of the data, interested readers can refer to the complete list of research institutions and authors in the replication package of this study. 

\section{Results - Characteristics and Focus of Research (RQ2)}\label{sec:characteristics}

As already introduced in Section~\ref{sec:extraction}, we identified a set of variables describing positioning and characterization of methods and techniques for cyber-physical systems security breaching and/or enforcing. With the purpose of evaluating what aspects of system are attacked or protected by an approach, in the following we indicate which application fields, points of view, security attributes, system components, plant models, state estimation and anomaly detection algorithms, controllers, communication aspects and network-induced imperfections are considered by each primary study. Furthermore, we give an account of the used time-scale models, attacks and their characteristics, attack and defense schemes, plant models used by an attacker, defense strategies and theoretical foundations, in order to understand how these methods and techniques are characterized.

In the remainder of this paper we will use area-proportional Euler diagrams \citeref{micallef2014euler} for visualizing the distribution over parameters with multiple values in which the discussion of their intersections is relevant for this study.

\subsection{CPS application field} \label{subsec:app-field}

\begin{figure}[!htbp]
	\centering
	\includegraphics[width=0.8\columnwidth]{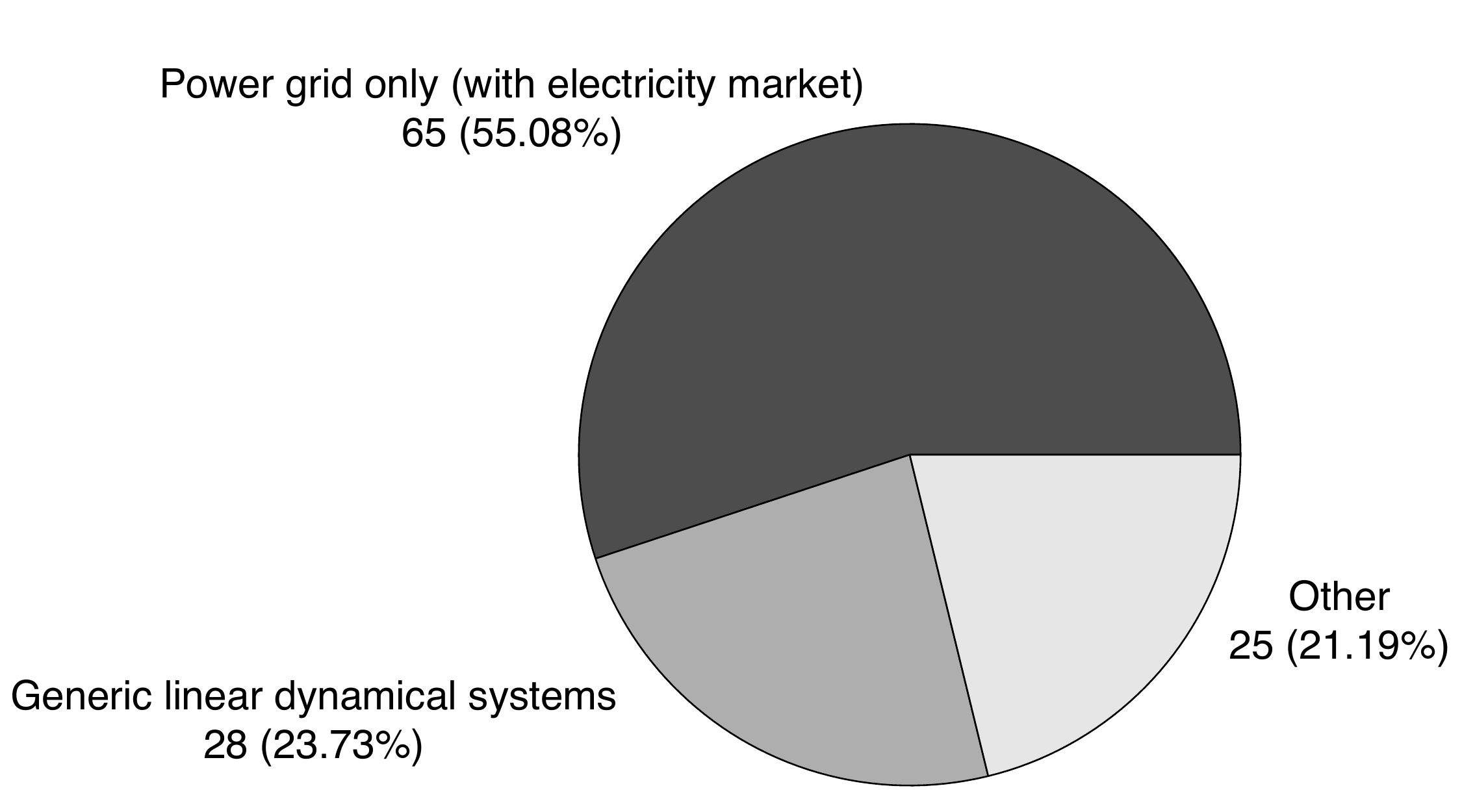}
	\caption{Distribution of primary studies by application area}
	\label{fig:application-fields}
\end{figure}

\begin{figure}[!htbp]
	\centering
	\includegraphics[width=\columnwidth]{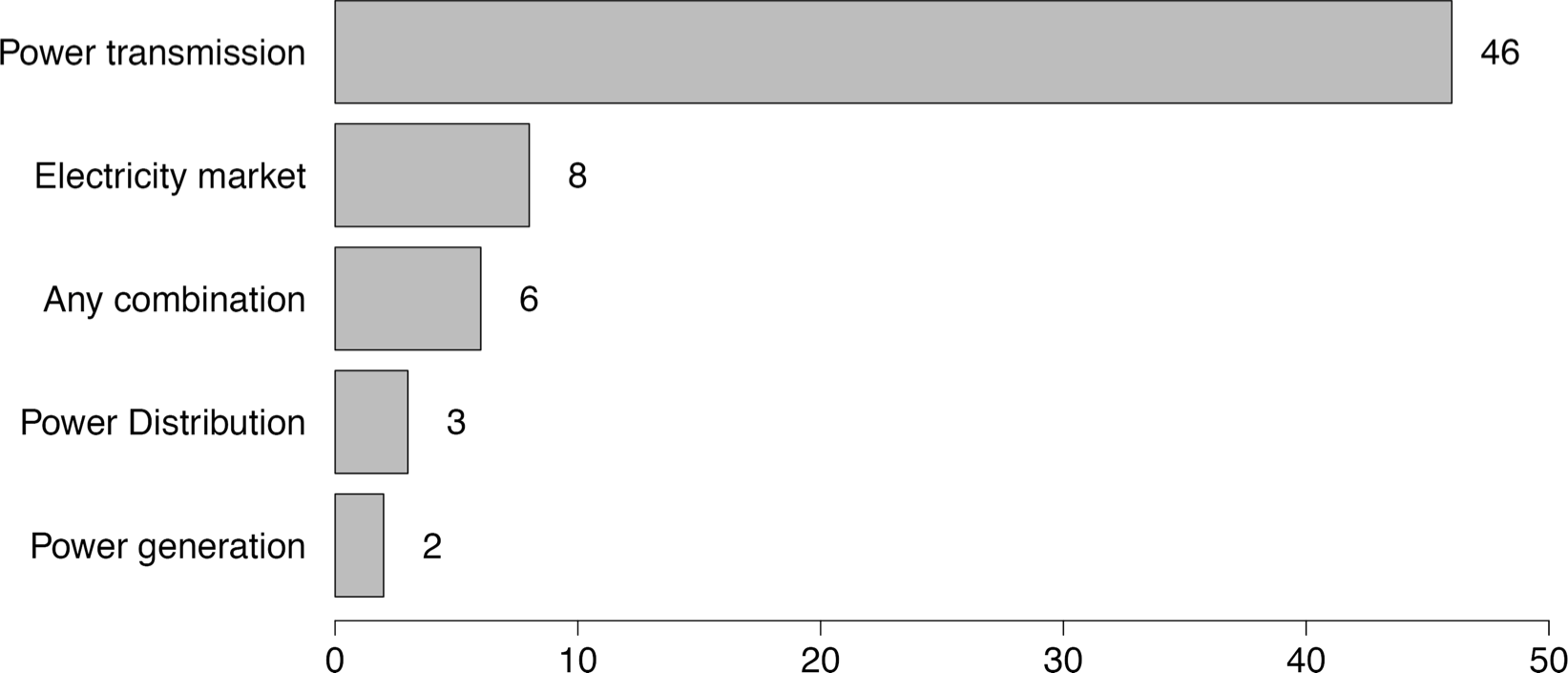}
	\caption{Distribution of primary studies applied in power grids}
	\label{fig:application-field-power-grids}
\end{figure}

As we can see from Figure~\ref{fig:application-fields}, from 65 out of 118 primary studies are focused exclusively on power grids, which corresponds to the 55.08\% of all selected studies. Among those, as shown in Figure~\ref{fig:application-field-power-grids}, 45 papers (i.e., 38.14\% of all the selected studies) deal exclusively with power transmission, 8 studies address the security aspects of the electricity market ([S061-S068]), 3 studies are focused on power distribution ([S018, S032, S056]), 2 studies on power generation ([S005, S024]), and the remaining 7 on any combination of the previous ones ([S002, S013, S028, S030, S049, S050, S059]).

The second largest group of publications in Figure~\ref{fig:application-fields} counts 28 works, i.e. 23.73\% of the whole set of primary studies of our research. All these papers study the security of generic linear dynamical systems. The proposed approaches can be used in any suitable application. However, these works do not provide examples of a particular application. 

\begin{figure}[!htbp]
	\centering
	\includegraphics[width=\columnwidth]{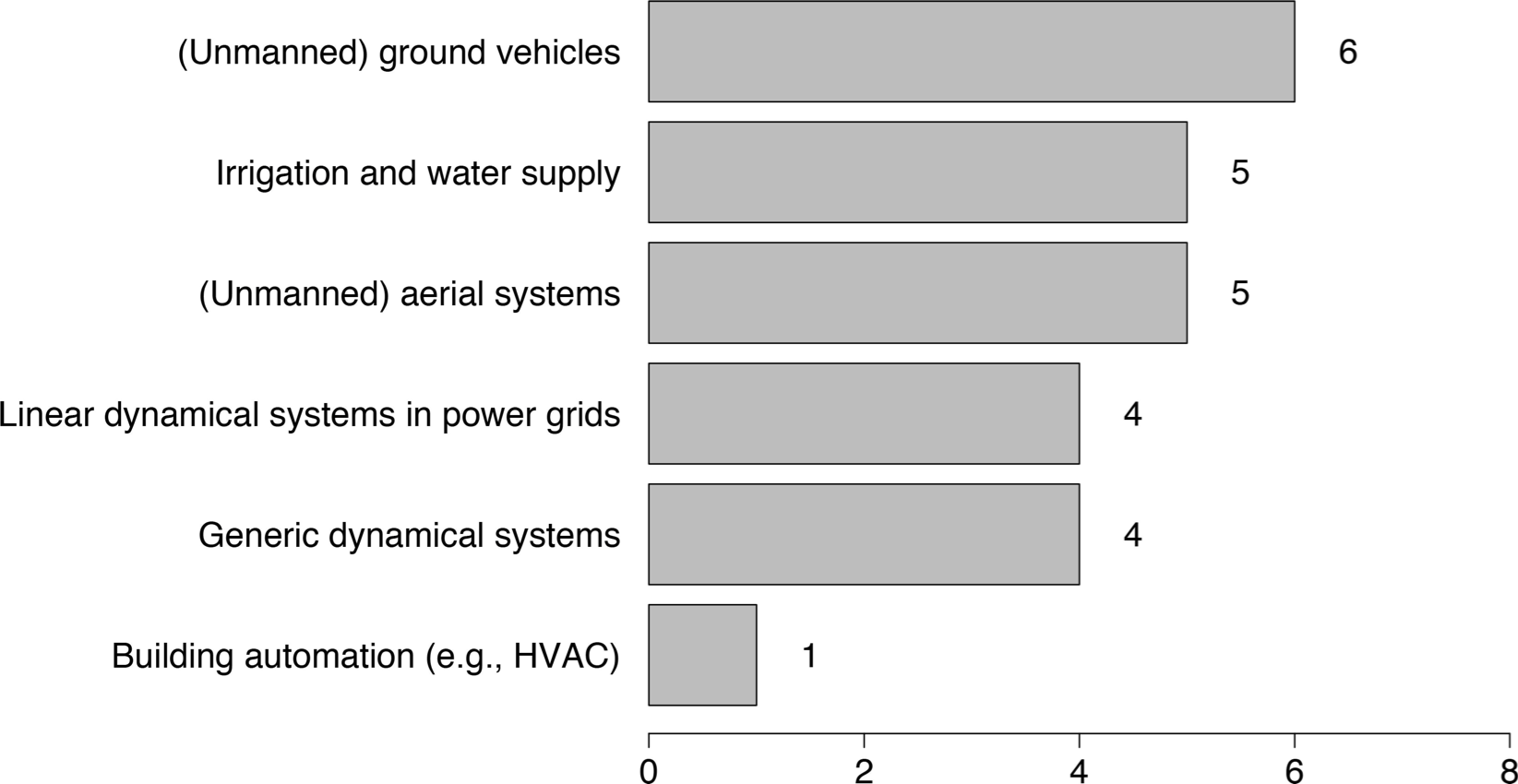}
	\caption{Distribution of primary studies by ``other'' application fields}
	\label{fig:application-field-others}
\end{figure}

The last group of the remaining 25 studies is detailed in Figure~\ref{fig:application-field-others}. 

These works are almost uniformly distributed among the following applications: (unmanned) ground vehicles (UGV) accounting for 6 of primary studies ([S084, S097, S099, S106, S111, S115]); (unmanned) aerial systems (e.g. unmanned aerial vehicles, air traffic management systems) and hydro-systems relying on automatic control, both considered in 5 papers ([S082, S090, S093, S108, S114] and [S010, S072, S078, S081, S083], respectively); generic (linear and non linear) dynamical systems and linear dynamical systems with applications to power grids, both found in 4 studies ([S035, S080, S096, S113] and [S048, S079, S100, S118], respectively). It is worth noting that UGV-based systems deal with the navigation and control of teleoperated and autonomous ground vehicles, together with their supervisory control and vehicle platooning. Finally, the security of building automation applications is investigated in one primary study ([S088]).

From the collected data, we can offer the following observations: 
\begin{itemize}
  \item the bulk of the selected works on security for cyber-physical systems is focused on power grids; this is not surprising, and may be due to the fact that smart grids are recognized as a driver for sustained economic prosperity, quality of life, and global competitiveness of a nation, attracting big research efforts to this area as a whole; also, the models used in this domain are well-known and the famous false data injection attack (FDIA) [S001] has been introduced in the context of power networks, giving traction to this kind of research applications. Moreover, the impressive market growth in renewable energy devices posed novel challenging problems in the design and management of power grids: as a consequence, the interest of energy providers on novel methods and technologies for optimizing network management with guaranteed performance, safety, and security provided a tremendous boost to academic research on these topics;
  \item only a small part of the selected papers presents the applications to the secure control of (unmanned) ground vehicles and aerial systems, and of heating, ventilation, and air-conditioning (HVAC), as well as lighting and shading, in large functional buildings; this application fields are relatively new for the approaches to the cyber-physical security, with the first studies appearing only in 2012; this result can be seen as indication of a potentially interesting direction for future research on CPS security;
  \item somehow surprisingly, we have not found any work focused on the cyber-physical security of medical CPS \citeref{Lee:2010:MCP:1837274.1837463}. We suppose that the topics of physiological close-loop control and patient modeling are seen as not mature enough to consider the security aspects specific to this important application field from the control-theoretic point of view. In any case, we expect that these topics will be considered and addressed in the near future.
\end{itemize}

\subsection{Point of view} \label{subsec:point-of-view}
\begin{figure}[!htbp]
	\centering
	\includegraphics[width=0.9\columnwidth]{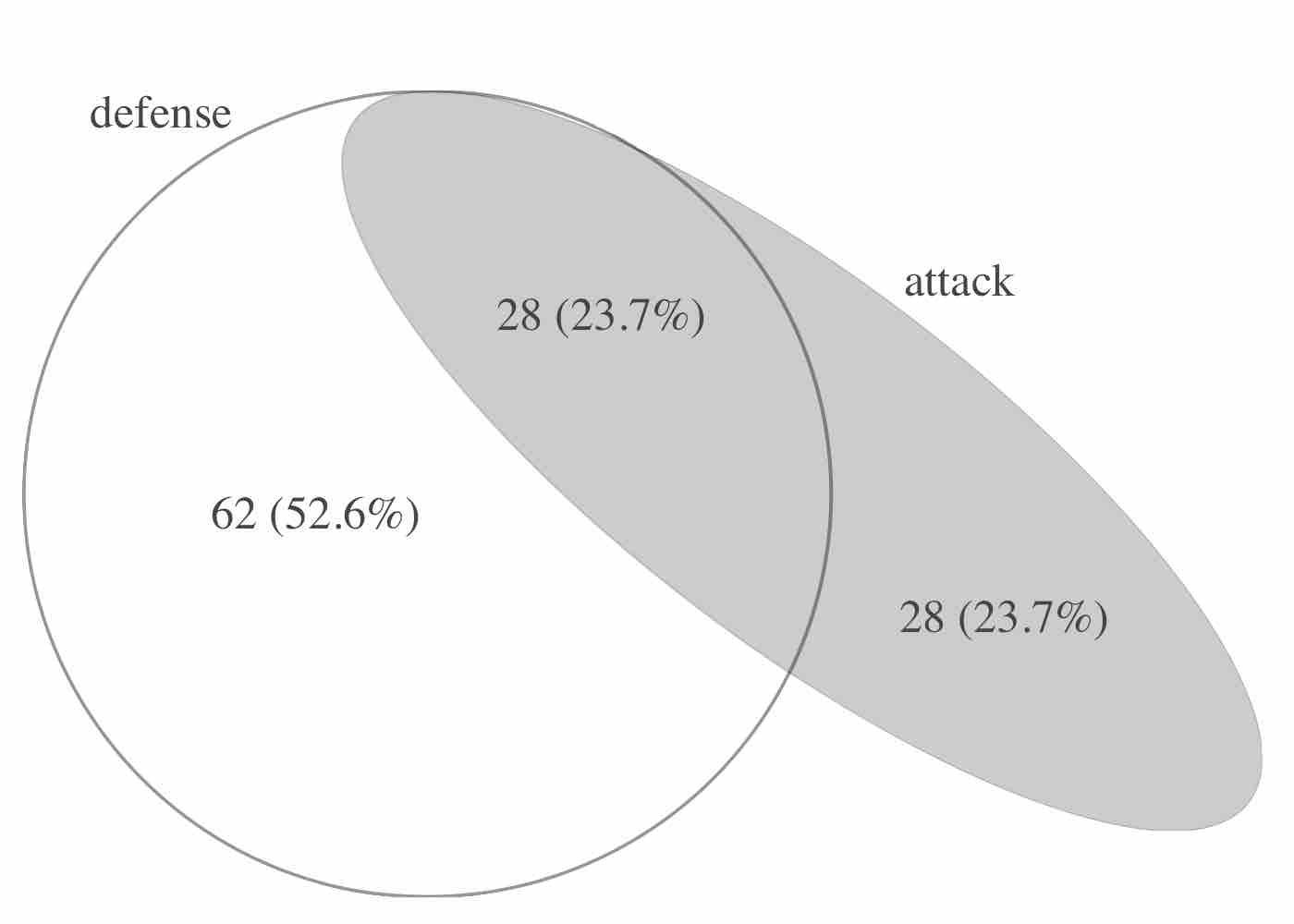}
	\caption{Distribution of primary studies by the adopted point of view}
	\label{fig:point-of-view}
\end{figure}

As reported in Figure~\ref{fig:point-of-view}, we distinguish primary studies based on whether they treat approaches for CPS security breaching (i.e. \textit{attack}) or enforcing via some kind of countermeasures (i.e. \textit{defense}), or both. From our analysis it emerged that 62 studies over 118 focus exclusively on the various countermeasures that a CPS may put in place in response to an attack, whereas 28 studies (i.e., 22.88\% of the total) focus exclusively on vulnerability analysis of CPS by proposing or improving an attack scheme using an adversary's point of view. They do not study the topic of the risk treatment, which is peculiar to the CPS designer's or operator's perspective. The remaining 28 works treat both attack and defense strategies.

From this result we can observe that the defense strategies are presented in most (76.27\%) of the selected studies, occupying the central spot of the research efforts on CPS security. A more detailed discussion of the various defense strategies proposed in research is provided in Section~\ref{subsec:defense-strategy}.

\subsection{Considered security attributes} \label{subsec:sec-attributes}
\begin{figure}[!htbp]
	\centering
	\includegraphics[width=0.9\columnwidth]{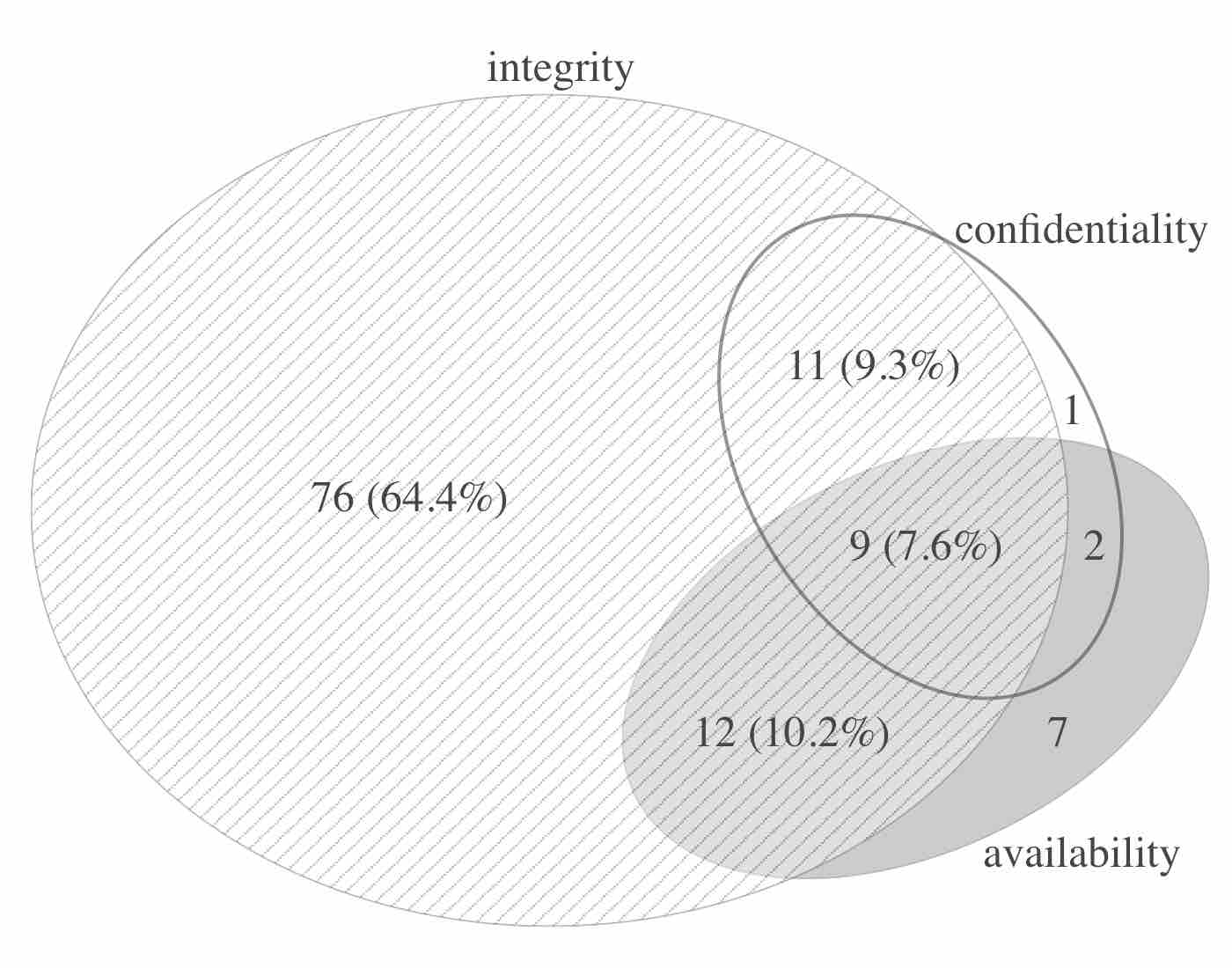}
	\caption{Distribution of primary studies by security attributes}
	\label{fig:security-attributes}
\end{figure}

Security can be seen as a composition of three main attributes, namely: confidentiality, integrity and availability \citeref{laprie}, 
Accordingly, we identified the security attributes considered by each primary study in order to understand how those attributes have been investigated by researchers on CPS security. Figure~\ref{fig:security-attributes} shows the distribution of the primary studies across confidentiality, integrity, and availability. 

The first thing that strikes the eye is that more than 90\% of the works are concerned with CPS \textit{integrity}, threatened by various types of deception attacks. Some of these works consider also the availability and/or confidentiality, together with integrity. On the contrary, only two studies ([S068, S105]) focus on the combination of solely \textit{availability and confidentiality}; those papers apply game theory to the design of countermeasures to intelligent jamming attacks, which have been published between the fall 2014 and 2015. For further discussion of security attributes, see Section~\ref{subsec:attacks}.

\subsection{System components} 
Each approach to security breaching or enforcing considers a particular set of system components to be compromised or protected. In our analysis we identified five main categories for describing the main system components to be compromised or protected, that are: sensors, actuators, network, controllers, plant. As an example, false data injection mainly targets a set of \emph{sensors}, while load altering can attack a set of \emph{actuators}. As for all deception and some disruption attacks, we should
   \textit{
   ``note that from a practical point of view, an attack on a sensor could either be interpreted as an attack on the node itself (making it transmit an incorrect signal), or it could also be interpreted as an attack on the communication link between the sensor and the receiver device; similarly an attack on an actuator could either be interpreted as an attack on the actuator itself, or on the communication link from the controller to the actuator'' [S079].
   }
Thus, we say that an approach considers a \emph{network} either when it does it implicitly by considering a denial-of-service (DoS) attack on communication links, or explicitly, by exploiting transmission scheduling, routing or some network-induced imperfections. Following the same line of reasoning, we say that the work takes into account a \emph{controller} when it proposes a novel one, whereas the \emph{plant} category comes into play with attacks at the physical layer and with eavesdropping.

\begin{figure}[!htbp]
	\centering
	\includegraphics[width=\columnwidth]{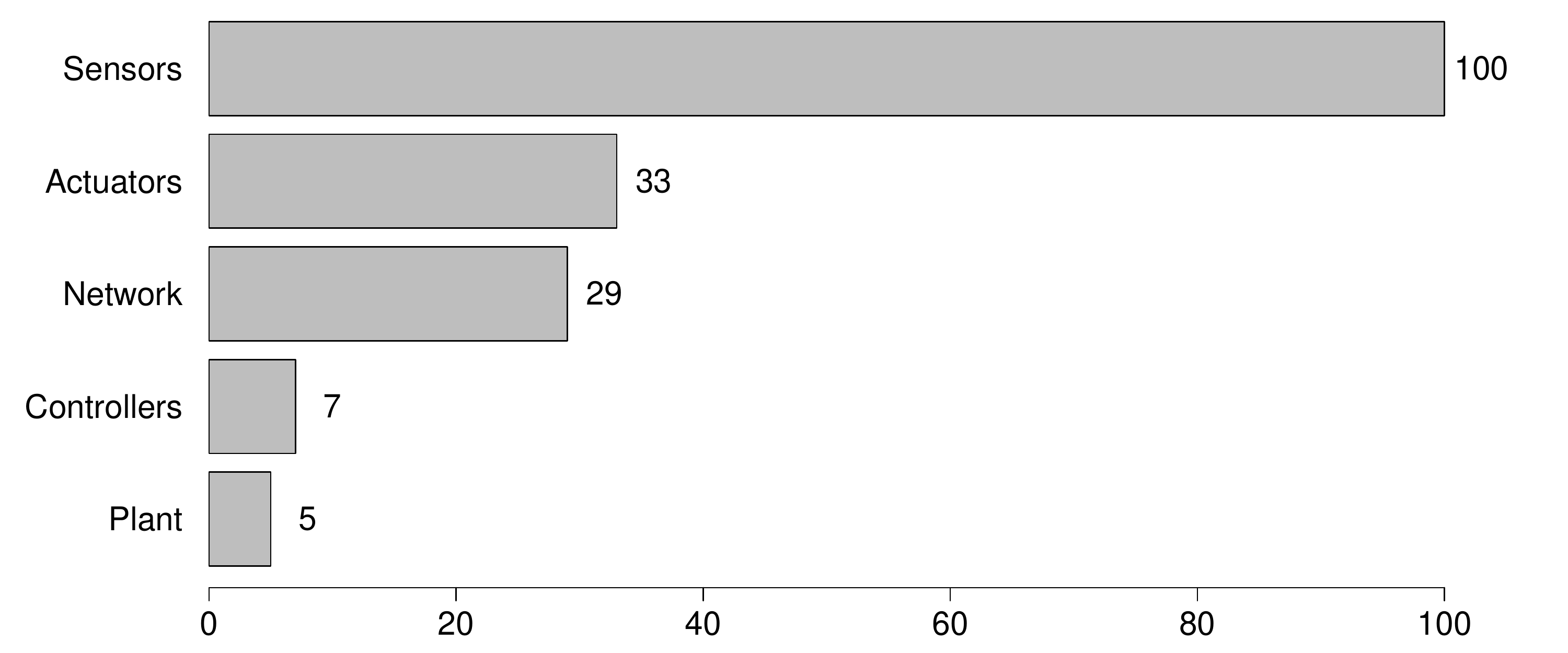}
	\caption{Distribution of primary studies by system components}
	\label{fig:system-components}
\end{figure}

Figure~\ref{fig:system-components} presents how system components have been considered among all the primary studies. Sensors were taken into account 100 (84.75\% of) times, 62 (52.54\% of) times alone and 27 (22.88\% of) times together with actuators. The actuators themselves were considered 33 (27.97\% of) times, while network was taken into account in 29 (24.58\% of) studies.

This data suggests that the approaches considering attacks on sensors and their protection completely dominate the scene. All the other system components have received much less attention, with a slight predominance of actuators and network. 


\subsection{Plant model}
We have seen in Section~\ref{subsec:app-field} that the application domain of research on CPS security is mainly divided between power grids and all the others. This result is reflected also in the choice of the mathematical models used to describe the physical domain. 

In particular, power transmission is traditionally studied via a power flow model, which is a set of equations that depict the energy flow on each transmission line of a power grid. An AC power flow model considers both real and reactive power and is formulated by nonlinear equations, where the state variables are voltage magnitudes and phase angles of the buses \citeref{abur2004power, wollenberg1996power}. However, state estimation using an AC power flow model can be computationally expensive and does not always converge to a solution. Thus, power system engineers sometimes use a linearized power flow model, DC power flow model, to approximate the AC power flow model [S001]. In DC model the reactive power is completely neglected and state variables only consist of voltage phase angles of the buses. 
As of power generation, the model based on equations describing the electromechanical swing dynamics of the synchronous generators \citeref{kundur1994power} is usually applied. In other application domains more general linear time invariant (LTI) or nonlinear dynamical models are used.

\begin{figure}[!htbp]
	\centering
	\includegraphics[width=\columnwidth]{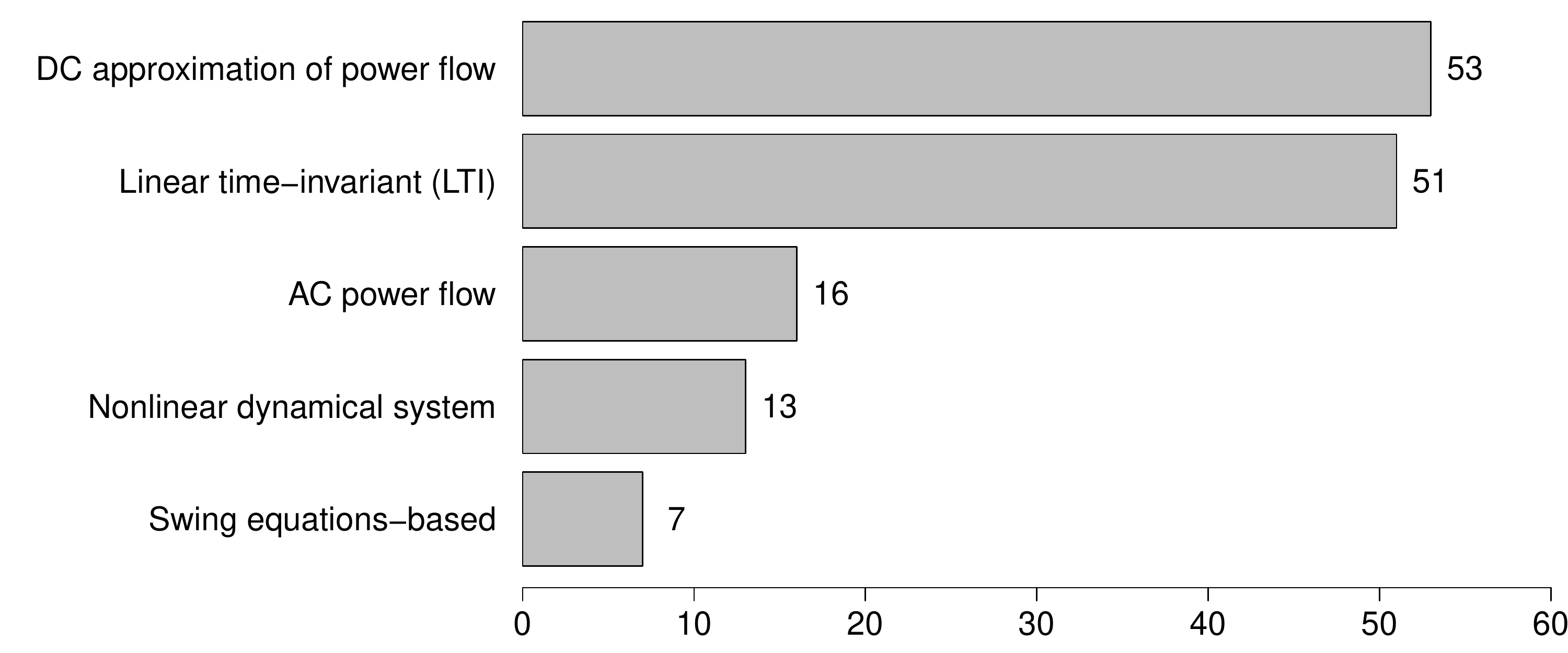}
	\caption{Distribution of primary studies by plant model}
	\label{fig:plant-model}
\end{figure}

Figure~\ref{fig:plant-model} shows how the above mentioned models have been used within the set of primary studies. The \textit{DC approximation of power flow} has been used in 53 works (44.92\% of whole set), while the more complicated and realistic \textit{AC power flow model} (which is capable to capture more subtleties) has been studied 16 (13.56\% of) times. In 6 studies both the AC power flow model and its linear DC approximation have been used ([S023, S028, S030, S051, S056, S057]). Other \textit{LTI models} were applied in 51 (43.22\% of) primary studies. \textit{Nonlinear dynamic} and \textit{swing-equation based models} were applied 13 (11.02\%) and 7 (5.93 \% of) times, respectively.

\subsection{Process noise}
To capture any deviation in the plant model from the real dynamics of the controlled physical system, the process noise is used; from the primary studies it emerged that it can be categorized into three main classes: \textit{Gaussian}, \textit{bounded (non-stochastic)}, and \textit{noiseless}. 

\begin{figure}[!htbp]
	\centering
	\includegraphics[width=0.9\columnwidth]{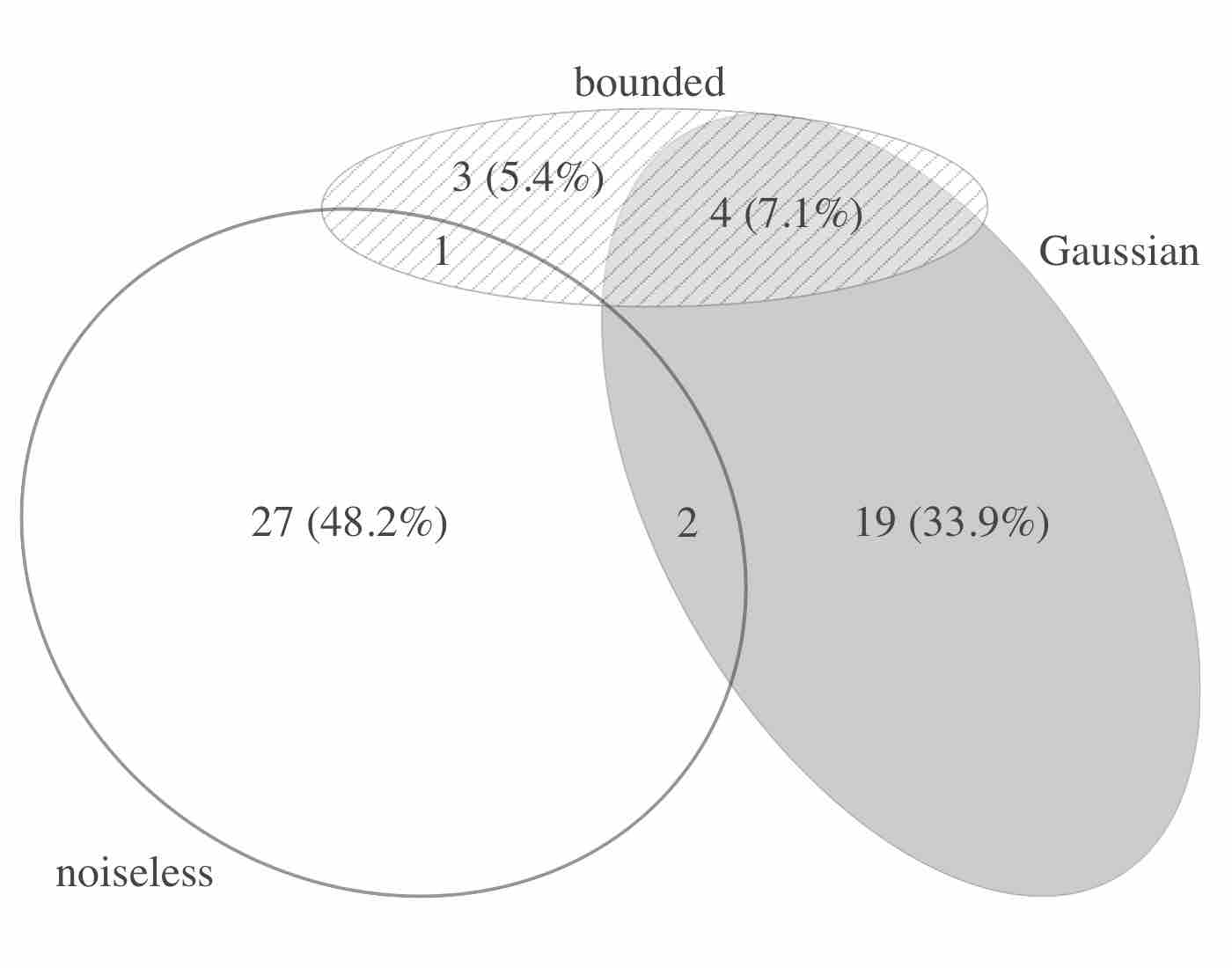}
	\caption{Distribution of primary studies by process noise}
	\label{fig:process-noise}
\end{figure}

The distribution of primary studies by process noise is reported in Figure~\ref{fig:process-noise}, where the studies considering the measurement model only (62, accounting for 52.54\% of the whole set of selected papers) were not included, since for them the facet of process noise is not applicable.

\begin{figure}[!htbp]
	\centering
	\includegraphics[width=\columnwidth]{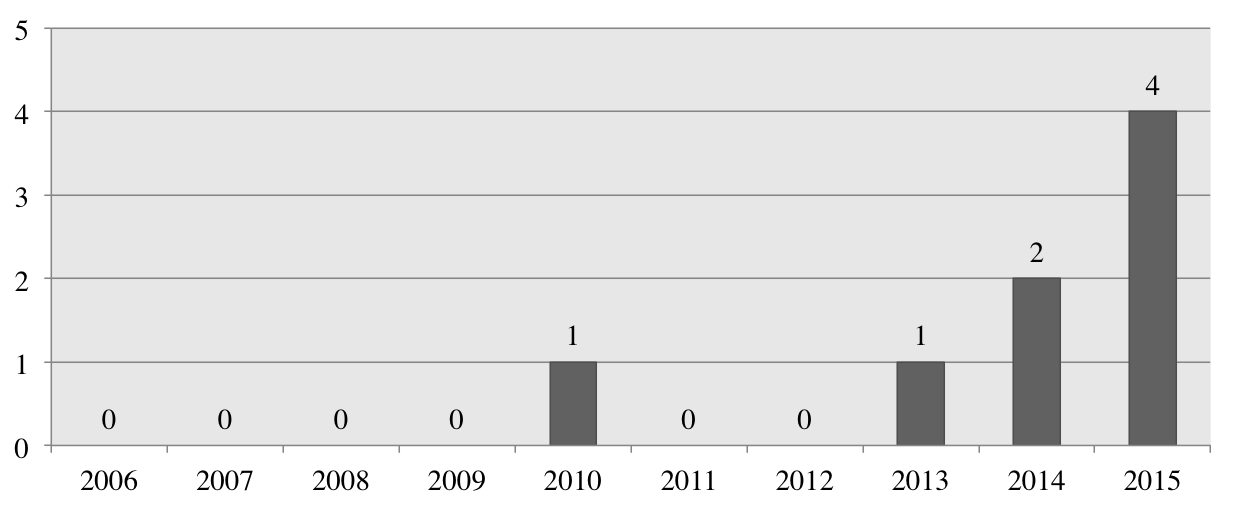}
	\caption{Distribution of primary studies with bounded process noise by year (partial data for 2015)}
	\label{fig:bounded-process-noise}
\end{figure}

We can see that the noiseless and Gaussian process noise models are the most used ones (accounted 30 and 25 times, respectively). As shown 
in Figure~\ref{fig:bounded-process-noise},
 the bounded non-stochastic model (used 8 times) is starting to receive a growing attention in the very last years.

\subsection{Measurement noise}
Depending on the assumptions on the noise, sensor measurement models can be broadly categorized into three classes: \textit{Gaussian}, \textit{bounded (non-stochastic)} and \textit{noiseless} [S116]. 

\begin{figure}[!htbp]
	\centering
	\includegraphics[width=0.9\columnwidth]{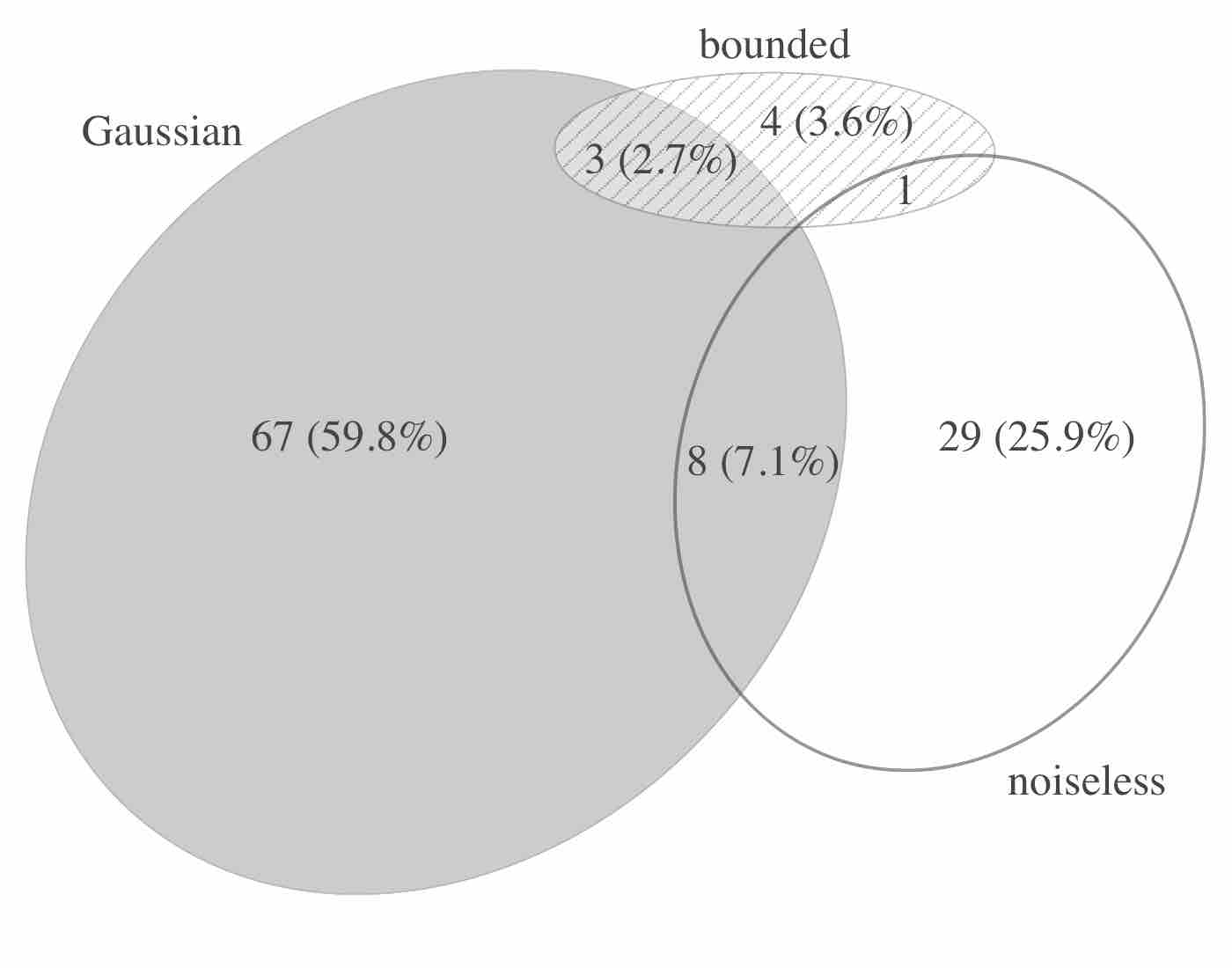}
	\caption{Distribution of primary studies by measurement noise}
	\label{fig:measurement-noise}
\end{figure}

\begin{figure}[!htbp]
	\centering
	\includegraphics[width=\columnwidth]{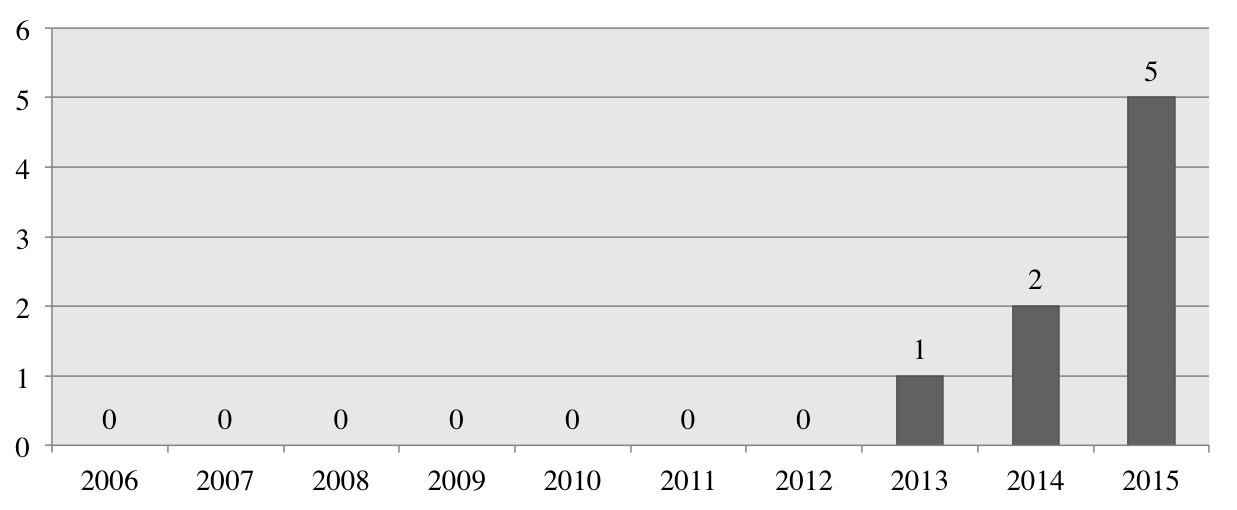}
	\caption{Distribution of primary studies with bounded measurement noise by year (partial data for 2015)}
	\label{fig:bounded-measurement-noise}
\end{figure}

As shown in Figure~\ref{fig:measurement-noise}, the majority of primary studies (78, i.e. 66.10\%) uses Gaussian measurement noise model; while 38 (32.20\% of all) works assume noiseless measurements. Only 8 works have used bounded (non-stochastic) assumptions. Similarly for the bounded process noise, the bounded measurement noise has started to gain attention only recently in the CPS security domain, as we can see from Figure~\ref{fig:bounded-measurement-noise}.

If a primary study does not consider the measurement model (e.g. when the work is not related to the secure state estimation against sensor attacks), we say that the measurement noise is not applicable. Among the selected primary studies there were 6 such works. 

\subsection{State estimation} \label{subsec:state-estimator}
\begin{figure}[!htbp]
	\centering
	\includegraphics[width=\columnwidth]{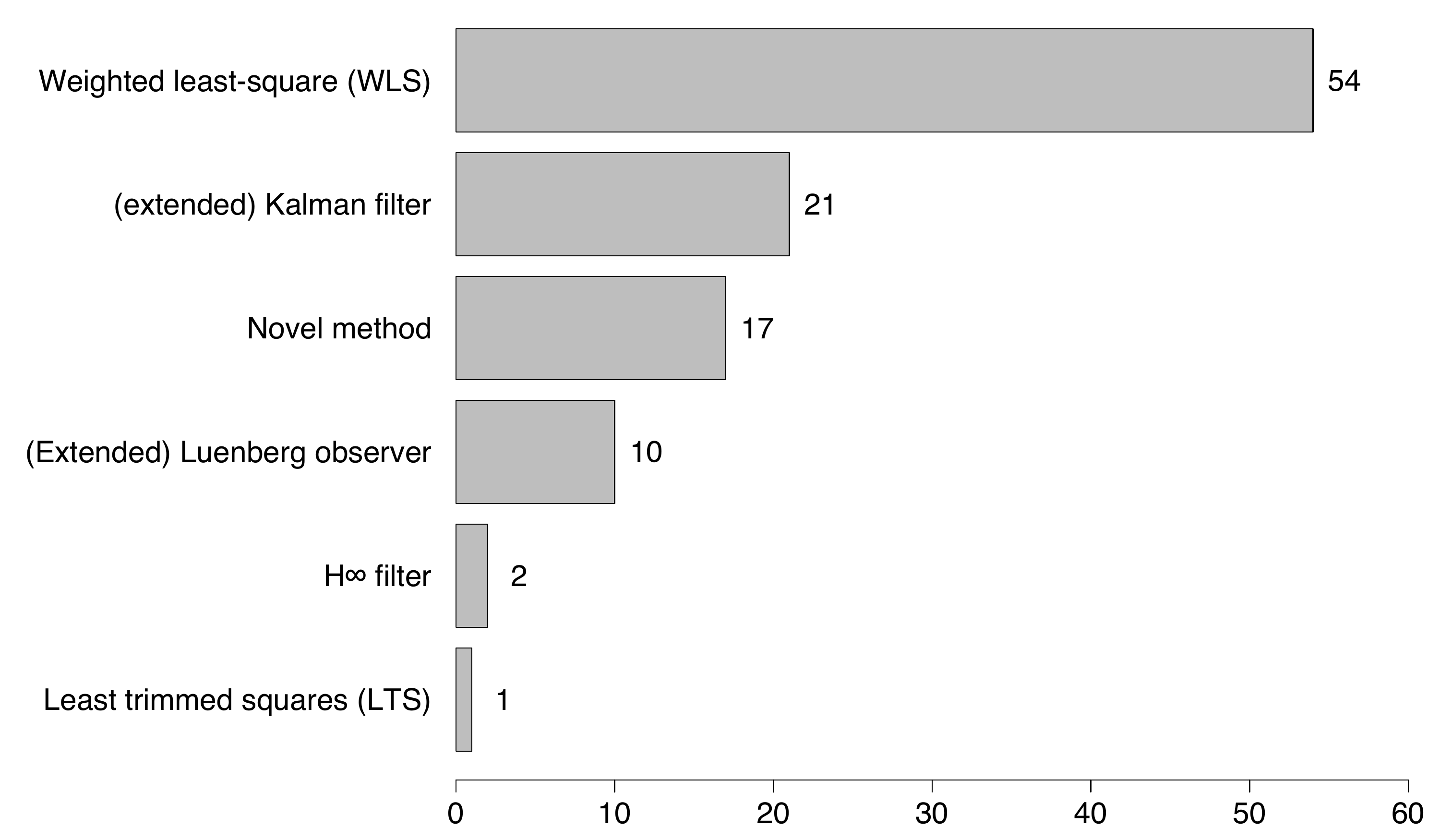}
	\caption{Distribution of primary studies by state estimation}
	\label{fig:state-estimation}
\end{figure}

For many situations, it may be unrealistic or unfeasible to assume that all the states of the system are measured. In fact, 89 studies were using some kind of state estimation, which corresponds to 75.42\% of all the primary studies (see Figure~\ref{fig:state-estimation}). The most used state estimation method is \textit{weighted least squares} (WLS), found in 54 (45.76\% of all) works (interestingly, all 54 studies were related to power grids). The WLS method for power system state estimation is optimal under Gaussian measurement noise [S057] and, in case of DC approximation of power flow, leads to an estimator identical to the one obtained with maximum likelihood or with minimum variance methods [S001]. The (extended) \textit{Kalman filter} was used in 21 studies (17.80\% of all primary studies), while the (extended) \textit{Luenberger observer} was used in 10 studies (8.47\%), the \textit{H$_\infty$ filter} in 2 studies ([S087, S093]) and the \textit{least trimmed squares} estimator in only one study ([S057]). Novel solutions for the state estimation were proposed in 17 (14.41\%) studies.

Novel methods range from application-specific solutions [S024, S072], distributed state estimation techniques for power networks [S011, S014, S025], to generic attack-resilient solutions inspired by Kalman filter [S091, S106, S116].

Within the domain of power grids, Giani et al. [S015] proposes state estimation based countermeasures to coordinated sparse attacks on power meter readings, that take advantage of graph-theoretic construct of \textit{observable islands}, which are disjoint subsets of buses sharing the same perceived change of state [voltage phase] under the attack. As a countermeasure to leverage point attacks against WLS state estimation in smart grid, Tan et al. [S049] introduces a modified robust Schweppe-Huber Generalized-M estimator. The WLS estimation method for power networks has been extended by Liu et al. [S054] by merging cyber impact factor matrix into the state estimation as a reasonable adjustment of the weight values, in order to create the abnormal traffic-indexed state estimation.

Regarding generic cyber-physical systems, to estimate the state of the plant despite attacks on sensors and actuators, Fawzi et al. [S079] propose an efficient state reconstructor inspired from techniques used in compressed sensing and error correction over the real numbers. Pajic et al. [S099] show that implementation issues such as jitter, latency and synchronization errors can be mapped into parameters of the state estimation procedure that describe modeling errors, and provides a bound on the state-estimation error caused by modeling errors. Mo and Sinopoli [S096] constructs an optimal estimator of a scalar state that minimizes the ``worst-case'' expected cost against all possible manipulations of measurements by the attacker, while Weimer et al. [S102] introduces a minimum mean-squared error resilient (MMSE-R) estimator for stochastic systems, whose conditional mean squared error from the state remains finitely bounded and is independent of additive measurement attacks.

Finally, for linear dynamical systems under sensor attacks, Shoukry and Tabuada [S111] present an efficient event-triggered projected Luenberger observer for systems under sparse attacks, and Shoukry et al. [S117] develop an efficient algorithm that uses a Satisfiability Modulo Theory (SMT) approach to isolate the compromised sensors and estimate the system state despite the presence of the attack.

Together, these results are an indication that the resilient state estimation under measurement attacks is a very active research topic within the area of CPS security,
making us reasonably confident about its future development and potential.

\subsection{Anomaly detector} \label{subsec:anomaly-detection}
Current state estimation algorithms use bad data detection (BDD) schemes to detect random outliers in the measurement data [S006]. Two of the most used BDD hypothesis tests are the \textit{performance index test} (also known in power system's community as \textit{J(\^{x})-test} or $\chi^2$\textit{-test}) and the \textit{largest normalized residual test} (often referred as $r^N_{max}$\textit{-test}) \citeref{abur2004power}.

\begin{figure}[!htbp]
	\centering
	\includegraphics[width=0.9\columnwidth]{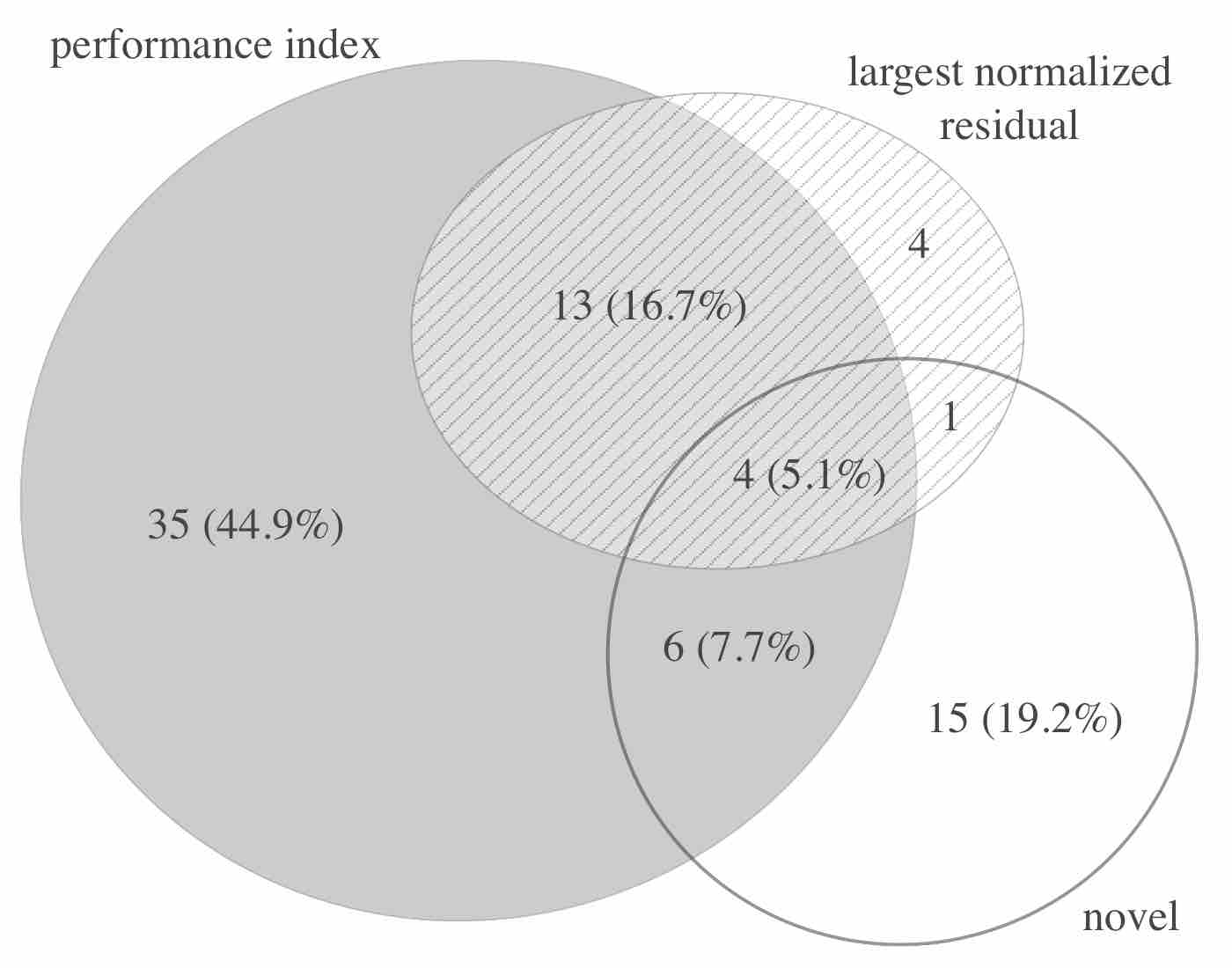}
	\caption{Distribution of primary studies by anomaly detection}
	\label{fig:anomaly-detection}
\end{figure}

As shown in Figure~\ref{fig:anomaly-detection}, among our primary studies there are 58 approaches considering performance index test, 22 approaches dealing with normalized residual test, and 13 considering both aforementioned hypothesis tests.

There are also two works considering an arbitrary anomaly detector implemented by the controller and deployed to detect possible deviations from the nominal behavior [S081, S101], while 36 (30.51\% of) primary studies do not deal at all with anomaly detection.

In an effort to minimize the detection delay, the change detection can be formulated as a quickest detection problem. Page's cumulative sum (CUSUM) algorithm \citeref{10.2307/2333009} is the best-known technique to tackle this type of problem. There are 5 selected primary studies, that propose or use a CUSUM-based attack detection schemes [S007, S016, S035, S060, S075]. There are also 26 (22.03\%) studies, that propose other novel anomaly detection approaches, either considering them together with the performance index test or normalized residual test.

The novel solutions for bad data detection cover the topics of distributed monitoring [S010, S011, S014, S029] and application-specific anomaly detection for multi-agent distributed flocking formation control [S024], automated cascade canal irrigation systems [S072], wireless control networks, \textit{``where the network itself acts as the controller, instead of having a specially designated node performing this task'' [S074],}
multi-hop control networks, \textit{``where the communication between sensors, actuators and computational units is supported by a (wireless) multi-hop communication network and data flow is performed using scheduling, routing and network coding of sensing and actuation data'' [S088],}
and air transportation systems [S108]. 

In the power system domain, Kosut et al. [S002] proposes a generalized likelihood ratio detector, that incorporates historical data and does not compute explicitly the residue error, while Gu et al. [S058] introduces a new method to detect false data injection attacks against AC state estimation by tracking the dynamics of measurement variations: the Kullback--Leibler distance (KL divergence, known also as relative entropy) is used to calculate the distance between two probability distributions derived from measurement variations. 

The KL divergence is adopted also by Mo et al. [S070, S112] in designing the optimal watermark signal in the class of stationary Gaussian processes, which is used to derive the optimal Neyman--Pearson detector of reply and covert attacks, respectively.

Valenzuela et al. [S031] use principal component analysis (PCA) \citeref{Lee:2007:NDR:1557216} to separate power flow variability into regular and irregular subspaces, with the analysis of the information in the irregular subspace determining whether the power system data has been compromised. Also Liu et al. [S033] views false data detection as matrix separation problem and, differently from the case of the PCA, proposes algorithms that exploit 

\textit{
``the low rank structure of the anomaly-free measurement matrix, and the fact that malicious attacks are quite sparse.''
}

Tiwari et al. [S097] propose an approach inspired by PCA, that uses an invariant 
\textit{
``–-- an over-approximation of the reachable states –-- of the system under normal conditions as the classifier'';
}
this set is called the safety envelope. An alarm is raised whenever the system state falls outside the safety envelope.

Security-oriented cyber-physical state estimation (SCPSE) for power grid, proposed in Zonouz et al. [S026], uses stochastic information fusion algorithms on
\textit{
``information provided by alerts from intrusion detection systems that monitor the cyber infrastructure for malicious or abnormal activity, in conjunction with knowledge about the communication network topology and the output of a traditional state estimator'',
}
in order to detect intrusions and malicious data, and to assess the cyber-physical system state.

Other novel anomaly detection methods in power grid comprise a detector implementing the Euclidean distance metric [S048], and a cosine similarity matching based approach [S055]. It is worth noting that the second one requires the usage of the Kalman filter as a source of estimated/expected data.

To contrast false data injection attacks, Sedghi and Jonckheere [S034] present a decentralized detection and isolation scheme based on the Markov graph of the bus phase angles, obtained via conditional mutual information threshold (CMIT) test, while Sou et al. [S020] introduces a scheme, 
that considers potentially compromised information from both the active and the reactive power measurements on transmission lines. In this second scheme, based on the novel reactive power measurement residual, 
\textit{
``the component of the proposed residual on any particular line depends only locally on the component of the data attack on the same line''.
}
Li and Wang [S040] presents the state summation detection using state variables' distributions, which tests hypothesis on true measurement square sum $S_x$ (assumed to follow normal distribution, given a large number of state variables) together with test on \textit{J(\^{x})}. Finally, Sanandaji et al. [S041] presents a heuristic for detecting abrupt changes in the system outputs based on the singular value decomposition of a history matrix built from system observations.

For dissipative or passive CPS, Eyisi and Koutsoukos [S098] propose energy-based attack detection monitor. 

To contrast stochastic cyber-attacks, Li et al. [S107] presents an algebraic detection scheme based on the frequency-domain transformation technique and linear algebra theory, together with sufficient and necessary conditions guaranteeing the detectability of such attacks.

Pasqualetti et al. [S010] characterizes fundamental monitoring limitations of descriptor systems from system-theoretic and graph-theoretic perspectives, and designs centralized and distributed monitors, which are complete, in the sense that they detect and identify every (detectable and identifiable) attack.

Finally, Jones et al. [S113] presents an automated anomaly detection mechanism based on inference via formal methods to develop an unsupervised learning algorithm, which constructs from data a signal temporal logic (STL) formula that describes normal system behavior. Trajectories that do not satisfy the learned formula are flagged as anomalous.

As a general comment, the literature described in this section appears quite fragmented, and a systematic high level view is still missing even within a specific application domain. The different results and methodologies are very difficult to relate each other and validate since both a comparison metric and a benchmark, neither academic nor industrial, have not been agreed and defined yet.

\subsection{Controller} \label{subsec:controller}

Considering the used controller, the first fact emerging from our analysis is that studies focusing on state estimation usually do not examine at all the controller. In fact, in 82 (69.49\% of 118 selected) studies the controller is not available. In the remainder of this section we will focus on the remaining 36 studies, some of which consider more than one controller at once.

\begin{figure}[!htbp]
	\centering
	\includegraphics[width=\columnwidth]{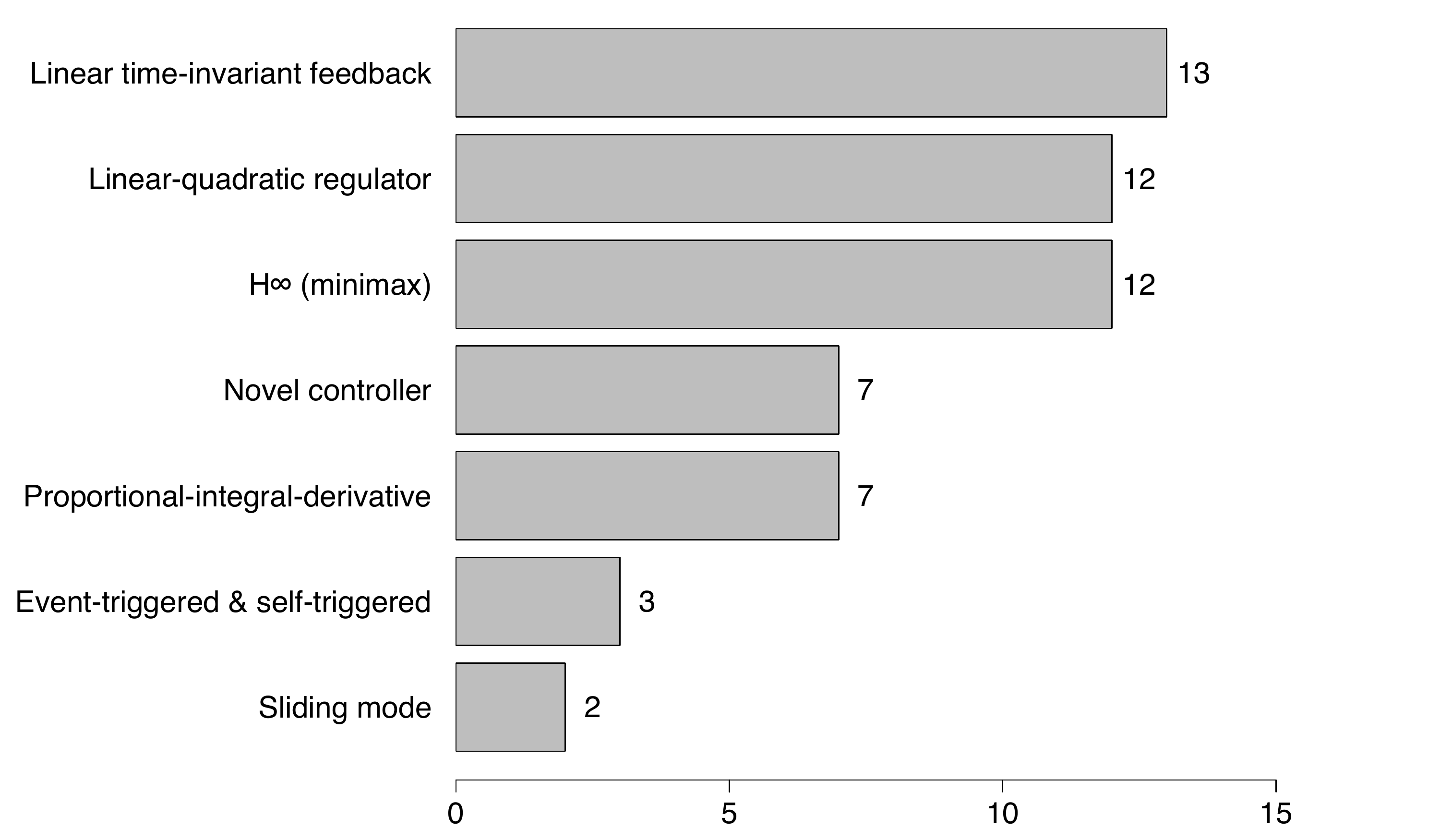}
	\caption{Distribution of primary studies by controller}
	\label{fig:controller}
\end{figure}

As shown in Figure~\ref{fig:controller}, the most considered controllers are generic state feedback or output feedback controllers with a control law restricted to be linear time invariant, found in 13 studies, together with linear quadratic regulators (LQR) and H$_\infty$ (minimax) controllers, each of which is seen in 12 works. The variations of proportional-integral-derivative (PID) controller are considered in 7 works, while the event-triggered and self-triggered controllers can be found in 3 studies [S085, S103, S111], and sliding mode controllers in 2 studies [S013, S115]. 

Interestingly, seven primary studies ([S024, S069, S073, S076, S077, S086, S093]) propose novel controllers.
More specifically, inspired by the analogy to flocking behavior, Wei and Kundur [S024] developed distributed hierarchical 
\textit{
``control methodologies that leverage cooperation between distributed energy resources and traditional synchronous machines to maintain transient stability in the face of severe disturbances''.
}
For a class of denial-of-service (DoS) attack models, Amin et al. [S069] presents an optimal minimax causal feedback control law, subject to the power, safety and security constraints. Gupta et al. [S073] studies a similar problem of optimal minimax control in the presence of an intelligent jammer with limited actions as dynamic zero-sum game between the jammer and the controller. Befekadu et al. [S076] introduces instead the 
\textit{
``measure transformation technique under which the observation and state variables become mutually independent along the sample-path (or path-estimation) of the DoS attack sequences in the system'',
}
thanks to which it derives the optimal control policy for the risk-sensitive control problem, under a Markov modulated DoS attack model.
Zhu and Mart\'{i}nez [S077] proposes a variation of the receding-horizon control law to deal with the replay attacks, 
while Zhu et al. [S086] provides a set of coupled Riccati differential equations characterizing feedback Nash equilibrium as the solution concept for the distributed control in the multi-agent system environment subject to cyber attacks and malicious behaviors of physical agents.
Finally, Kwon and Hwang [S093] proposes 
\textit{
``a hybrid robust control scheme that considers multiple sub-controllers, each matched to a specific type of cyber attacks'',
}
together with a method for designing the corresponding secure switching logic.

As a general comment, the literature described in this section derives interesting theoretical results, but there is still a lot of work to do for addressing the practical challenges in CPS security.

\subsection{Communication aspects and network-induced imperfections}
The introduction of the communication network in a control loop modifies the external signals of the plant and the controller due to the 
network-induced imperfections \citeref{levine2010control}, which in turn depend on some communication aspects, such as transmission scheduling and routing. 

\begin{figure}[!htbp]
	\centering
	\includegraphics[width=\columnwidth]{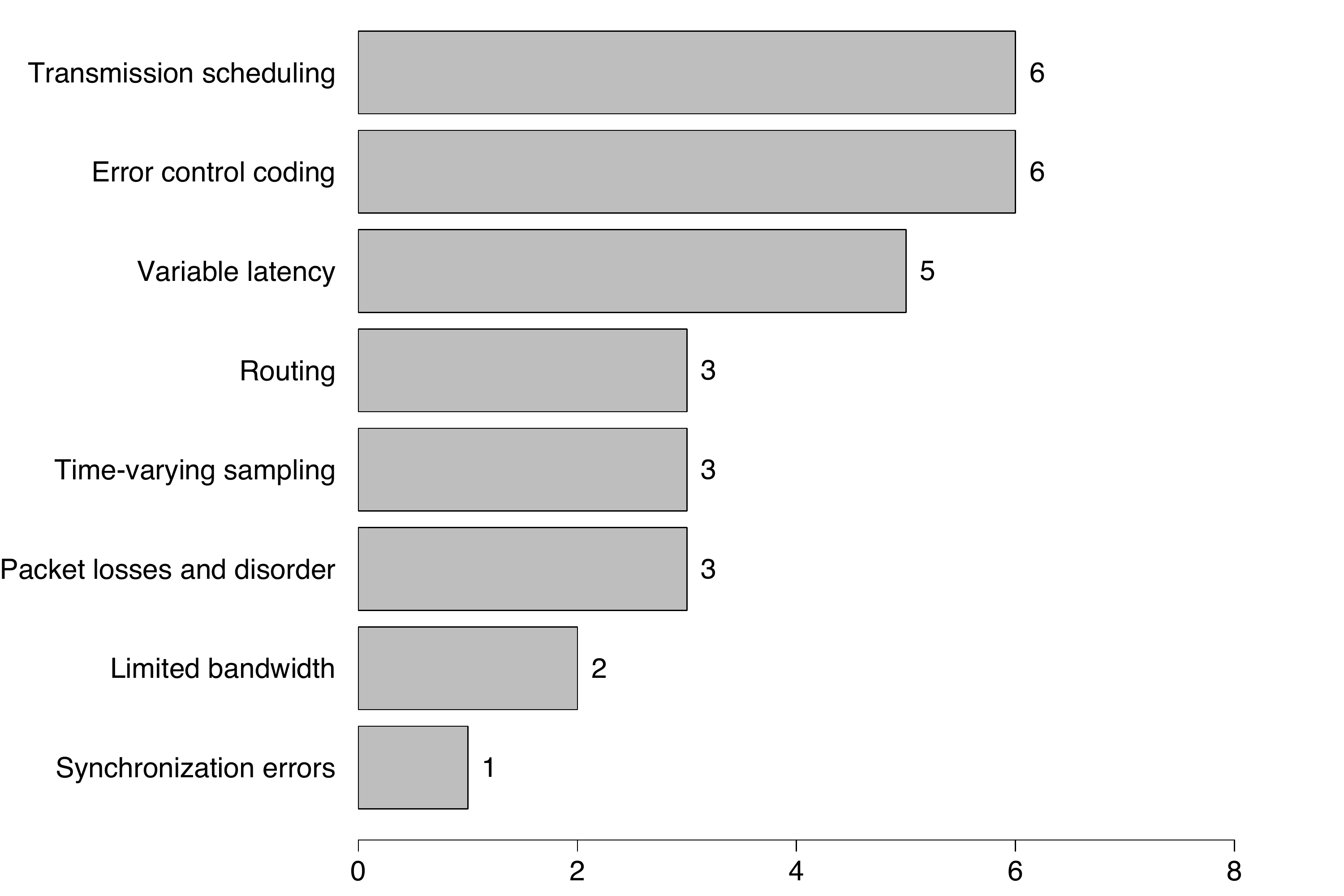}
	\caption{Distribution of primary studies by communication aspects and network-induced imperfections}
	\label{fig:communication-aspects}
\end{figure}

When analyzing the primary studies on the basis of this facet we got a surprise: 100 out of 118 studies (i.e., 84.75\%) do not explicitly consider any communication aspect or imperfection, while only 6 studies (i.e. 5.08\%) address more than one aspect. The total number of times each communication aspect was addressed within the set of the primary studies is shown in Figure~\ref{fig:communication-aspects}.

Synchronization errors are considered only by Pajic et al. [S099], where also variable latency and time-varying sampling are mapped into parameters of the state estimation procedure that describe modeling errors. Time-varying sampling is taken into account also by Yilmaz and Wang [S060] and, together with transmission scheduling, by De Persis and Tesi [S103]. Limited bandwidth is considered together with error control coding by Gupta et al. \citeref{6161475} (which is related to [S073]), and by Sundaram et al. [S074], in which \textit{``nodes in a network transmit linear combinations of incoming packets rather than simply routing them''}. Packet losses and disorder alone is taken into consideration in two works ([S091, S118]) and together with variable latency and transmission scheduling in another one ([S087]). Routing by itself is examined by Vukovi\'{c} et al. [S022], and together with error control coding, transmission scheduling and variable latency, by D'Innocenzo et al. [S088]. Only variable latency is considered by Miao and Zhu [S094] and by Jones et al. [S113]. Both error control coding and transmission scheduling by themselves are taken into account in 3 works ([S079, S109, S110] and [S085, S095, S104], respectively).

Surprisingly, very few papers (attempt to) provide non-trivial mathematical models of the communication protocol, which indeed is a fundamental actor of almost any CPS. In particular, only in D'Innocenzo et al. [S088] a specific standard for communication, i.e. WirelessHART and ISA-100, is explicitly considered in the CPS mathematical model.

\subsection{Time-scale model}
The dynamic system behavior can be modeled via different time-scale models, such as continuous, discrete and hybrid. In the case of the (quasi-)steady state assumption, the system is treated as (quasi-)static, and the time-scale model is named accordingly. 

\begin{figure}[!htbp]
	\centering
	\includegraphics[width=0.9\columnwidth]{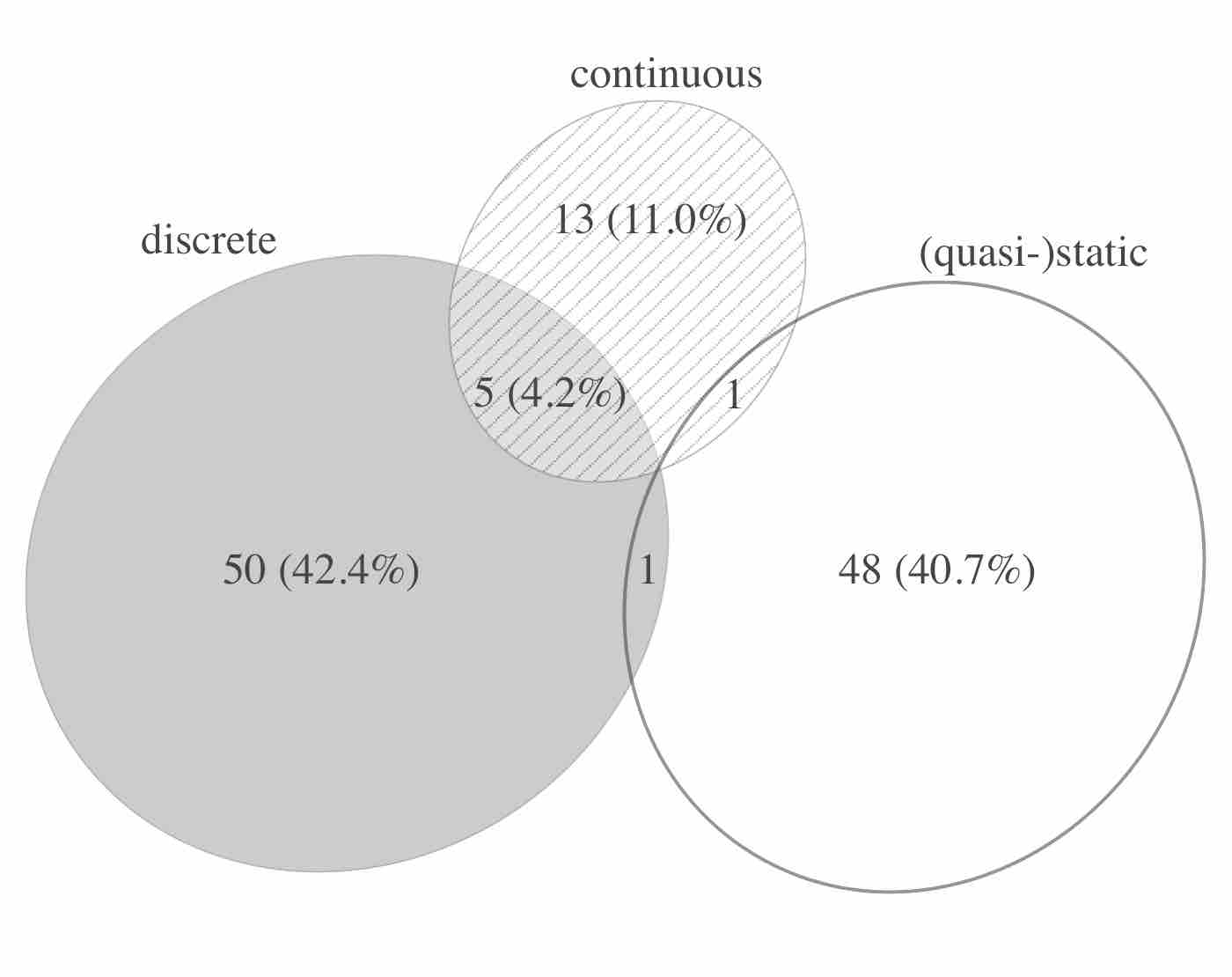}
	\caption{Distribution of primary studies by time-scale model}
	\label{fig:time-scale}
\end{figure}

As shown in Figure~\ref{fig:time-scale}, the quasi-static model is used in 48 studies (40.68\%), all of them concerned with power systems state estimation, while there are 13 studies (11.02\%) considering continuous time, 50 (42.37\%) discrete time, and only 5 considering both continuous and discrete time ([S080, S083, S086, S103, S113], only 3 of which actually using hybrid time [S080, S086, S113]).
There is also one work with both continuous time and quasi-static model ([S015]), and one with both discrete time and quasi-static model ([S016]).

In particular, quasi-static analysis is mostly chosen for addressing control architectures like SCADA, which provide steady-state set-points to inner control loops.

\subsection{Attacks and their characteristics} \label{subsec:attacks}
Regardless of the adopted point of view (see Section~\ref{subsec:point-of-view}), every study on CPS security deals with attacks in order to either implement or to counteract them. 
Each attack threats one or more primary security attributes (see Section~\ref{subsec:sec-attributes}). More specifically, the best known attack on availability is the \emph{denial of service} (DoS) attack, that renders inaccessible some or all the components of a control system by preventing transmissions of sensor or/and control data over the network. \textit{``To launch a DoS an adversary can jam the communication channels, compromise devices and prevent them from sending data, attack the routing protocols, flood with network traffic some devices, etc.'' [S069].} 
Attacks on data integrity are known as \emph{deception} attacks and represent the largest class of attacks on cyber-physical systems, including false data injection attacks. 
The attacks on confidentiality alone are often referred to as \emph{disclosure} attacks, i.e. \emph{eavesdropping}, which is discussed only in two studies [S081, S084].

\begin{figure}[!htbp]
	\centering
	\includegraphics[width=\columnwidth]{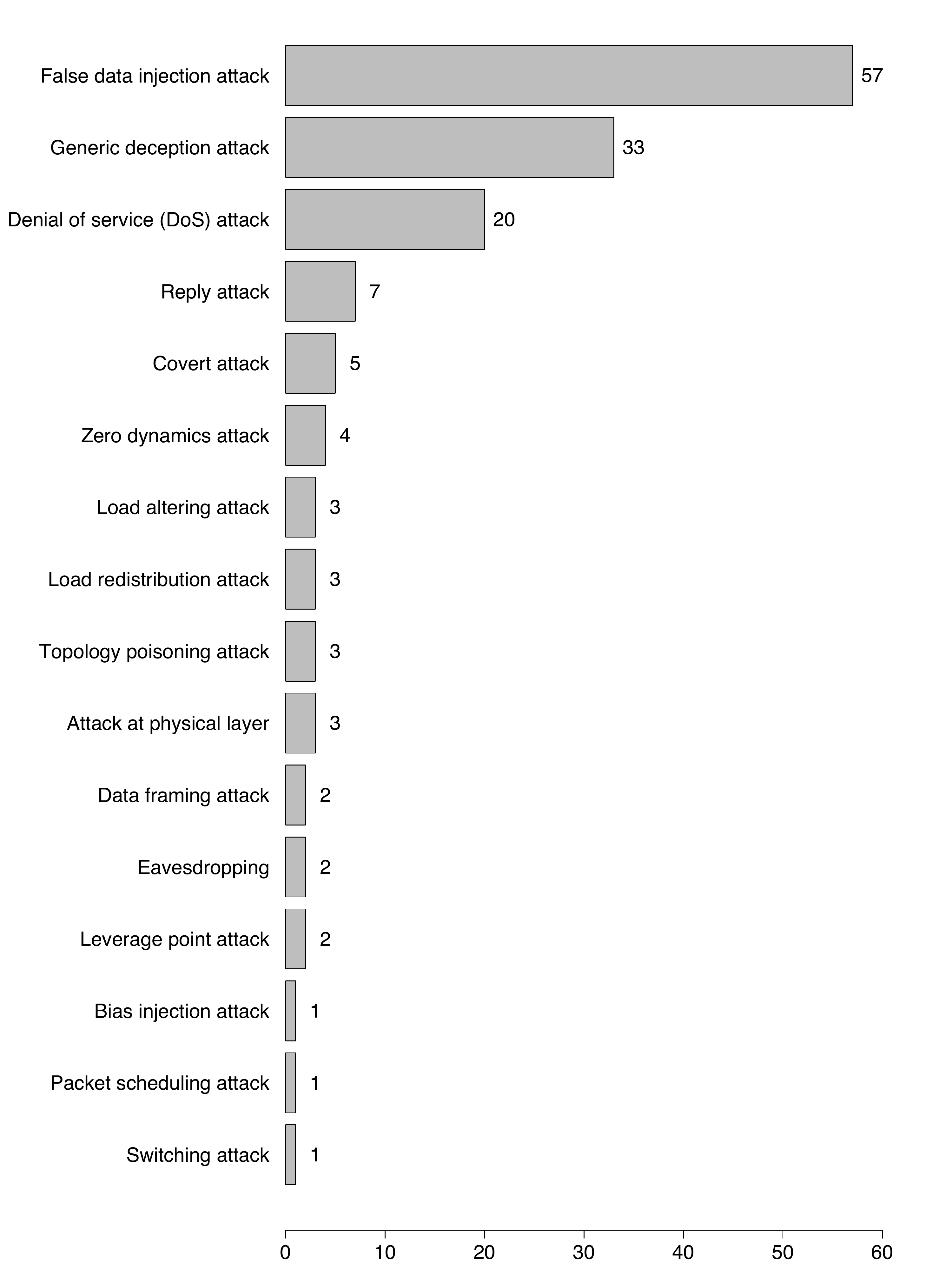}
	\caption{Distribution of attacks considered by primary studies}
	\label{fig:attack-names}
\end{figure}

\vspace*{-2mm}
Figure~\ref{fig:attack-names} shows the distribution of attacks within the set of our primary studies.
The false data injection, together with generic deception and DoS, with 57, 33 and 20 occurrences respectively, accounts for 74.8\% of all considered attacks, while the variable structure switching, the packet scheduling, and the bias injection attacks are considered only once.

\noindent
\textbf{Characterization of the attacks.} Generally speaking, an attack on control systems can be characterized by the amount of available resources and knowledge [S081]. The resources of an adversary can be split in \emph{disclosure} resources, which enable her to obtain sensitive information about the system during the attack by violating data confidentiality, and \emph{disruption} resources, that affect the system operation by compromising the integrity and/or availability. The amount of a priori \emph{knowledge} regarding the control system is another core component of the adversary model, as it may be used, for instance, to render the attack undetectable. In the rest of Section \ref{subsec:attacks} we describe the characteristics of each type of attack individually.

In the \textbf{bias injection} attack, considered only by Teixeira et al. [S081], the adversary's goal is to inject a constant bias in the system without being detected. \emph{No disclosure} capabilities are required for this attack, since the attack policy is open-loop. The data corruptions may be added to both the actuator and sensor data, and the amount of disruption resources should be \emph{above the threshold of undetectability}\footnote{In other words, the attacker should have enough resources to construct an unobservable attack; a good example of the amount of disruption resources above the threshold of undetectability in the context of power transmission networks is given by the security index \citeref{sandberg2010security}, defined as minimum number of measurements an attacker needs to compromise, in order to attack measurement $k$ without being detected.}. Furthermore, the open-loop attack policy requires an extensive knowledge of the parameters of considered closed-loop system and anomaly detector.

In the coordinated \textbf{variable structure switching} attack and its extension to multi-switch attack considered in the work of Liu et al. [S013], an opponent controls multiple circuit breakers within a power system, and employs a local model of the system and local state information (i.e. some knowledge of the target generator states, which are rotor angle and frequency) to design a state-dependent breaker switching sequence, that destabilizes target synchronous generators.  

The attack on the \textbf{scheduling} algorithm influences the temporal characteristics of the network, as 
\textit{
``it results in time-varying delays and data packets possibly received out-of-order''
} [S087].
To remain stealthy, the attacker is not able to delay the packets beyond a maximum allowable delay consistent with the network protocol in place. 
On the system level, this attack does not require any a priori knowledge of the system model, nor any disclosure resources. 

The \textbf{false data injection} is a specific deception attack on state estimation, introduced in the context of electric power grids by Liu et al. [S001]. This attack on cyber-physical systems is the most studied one. To perform it, an adversary with some knowledge of the system topological information manipulates sensor measurements in order to change the state variables, while bypassing existing bad data detection schemes. This attack is based on the open-loop policy and does not require any disclosure resources. To construct the attack vectors, a common assumption in most works on false data injection attacks on power system state estimation is that the attacker has complete knowledge about the power grid topology and transmission-line admittances. This information is abstracted in the Jacobian matrix \textbf{H} \citeref{6279588, abur2004power}, known also as  measurement or (power network) topology matrix. By contrast, Teixeira et al. [S006] assumes the attacker only possesses a perturbed model of the power system,
\textit{
``such a model may correspond to a partial model of the true system, or even an out-dated model''
}[S006].
In this way it quantifies a trade-off between the accuracy of the model known by adversary and possible attack impact for different BDD schemes, showing that
\textit{
``the more accurate model the attacker has access to, the larger deception attack he can perform undetected''
} [S006].
Similarly, Rahman and Mohsenian-Rad [S027] argues that
\textit{
``a realistic false data injection attack is essentially an attack with incomplete information due to the attackers lack of real-time knowledge with respect to various grid parameters and attributes such as the position of circuit breaker switches and transformer tap changers and also because of the attacker's limited physical access to most grid facilities'',
}
and presents a vulnerability measure for topologies of power grids subject to attacks based on incomplete information.
%
On the same line, Bi and Zhang [S017] derives a necessary and sufficient condition to perform undetectable false data injection attack with partial topological information and develops a min-cut method to design the optimal attack, which requires the minimum knowledge of system topology.
Finally, the problem of constructing a \emph{blind} false data injection attacks without explicit prior knowledge of the power grid topology is studied by Esmalifalak et al. [S012], Kim et al. [S051], and Yu and Chin [S052]. In Esmalifalak et al. [S012] attackers  try to make inferences through phasor observations applying linear independent component analysis (ICA) technique. However, such technique requires that loads are statistically independent and non-Gaussian, and the technique need full sensor observations [S051]. Kim et al. [S051] instead proposes subspace methods, which requires no system parameter information. In this case the attack can be launched with only partial sensor observations. Yu and Chin [S052] proposes to use principal component analysis (PCA) approximation method without the assumption regarding the distribution of state variables, to perform the same task of making inferences from the correlations of the line measurements, in order to construct the blind false data injection attack. Differently from the works on undetectable false data injection attacks on power grids summarized up to here, Qin et al. [S036] presents an \emph{unidentifiable} version of this attack, in which the control center can detect that there are bad or malicious measurements, but it cannot identify which meters have been compromised.

A special type of false data injection attack on electric power grid is the \textbf{load redistribution} attack, in which only load bus power injection and line power flow measurements are attackable [S008]. It consists in increasing load at some buses and reducing loads at other buses, while maintaining the total load unchanged, in order to hide the attack from bad data detection. The construction of load redistribution attack relies on topological information of the network, that can be derived from the Jacobin matrix \textbf{H}. Considering the practical issue that an attacker can only obtain the parameter information of a limited number of lines, Liu et al. [S043] presents a strategy to determine optimal local attacking region, that requires the minimum network parameter information. The undetectability is obtained by 
\textit{
``making sure that the variations of phase angles of all boundary buses connected to the same island of the nonattacking region are the same''
} [S043].

The \textbf{data framing} attack is a deception attack on power system state estimation that exploits current bad data detection and removal mechanisms. It purposely triggers the bad data detection mechanism and frames some normally operating meters as sources of bad data such that their data will be removed. After such data removal, although the remaining data appear to be consistent with the system model, the resulting state estimate may have an arbitrarily large error [S037]. Also this attack does not require any disclosure resources, since the attack policy is open-loop. By applying the subspace methods presented in 2015 by Kim et al. [S051] to learn the system operating subspace from measurements, the data framing can be performed without knowledge of the Jacobian matrix \textbf{H}. A limited a priori knowledge required consists of a basis matrix \textbf{U} of a subspace of all possible noiseless measurements $\mathcal{R}$ of \textbf{H}.

The \textbf{leverage point} attack is a deception attack which creates leverage points within the factor space of the (power system) state estimation regression model [S049]. The residual of the measurement corresponded with the leverage point is very small even when it is contaminated with a very large error. Thus the adversary can freely introduce arbitrary errors into the meter measurements without being detected. This attack is based on an open-loop policy and thus does not require disclosure resources. However, to be fully effective, it requires a complete knowledge of the Jacobian matrix \textbf{H} and amount of disruption resources above the threshold of undetectability [S057]. 

The \textbf{load altering} attack against power grid's demand response and demand side management programs can bring down the grid or cause significant damage to the power transmission and user equipment. It consists in an attempt to control and change (usually increase) certain load types in order to damage the grid through circuit overflow or disturbing the balance between power supply and demand [S018]. The \emph{static} load altering is mainly concerned in changing the volume of the load. Here the attacker without any prior knowledge of the plant model uses some historical data to impose a pre-programmed trajectory to the victim load (an open-loop policy). In the more advanced \emph{dynamic} load altering attack, presented in 2015 by Amini et al. [S050], the adversary
\textit{
``constantly monitors the grid conditions through the attacker's installed sensors so that it can adjust the attack trajectory based on the current conditions in the power grid''
} [S050].
With this closed-loop policy, the attacker having a complete knowledge of the plant's model controls the victim load based on a feedback from the power system frequency and can make the power system unstable, without the need for increasing the scope or volume of the attack, compared to a static scenario.

The attacks at \textbf{physical layer} range from attacks that affect both the physical infrastructure and the control network (of power grids) [S053] to attacks through physical layer interactions, such as an attack on vehicle platoon traveling at a constant speed, presented by Dadras et al. [S115]. 
The attack studied by Soltan et al. [S053] physically disconnects some power lines within the attacked zone (which is defined as a set of buses, power lines, phasor measurement units (PMUs) and an associated phasor data concentrator (PDC) \citeref{6279588}) and disallows the information from the PMUs within the zone to reach the control center. This attack does not require any knowledge of the plant model, nor disclosure resources. 
%
The attack on vehicle platoons [S115] is carried out by a maliciously controlled vehicle, who attempts to destabilize or take control of the platoon by combining changes to the gains of the associated law with the appropriate vehicle movements. This closed-loop attack 
\textit{
``bears some resemblance to an insider version of the replay attack of [S010], in that the attacker is part of the CPS and is therefore able inject control inputs legitimately''.
}

In \textbf{topology poisoning} attack an adversary covertly alters data from certain meters, network switches and line breakers to mislead the control center with an incorrect network topology. Kim and Tong [S028] shows that under certain conditions even in a local information regime, where the attacker has only local information from those meters it has gained control, undetectable topology poisoning attacks exist and can be implemented easily based on simple heuristics.
Deka et al. [S039] proves that grids completely protected by secure measurements are also vulnerable to hidden topology poisoning attacks, if the adversary armed only with generic information regarding the grid structure can corrupt the breaker statuses on transmission lines and jam the communication of flow measurements on the attacked lines.

The \textbf{zero dynamics} attack, first considered in \citeref{5605238, 5779706}, is one in which an adversary constructs an open-loop policy such that the attack signal produces no output. In other words, \textit{``these attacks are decoupled from the plant output $y_k$, thus being stealthy with respect to arbitrary anomaly detectors''} [S081]. For an attacker with limited disruption resources, zero dynamics attacks are based on the perfect (local) knowledge of the plant dynamics. In this setting, Teixeira et al. [S083] shows that zero-dynamics attacks may not be completely stealthy since they require the system to be at a non-zero initial condition; however for the subset of attacks exciting unstable zero-dynamics, the effect of initial condition mismatch in terms of the resulting increase in the output energy can be made arbitrarily small while still affecting the system performance. We should notice that an adversary capable of changing all the measurements can, of course, force the system's output to zero without any knowledge of the model, initial state and nominal input. Furthermore, for a linear not left-invertible system, the knowledge of the initial state is not required, because an attacker can exploit the kernel of the transfer matrix and the linearity of the system.

With the covert attack, also known as a \textbf{covert misappropriation} of the plant [S078], an adversary can gain control of the plant in a manner that cannot be detected by the controller. This attack requires high levels of system knowledge and the ability of attacker to both read and replace communicated signals within the control loop, indeed
\textit{
``the covert agent is assumed to have the resources to read and add to both the control actuation commands and the output measurements. In practice, this could also be accomplished by augmenting the physical actuators or modifying the sensors. Examples of such modifications include installing a controlled-flow bypass around a sluice gate in an irrigation system and connecting a controlled voltage source between a voltage measuring device and its intended connection point in an electrical network. Another potential mode of attack would involve corrupting the PLCs used by the nominal controller to implement the control and sensing operations''
} [S078].
Pasqualetti et al. [S010] observe that the covert attack can be seen as a feedback version of the replay attack, while Smith [S078] examines also the effects of lower levels of system knowledge and nonlinear plants on the ability to detect a covert misappropriation of the plant.

The \textbf{replay} attack is a deception attack (possibly combined with a physical attack), in which an adversary first gathers sequences of measurement and/or control data, and then replays the recorded data while injecting an exogenous signal into the system [S081]. The adversary requires no knowledge of the system model to generate stealthy outputs. However, the attacker needs to have 
\textit{
``enough knowledge of the system model to design an input that may achieve its malicious objective, such as physically damaging the plant'' [S070].
}
The model of this attack is inspired by the Stuxnet \citeref{5742014} example. 

A \textbf{generic deception} attack is an attack on data integrity, where an adversary sends false information from (one or more) sensors or/and controllers in order to deceive a compromised system's component into believing that a received false data is valid or true [S071]. Usually it is modeled as an arbitrary additive signal injected to override the original data. Since generic deception attacks can be used to represent also other, more specialized deception attacks, they are considered mostly in the studies adopting the defender's point of view, presented in Section~\ref{subsec:point-of-view}. There are 23 (19.49\% of all) studies using a generic deception attack model only to develop some defense strategy. The remaining 10 primary studies present (generic) deception attacks, that are different from any other attack considered above. Vrakopoulou et al. [S005] deals with a cyber-attack on the automatic generation control (AGC) signal in multi-area power system as a controller synthesis problem, where the objective is to drive the system outside the safety margins. It investigates two cases according to whether the attacker has perfect model knowledge or not, and provides different alternatives for attack synthesis, ranging from
\textit{
``open loop approaches, based on Markov Chain Monte Carlo (MCMC) optimization, to close loop schemes based on feedback linearization and gain scheduling''
} [S005].
Always within power grids' application domain, Vukovi\'{c} and D\'{a}n [S029] consider a sophisticated adversary, that knows the system model and 
aims to disable the state-of-the-art distributed state estimation by preventing it from converging. To this end, he or she compromises the communication infrastructure of a single control center in an interconnected power system, in order to manipulate the exchanged data (i.e. state variables) used as an input to the state estimator. The stealthy cyber attacks that maximize the error in unmanned aerial systems' state estimation are studied in Kwon et al. [S082]. To consider the worst-case security problem, this study assumes the attacker has the perfect knowledge on the system model and can compromise sensors and/or actuators. The attacks on both sensors and actuators by the adversary with a perfect knowledge of the static parameters of a CPS (modeled as a discrete LTI system equipped with a Kalman filter, LQG controller and $\chi^2$ failure detector) are considered also by Mo and Sinopoli [S071], where the adversary's strategy is formulated as a constrained control problem. Djouadi et al. [S100] instead present optimal sensor signal attacks for the observer-based finite and infinite horizon linear quadratic (LQ) control in terms of maximizing the corresponding cost functions. Also this study assumes full-information, i.e. the system parameters are known to the adversary. 
Zhang et al. [S104] studies stealthy deception attacks on remote state estimation with communication rate constraints. Here the deception attacker intrudes the sensor, learns its online transmission strategy and then modifies the event-based sensor transmission schedule, in order to degrade the estimation quality. For the domain of electricity market, Jia et al. [S062] studies the average relative perturbation of the real-time locational marginal price as an optimization problem; the adversary is assumed to have not only the perfect knowledge of the system model, but also the possibility to access the measurement values in real-time, in order to inject bad data that is state independent, partially adaptive, or even fully adaptive. A stealthy deception scheme capable of compromising the performance of the automated cascade canal irrigation systems is presented by Amin et al. [S072]. This attack scheme is based on approximate knowledge of canal hydrodynamics and is implemented via switching the linearized shallow water partial differential equation parameters and proportional boundary control actions, to withdraw water from the pools through offtakes. Similarly, the stealthy deception attacks on process control systems performed by a very powerful adversary with knowledge of the exact linear model of the plant, the parameters of anomaly detector and control command signals, are presented by C\'{a}rdenas et al. [S075]. In the most sophisticated attack considered in this study, adversaries
\textit{
``try to shift the behavior of the system very discretely at the beginning of the attack and then maximize the damage after the system has been moved to a more vulnerable state''
} [S075].
Finally, for a single-input single-output plant, Bai et al. [S101] analytically characterizes an optimal stealthy attack strategy, that maximizes the estimation error of the Kalman filter by tampering with the control input, as a function of the system parameters, noise statistics and information available to the attacker.

From such literature a systematic characterization of ``types'' of attack is emerging, even if the ``generic deception attack'' and ``false data injection attack'' have been primarily addressed.

\subsection{Attack scheme} \label{subsec:attack-schemes}
\begin{figure}[!htbp]
	\centering
	\includegraphics[width=0.9\columnwidth]{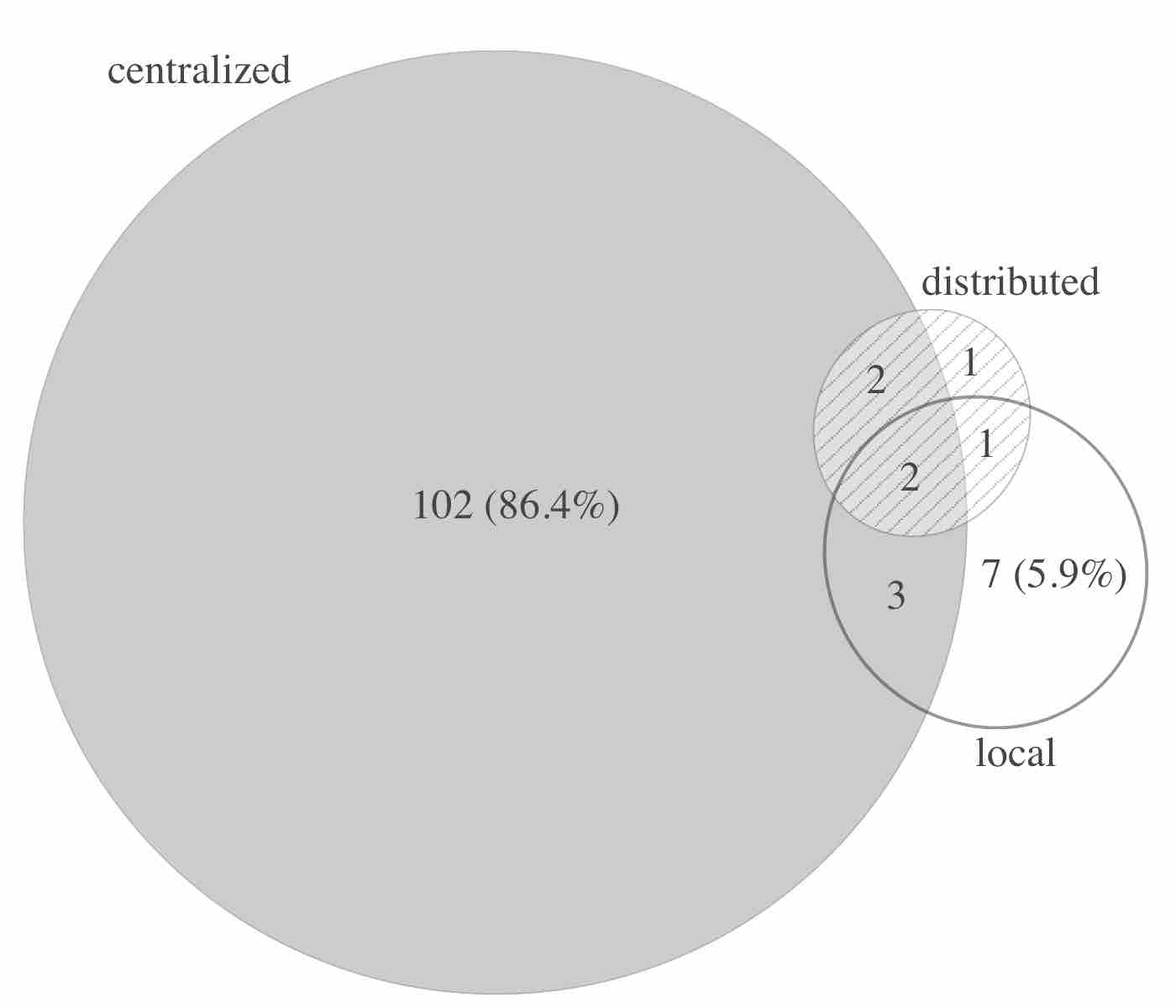}
	\caption{Distribution of primary studies by attack scheme}
	\label{fig:attack-scheme}
\end{figure}

In this section we distinguish the selected studies based on whether they consider centralized, distributed or local attack strategies. The distribution of studies based on this facet is shown in Figure~\ref{fig:attack-scheme}.

The overwhelming majority of primary studies (102, 86.44\%) considers only near omniscient adversary, capable of compromising several system components in a centralized fashion, while there are only 6 (5.17\%) studies that study distributed attacks ([S010, S013, S014, S024, S025, S086]), and 13 (11.02\%) studies dealing with local attacks ([S005, S013, S019, S025, S028, S029, S043, S044, S074, S084, S086, S104, S115]).

It is clear from this data that distributed and local solutions require more attention.

\subsection{Plant model used by the attacker} \label{subsec:plant-attacker}
\begin{figure}[!htbp]
	\centering
	\includegraphics[width=\columnwidth]{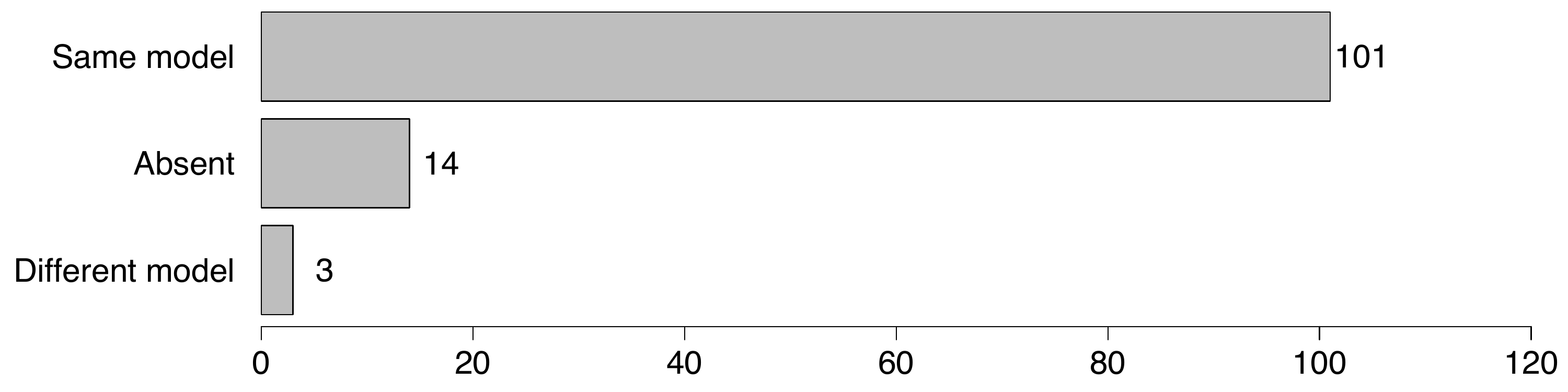}
	\caption{Distribution of primary studies by plant model used by an attacker}
	\label{fig:plant_model-attacker}
\end{figure}

This facet characterizes a modeling framework used by an adversary to design an attack on a CPS. Since attacker's knowledge of the control system and plant model can be limited or absent, an adversary may rely on a model of plant that is different from the actual model used by a system operator. Here our focus is on such cases, Figure~\ref{fig:plant_model-attacker} shows the distribution of the primary studies by plant model used by an attacker.

In 101 studies (85.59\%) it is assumed that the attacker uses the same model of the plant as the system operator, while in 14 studies (11.86\%) the adversary does not use any model of plant. In the remaining 3 studies (2.54\%) the attacker uses a model of plant that is simpler than the one used by operator. In particular, in the works of Kim, Tong and Thomas [S037, S051] data framing attacks on power transmission system are designed using a linearized system. It is shown that such attacks can successfully perturb a nonlinear 
\textit{
``state estimate, and the attacker is able to control the degree of perturbation as desired'' [S037].
}
This is an answer on the question on 
\textit{
``whether attacks constructed from a linear model is effective in a nonlinear system'' [S051].
}
Liang, Kosut and Sankar [S044] studies both DC and AC attack models to construct the false data injection in AC state estimation, showing that the DC attack is detectable when the injected values are too large, while the AC attack model permits to \textit{``hide the attack completely''} [S044].


\subsection{Defense scheme} \label{subsec:defense-scheme}
Similarly to attack schemes, we differentiate the studies also based on whether the proposed approach to defend a CPS focuses on the local or global scale of the system. In case of the global scale, this dimension also specifies whether a defense mechanism uses centralized or distributed coordination model. 

\begin{figure}[!htbp]
	\centering
	\includegraphics[width=0.9\columnwidth]{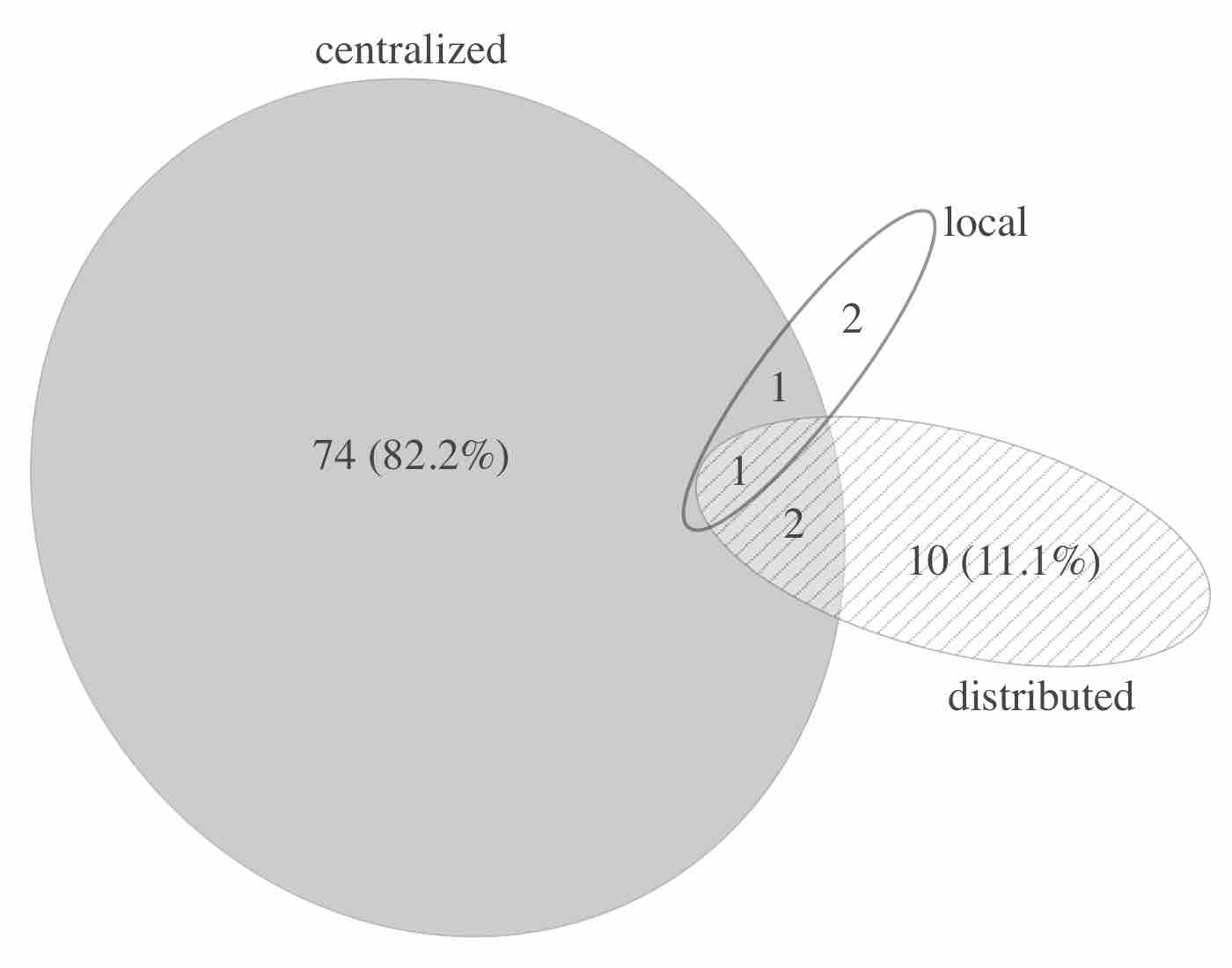}
	\caption{Distribution of primary studies by defense scheme}
	\label{fig:defense-scheme}
\end{figure}

We recall from Section~\ref{subsec:point-of-view} that there are 28 primary studies adopting only an adversary's point of view and not concerned with countermeasures against attacks. We say that for them the defense schemes are not available.
The distribution of remaining (90, i.e. 76.27\% of all) primary studies by defense scheme is shown in Figure~\ref{fig:defense-scheme}. 

Most of the studies (74) on defense mechanisms uses only centralized scheme, while the local scale is considered only in 4 works (\citeref{6426257} and \citeref{Liu:2012:CVS:2185505.2185509}, related to [S010] and [S013], respectively, together with [S020], where also the centralized scheme is taken into account, and [S105]). Distributed approaches are examined in 13 works (alone in [S011, S014, S024, S029, S034, S084, S086, S100, S108, S110] and together with centralized ones in [S010, S025, S060]). We must point out that according to our selection strategy we do not consider the studies focused on the typical distributed problem of reaching consensus in the presence of malicious agents \citeref{5779706, 5605238}; this is because in these works the dynamics is part of the consensus algorithm and can be specifically designed, rather than being given as in a physical system [S058].

This data suggests that distributed and local defense solutions require more attention.

\subsection{Defense strategy} \label{subsec:defense-strategy}
We have already anticipated in Section~\ref{subsec:point-of-view} that countermeasures against attacks, i.e. actions minimizing the risk of threats, are presented in more than three-fourth of primary studies, and occupy the central spot of the research efforts. The defense strategies can be classified as prevention, detection, and mitigation \citeref{7011201}; following the line of the fault diagnosis literature \citeref{5282515}, we advocate isolation as a further defense strategy extending detection approaches. 

\begin{figure}[!htbp]
	\centering
	\includegraphics[width=\columnwidth]{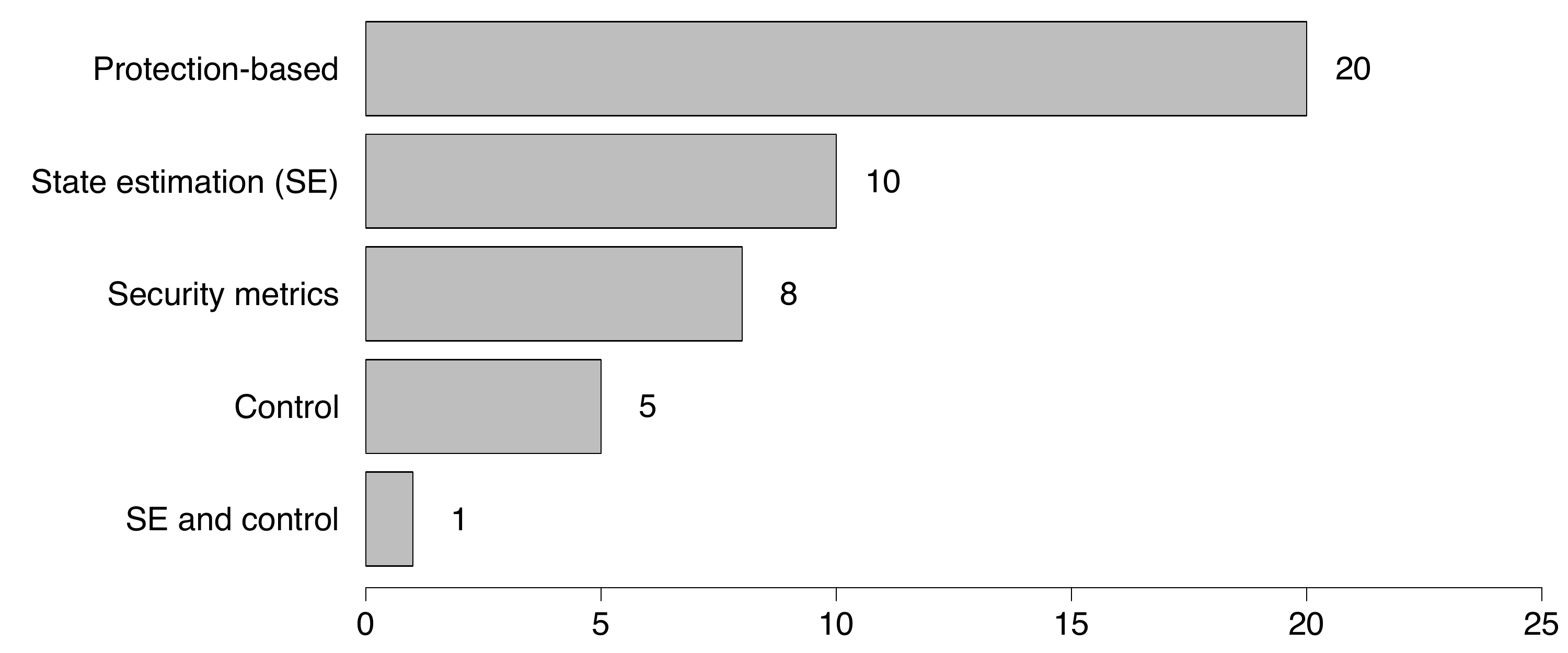}
	\caption{Distribution of primary studies by prevention approach}
	\label{fig:prevention-defenses}
\end{figure}

\textbf{Prevention} aims at decreasing the likelihood of attacks by reducing the vulnerability of the system \citeref{7011201}. It brings together all the actions performed \textbf{offline}, before the system is perturbed or attacked. There are 43 studies (36.44\%) studying prevention mechanisms. These studies range from security metrics for the vulnerability analysis of systems or their critical components to design and analysis of resilient state estimators and controllers capable to withstand some attacks, and protection-based approaches aiming to identify and secure some strategic distributed components. Figure~\ref{fig:prevention-defenses} shows the distribution of the primary studies focussing on prevention.

Twenty studies present \textbf{protection-based} approaches.
Among them, 6 studies discuss the \emph{secure sensor allocation} against undetectable false data injection attacks in power transmission networks. More specifically, Bobba et al. [S003] show that it is necessary and sufficient to protect a set of basic measurements (in number equal to number of all the unknown state variables in the state estimation problem) to ensure that no such attack can be launched, while Giani et al. [S015] proof that placing $p+1$ secure phasor measurement units (PMUs) at carefully chosen buses are sufficient to neutralize any collection of $p$ sparse attacks, and Kim and Tong [S028] present a so-called cover-up protection that identifies the set of meters that need to be secured so an undetectable attack does not exist for any target topology. Also Yang et al. [S016] identify the critical meters to protect and observes that the meters measuring bus injection powers play a more important role than the ones measuring the transmission line power flows, since they are essential in determining a specific state variable, while the measurements of line power flows are redundant to improve the accuracy of state estimation. As finding the minimum number of protected sensors such that an adversary cannot inject false data without being detected is NP-hard\footnote{since this problem is reducible to the \emph{hitting set problem}} [S003], Kim and Poor [S009] and Deka et al. [S038] present greedy algorithms to select a subset of measurements to be protected. To validate the correctness of customers' energy usage by detecting anomaly activities at the consumption level in the power distribution network, Lo and Ansari [S032] present
\textit{
``a hybrid anomaly intrusion detection system framework, which incorporates \textit{power information} and \textit{sensor placement} along with grid-placed sensor algorithms using graph theory to provide network observability.''
}
To reveal zero-dynamics attacks, Teixeira et al. [S083] provide necessary and sufficient conditions on modifications of the CPS's structure and presents an algorithm to deploy additional measurements to this end, while Bopardikar and Speranzon [S089] develop design strategies that can prevent or make stealth attacks difficult to be carried out; the proposed modifications of the legacy control system include optimal allocation of countermeasures and design of augmented system using a Moore-Penrose pseudo-inverse. Mohsenian-Rad and Leon-Garcia [S018] discuss the defense mechanisms against static load altering attacks and presents a cost-efficient load protection design problem minimizing the cost of protection while ensuring that the remaining unprotected load cannot cause circuit overflow or any other major harm to the electric grid. For electricity market domain, Esmalifalak et al. [S065] use a two-person zero-sum game model to obtain an equilibrium solution in protecting different measurements against false data injection attacks impacting locational marginal price (LMP). Within the same domain, Ma et al. [S068] consider a multiact dynamic game where the attacker can jam a reduced number of signal channels carrying measurement information in order to manipulate the LMP creating an opportunity for gaining profit, and the defender is able to guarantee a limited number of channels in information delivery. Other protection-based approaches include, for instance
\textit{
``intentionally switch on/off one of the selected transmission lines by turns, and therefore change the system topology'' [S042];
}
dynamically change the set of measurements considered in state estimation and the admittances of a set of lines in the topology in a controlled fashion [S047], that is an application of a moving target defense (MTD) paradigm;
use covert topological information by keeping the exact reactance of a set of transmission lines secret, possibly jointly with securing some meter measurements [S017];
use an algebric criterion to reconfigure and partition a Jacobian matrix \textbf{H} into two sub-matrices, on each of which to perform a corresponding residual test [S021];
use graph partition algorithms to decompose a power system into several subsystems, where false data do not have enough space to hide behind normal measurement errors [S030];
or even use voltage stability index \citeref{Chakravorty2001129} to identify nodes in power distribution networks with similar levels of vulnerabilities to false data injection attacks via a hybrid clustering algorithm [S056]; 
\textit{
``employ a coding matrix to the original sensor outputs to increase the estimation residues, such that the alarm will be triggered by the detector even under intelligent data injection attacks'' [S109],
}
under the assumption that the attacker does not know the coding matrix yet.
Finally, in order to detect and isolate the disconnected lines and recover the phase angles, in front of the joint cyber and physical attack [S053] outlined in Section~\ref{subsec:attacks}, Soltan et al. [S053] present an algorithm that partitions the power grid into the minimum number of attack-resilient zones, ensuring the proposed online methods are guaranteed to succeed.

Then, the four over five \textbf{resilient controllers} [S069, S073, S076, S077] and nine over ten \textbf{state estimators} [S015, S049, S054, S091, S096, S099, S102, S111, S116] presented in the primary studies were already described in the end of Sections~\ref{subsec:controller} and \ref{subsec:state-estimator}, respectively. The only works not discussed there are Bezzo et al. [S114] and Mishra et al. [S110]. The first one builds an algorithm that leverages the theory of Markov decision processes to determine the optimal policy to plan the motion of unmanned vehicles and avoid unsafe regions of a state space despite the attacks on sensor measurements, when 
\textit{
``the system is fully observable and at least one measurement (however unknown) returns a correct estimate of a state'' [S114],
}
while in the second study the state estimation is performed in a private and secure manner across multiple computing nodes (observers) with an approach inspired by techniques in cryptography, i.e. decoding Reed-Solomon codes, and results from estimation theory, such as Cramer-Rao lower bound, as a guarantee on the secrecy of the plant's state against corrupting observers [S110]. Finally, Shoukry et al. [S087] present a minimax \textbf{state estimator and controller} design as a defense against packet scheduling attacks.

There are 8 works presenting \textbf{security metrics}, such as \emph{security indices} defined in the context of power networks as a minimum number of meters to perform an unobservable attack whether including [S004] or not [S002] a given meter, and \emph{$\epsilon$-stealthiness}, which is a notion that quantifies the difficulty to detect an attack when an arbitrary detection algorithm is implemented by the controller [S101]. A vulnerability measure for topologies of power grids subject to false data injection attacks based on incomplete information is presented by Rahman and Mohsenian-Rad [S027], while the vulnerability of the power system state estimator to attacks performed against the communication infrastructure is analyzed by Vukovi\'{c} et al. [S022] via security metrics that quantify the importance of individual substations and the cost of attacking individual measurements in terms of number of substations that have to be attacked. For the domain of electricity market, Jia et al. [S062] introduces the \emph{average relative price perturbation} as a measure of a system-wide price perturbation resulting from a deception attack described in Section~\ref{subsec:attacks}. In the context of canonical double-integrator-network (DIN) model of autonomous vehicle networks, 
to reflect the quality of the adversary's estimate of the desired nonrandom statistics
Xue et al. [S084] defines 
\textit{
``the error covariance for a minimum-variance-unbiased estimate of the initial-condition vector as the \emph{security level matrix}''
}
and considers its scalar measures as security levels characterizing the confidentiality of network's state.
Finally, Kwon and Hwang [S090] consider the \emph{dynamic behavior cost} and \emph{estimation error costs} to analytically test the behavior of unmanned aerial systems under various deception attacks and quantify their severity accordingly.



\begin{figure}[!htbp]
	\centering
	\includegraphics[width=0.9\columnwidth]{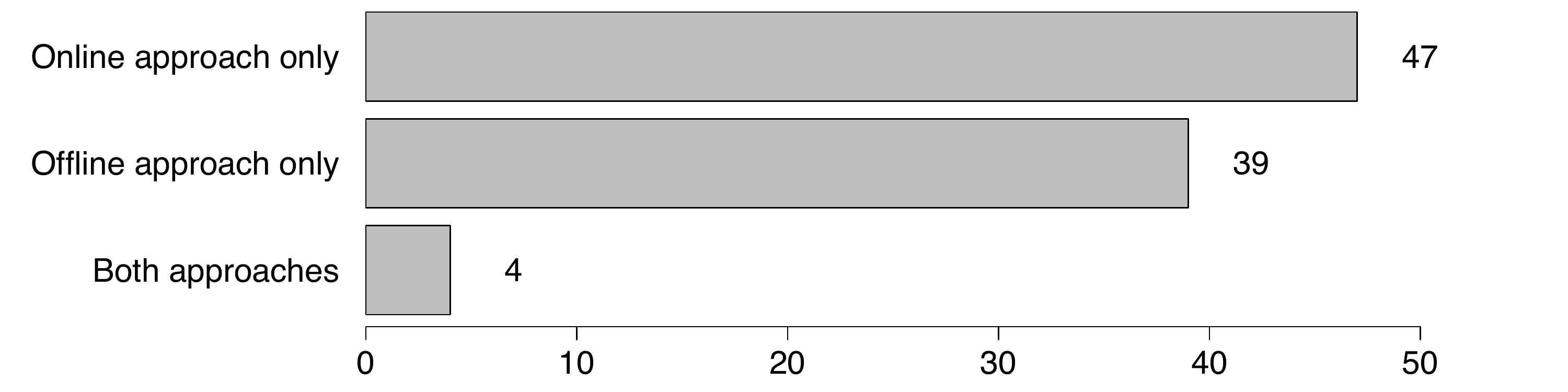}
	\caption{Distribution of primary studies between defense strategies}
	\label{fig:defense-strategies}
\end{figure}

\begin{figure}[!htbp]
	\centering
	\includegraphics[width=0.9\columnwidth]{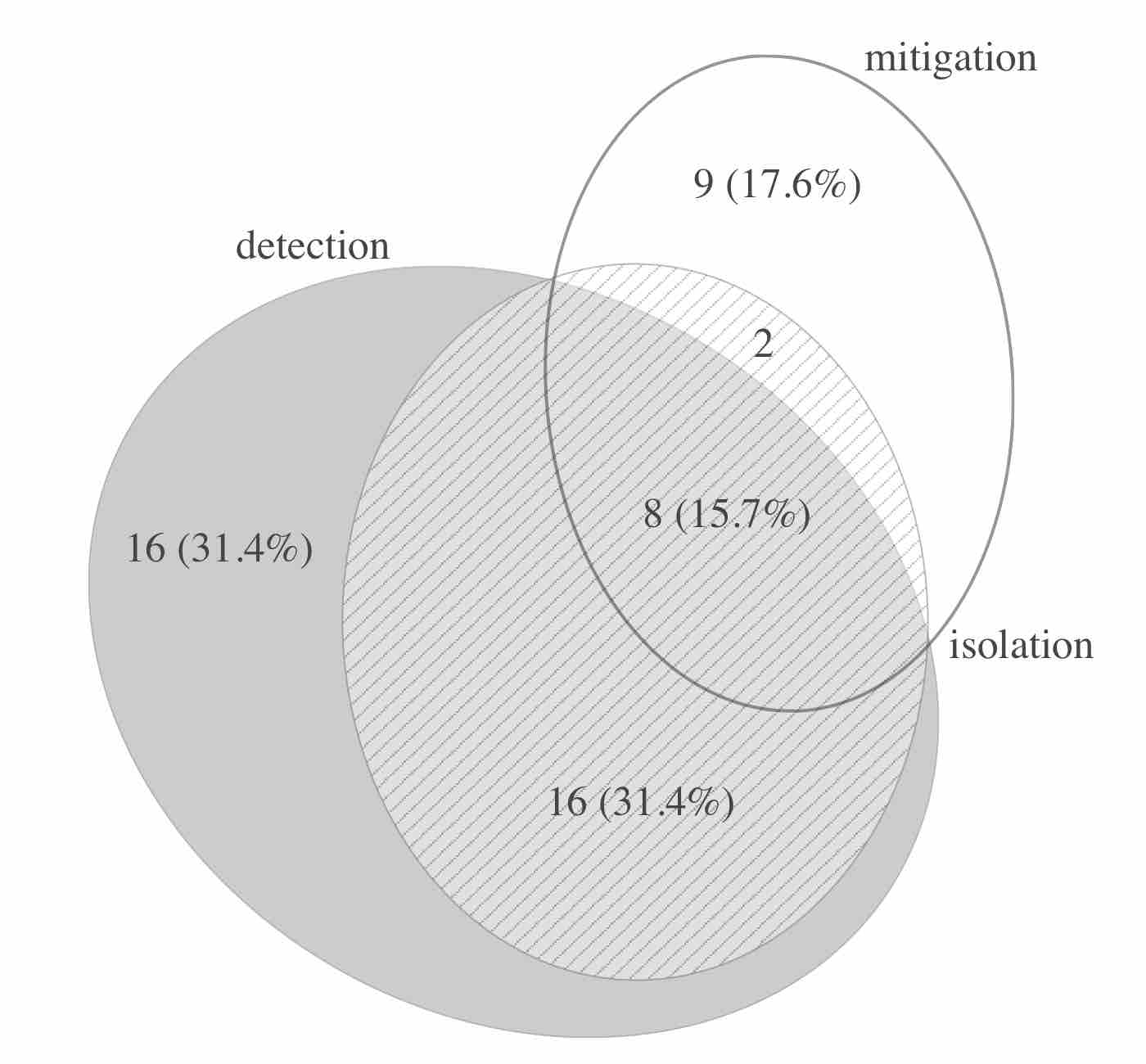}
	\caption{Distribution of primary studies by online defense strategy}
	\label{fig:online-defense}
\end{figure}

The distribution of primary studies between offline and online defense strategies is shown in Figure~\ref{fig:defense-strategies}, while the distribution of studies by online defense strategy is reported in Figure~\ref{fig:online-defense}.

The \textbf{online} approaches come into play after adversarial events happen [S080]. \emph{Detection} is an online approach in which the system is continuously monitored for anomalies caused by adversary actions \citeref{7011201}, in order to decide whether an attack has occurred. Attack \emph{isolation} is one step beyond attack detection, since it distinguishes between different types of attacks \citeref{5282515}, and requires also that the exact location(s) of the compromised components(s) be identified [S020]. Once an anomaly or attack is detected (and isolated), \emph{mitigation} actions may be taken to disrupt and neutralize the attack, thus reducing its impact \citeref{7011201}.

Among the 51 studies concerned with online defenses, 16 are focused on detection only, other 16 on detection and isolation, while 8 on detection, isolation and mitigation. There are 9 works studying mitigation only, and two works on isolation and mitigation [S036, S085].

To contrast unidentifiable false data injection, Qin et al. [S036] present an algorithm to enumerate all feasible cases and proposes a mitigation strategy to minimize the average damage to the system. Another work on \textbf{isolation and mitigation} is Foroush and Mart\'{i}nez [S085], which introduces joint identification and control strategy, that renders the system asymptotically stable in front of unknown periodic DoS in form of pulse-width modulated jamming attacks.

Three of the works focused on \textbf{mitigation} were already described in previous Sections (i.e. [S079] in \ref{subsec:state-estimator}, [S086] and [S093] in \ref{subsec:controller}). Here we spend some words on the remaining 5 studies. Liu et al. [S013] recalls their study of strategies to be 
\textit{
``employed by a power system operator in the face of a switching attack to steer the system to a stable equilibrium through persistent co-switching and by leveraging the existence of a stable sliding mode''
} \citeref{Liu:2012:CVS:2185505.2185509}.
Zhu and Ba\c{s}ar [S080] presents a cross-layer, hybrid dynamic game-theoretic model that captures the coupling between the cyber and the physical layers of the system dynamics, extending the control and defense strategy designs
\textit{
``to incorporate post-event system states, where resilient control and cyber strategies are developed to deal with uncertainties and events that are not taken into account in pre-event robustness and security designs''
} [S080].
The overall optimal design of the cyber-physical system is characterized here by a Hamilton-Jacobi-Isaacs equation, together with a Shapley optimality criterion. Yuan et al. [S118] uses this model to construct a hierarchical Stackelberg game, in order to design a control strategy resilient to DoS launched by the intelligent attacker, which adjusts its strategy according to the knowledge of the defender's security profile. Also Barreto et al. [S092] studies a game-theory problem (via differential games and heuristic stability games) where the actions of the players are the control signals each of them has access to. It focuses on reactive security mechanisms, which change the control actions in response to attacks. Another game-theoretic study is Liu et al. [S105], in which the objective of the defender is to guarantee the dynamic performance of the networked control system (NCS) by transmitting signals with higher power levels than that of jammer's noisy signals. The cost function of the proposed two-player zero-sum stochastic game includes
\textit{
``not only the resource costs used to conduct cyber-layer defense or attack actions, but also the dynamic performance (indexed by quadratic state errors) of the NCS''
} [S105].
To contrast the DoS attacks characterized by their frequency and duration, De Persis and Tesi [S103] determines suitable scheduling of the transmission times achieving input-to-state stability (ISS) of the closed-loop system. It considers periodic, event-based and self-triggering 
implementation of sampling logics, all of which adapt the sampling rate to the occurrence of DoS and, sometimes, to the closed-loop behavior. 

Regarding \textbf{detection} mechanisms, most of all related works were already described in Section~\ref{subsec:anomaly-detection}. Here we introduce the remaining ones.

In order to detect a zero dynamics attack, Keller et al. [S091] proposes to destroy the stealthy strategy of the attacker by triggering data losses on the control signals corrupted by the attack and to use the (augmented state version of) intermittent unknown input Kalman filter. For a system equipped with multiple controllers/estimators/detectors, such that each combination of these components constitute a subsystem, Miao and Zhu [094] presents a moving-horizon approach to solve a zero-sum hybrid stochastic game and obtain a saddle-point equilibrium policy for balancing the system's security overhead and control cost, since each subsystem has a probability to detect specific types of attacks with different control and detection costs. In the power systems domain, Hao et al. [S046] takes advantage of the sparse and low rank properties of the block measurements for a time interval to make use of robust PCA with element-wise constraints to improve both the error tolerance and the capability of detecting false data with partial observations. 

The \textbf{detection and identification} of false data injection attacks on power transmission systems is considered by Davis et al. [S019], which outlines an ``observe and perturb methodology'' to compare the expected results of a control action with the observed response of the system, while
Ozay et al. [S025] use a modified version of normalized residual test coupled with proposed state vector estimation methods against sparse attacks.
Assuming the attack signal enters through the electro-mechanical swing dynamics of the synchronous generators in the grid as an unknown additive disturbance, Nudell et al. [S059] divide the grid into coherent areas via
\textit{
``phasor-based model reduction algorithm by which a dynamic equivalent of the clustered network can be identified in real-time'',
}
and localizes which area the attack may have entered using relevant information extracted from the phasor measurement data.

\subsection{Theoretical foundation} \label{subsec:theoretical-foundations}
\begin{figure}[!htbp]
	\centering
	\includegraphics[width=\columnwidth]{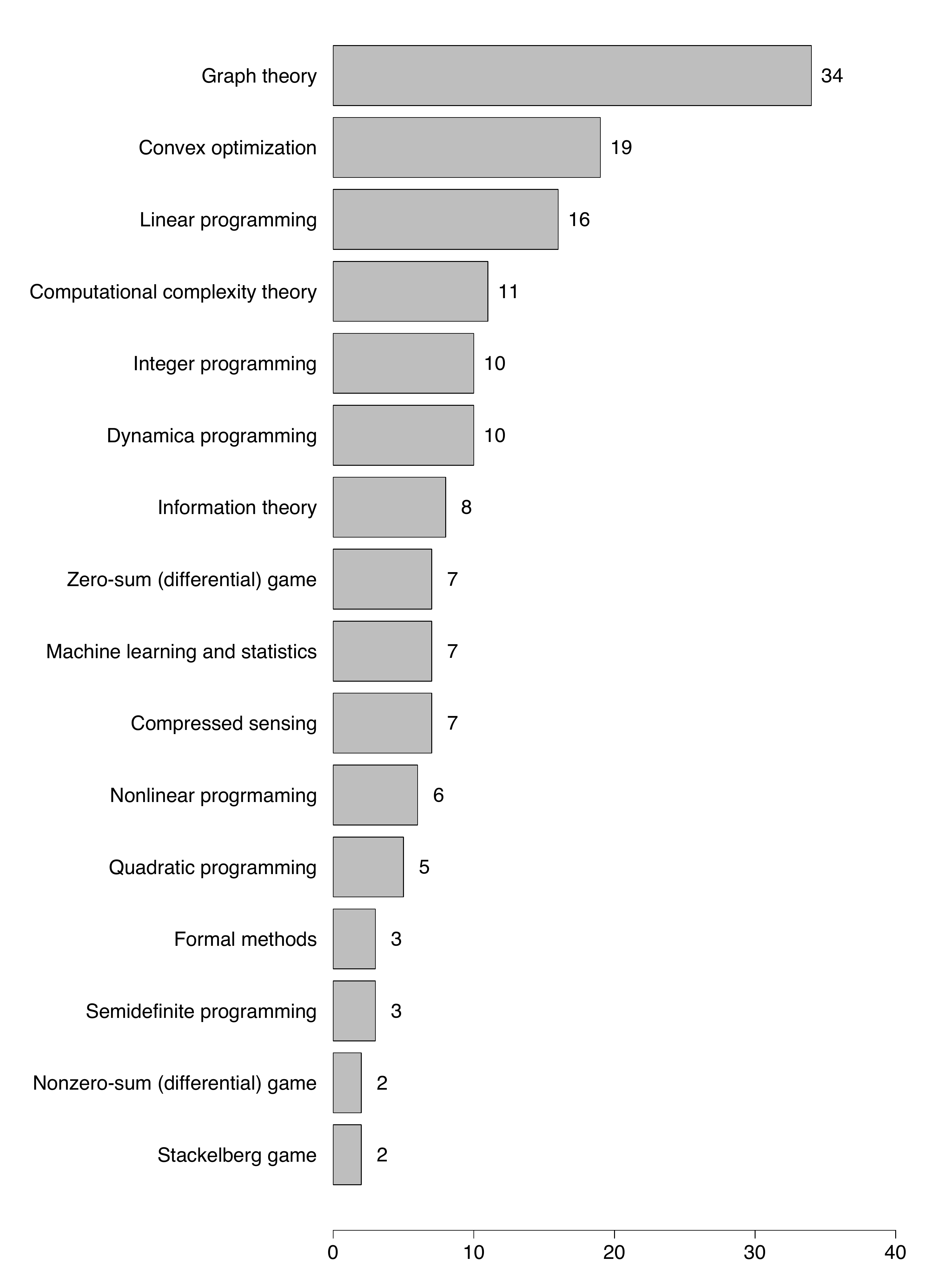}
	\caption{Distribution of theoretical backgrounds considered by primary studies}
	\label{fig:theoretical-foundations}
\end{figure}

Because of the intrinsic multidisciplinary nature of cyber-physical systems, we payed attention also on the theoretical background on which primary studies are built upon. Since the control systems are at the heart of CPS, it isn't a surprise that control theory is used in every study considered in our mapping study. The distribution of other theoretical backgrounds considered by primary studies is presented in Figure~\ref{fig:theoretical-foundations}.

The study of graphs
\citeref{kleinberg2006algorithm, bondy1976graph} is the most used theoretical foundation, found in 34 studies (28.81\%), that are [S002, S004, S009-S011, S014-S017, S020, S022-S024, S026-S030, S032, S034, S037-S039, S042, S045, S051, S053, S059, S062, S074, S084, S086, S088, S100]. \emph{Graph theory} is well suited to represent any kind of networks, and, in fact, it was used in 26 studies on security of power transmission networks. 

To asymptotically analyze the intrinsic difficulty of problems and algorithms and to decide which of these are likely to be tractable, \emph{computational complexity theory} \citeref{horst1995globalopt, kleinberg2006algorithm} is employed in 11 works, all within the field of power transmission ([S001-S004, S010, S015, S016, S032, S038, S039, S053]).

\emph{Information theory} \citeref{cover2006elements} is used in 8 works (\citeref{6161475}, related to [S073], and [S018, S024, S074, S079, S101, S110, S116]), most of which treating the security of generic linear dynamical systems.

The methods of dimensionality reduction (such as principal component analysis) and of latent variable separation (e.g. independent component analysis) from \emph{machine learning} and \emph{statistics} provide a way to understand and visualize the structure of complex data sets \citeref{Lee:2007:NDR:1557216} and are used in 7 works ([S012, S031, S033, S052, S056, S097, S113]). Their application domain is power grids and generic dynamical systems.

Other methods of linear dimensionality reduction are used for simultaneous sensing and compression of finite-dimensional vectors. Providing means for recovering sparse high-dimensional signals from highly incomplete measurements by using efficient algorithms \citeref{eldar2012compressed}, \emph{compressed sensing} is applied in 7 works on power grids and linear dynamical systems ([S004, S009, S025, S033, S046, S079, S111]).

Starting from 2014, typical \emph{formal methods} concepts of signal temporal logic (STL, which is a rigorous formalism for specifying desired behaviors of continuous signals \citeref{maler2004stl}) and satisfiability modulo theories (SMT) \citeref{Barrett2009SMT} have found their way in 3 studies on CPS security ([S113] and [S047, S117], respectively), with applications to anomaly detection and resilient state estimation in generic cyber-physical systems and power grids.

The mathematical \emph{optimization} \citeref{horst1995globalopt, rao2009engineering} is used in several studies and application areas. The sub-fields of optimization found in primary studies include \emph{convex optimization} (19 studies), \emph{linear programming} (16 studies), \emph{dynamic programming} and \emph{integer programming} (both appeared in 10 studies), \emph{nonlinear programming} (6 studies), \emph{quadratic programming} (adopted in 5 works) and \emph{semidefinite programming} (3 studies).

The most used sub-field of \emph{game theory} \citeref{basar1999dynamic}, found in 7 primary studies, is zero-sum game, which do not allow for any cooperation between the players, since what one player gains incurs a loss to the other player ([S065, S068, S073, S080, S087, S094, S105]). Both non-zero sum games and Stackelberg games are formulated in 2 works ([S086, S092] and \citeref{5991463}, related to [S077], together with [S118], respectively). As expected, all these games belong to a class of continuous-time infinite dynamic games, also known as \emph{differential games}, wherein the evolution of the state is described by a differential equation and the players act throughout a time interval.

\section{Results - Validation Strategies (RQ3)}\label{sec:validation}
We determined the research type and related research methods of each primary study, simulation models, simulation test systems and experimental testbeds used, repeatability and availability of replication package. In the following we describe the main facts emerging from the collected data.

\subsection{Research type and related research methods}\label{subsec:research-types-and-methods}
\begin{figure}[!htbp]
	\centering
	\includegraphics[width=\columnwidth]{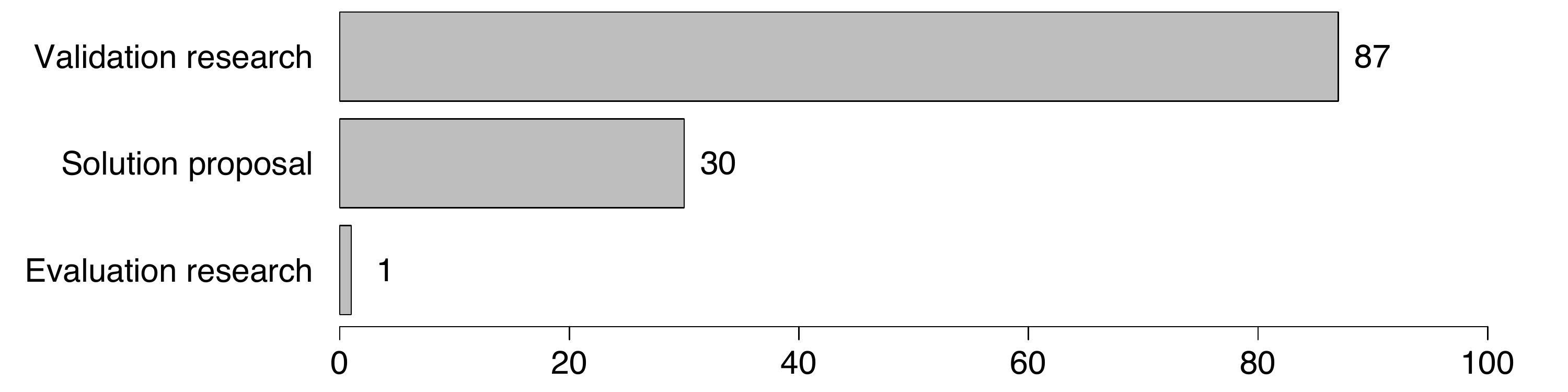}
	\caption{Distribution of primary studies by research type}
	\label{fig:research-type}
\end{figure}

Following the guidelines of systematic mapping studies~\citeref{petersen2015guidelines}, we reuse the classification of research approaches proposed by Wieringa et al.~\citeref{wieringa}, applying the research type classification presented in Petersen et al.~\citeref{petersen2015guidelines}. It is worth noting that our selection strategy (see Section~\ref{sec:selection}) focusses on studies \textit{proposing a method or technique} for cyber-physical
system security, so the \textit{philosophical papers}, \textit{opinion papers} and \textit{experience papers} are not considered in our study. The distribution of primary studies by research type is presented in Figure~\ref{fig:research-type}.

\begin{figure}[!htbp]
	\centering
	\includegraphics[width=\columnwidth]{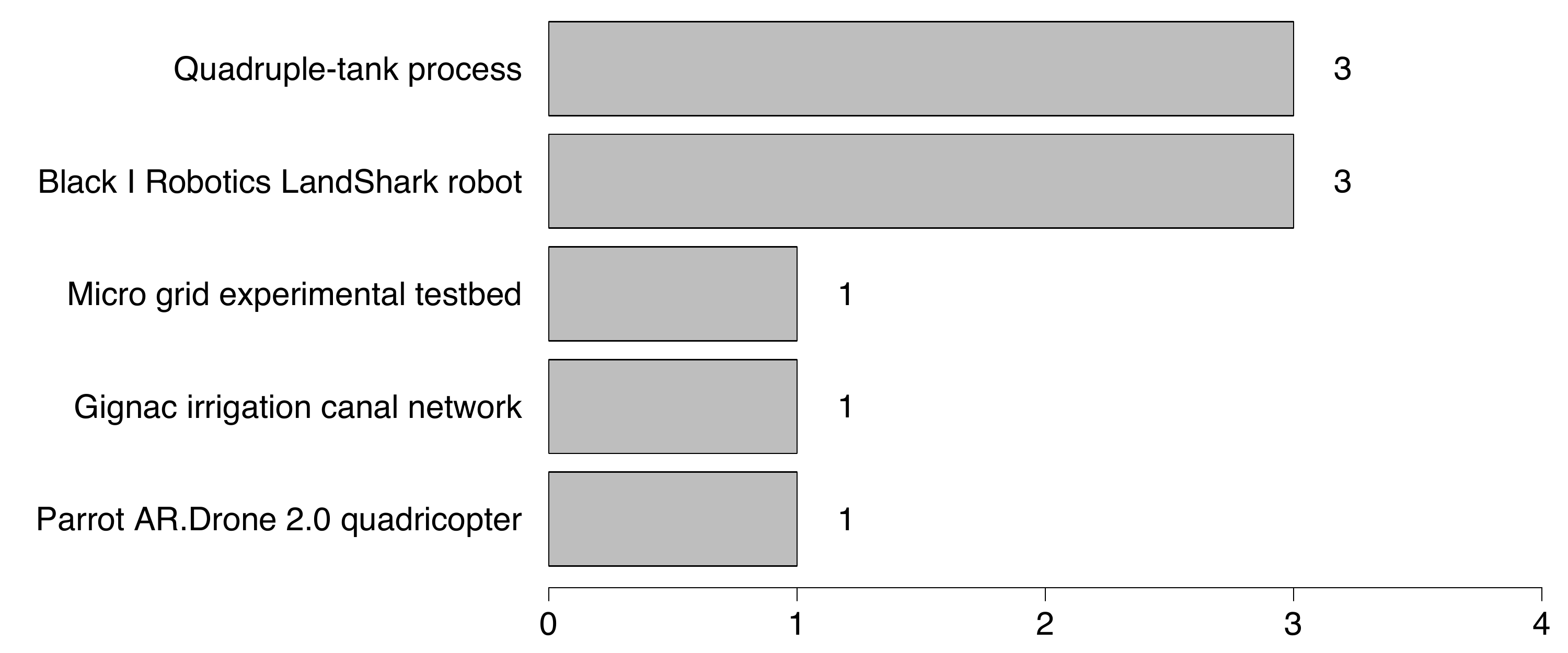}
	\caption{Distribution of experimental testbeds found in primary studies}
	\label{fig:experimental-testbeds}
\end{figure}

\emph{Validation research} is applied in 87 studies (73.73\%), where the techniques investigated are novel and have not yet been implemented in practice; the research methods used are formal mathematical proofs, case studies and lab experiments, together with simulations as a means for conducting an empirical study. In particular, formal \emph{mathematical proofs} are used in 63 studies (53.39\%), in 5 of which as the only validation method adopted. There are 18 primary studies providing both mathematical proofs and illustrative numerical examples, and 14 works illustrating formal mathematical proofs and examples applied to simulation test systems. \emph{Case studies} via simulation, understood as empirical inquiries that draw on multiple sources of evidence to investigate contemporary phenomena in their real-life context, especially when the boundary between phenomenon and context cannot be clearly specified \citeref{wohlin2012experimentation}, are employed in 4 studies, twice as validation of a good line of argumentation [S026, S114], and twice as a follow up of formal mathematical reasoning [S082, S088]. It is worth noting that in Bezzo et al. [S114] also a hardware evaluation on a remotely controlled flying quadricopter is performed, while the case study of D'Innocenzo et al. [S088] is extracted from its previous work cited therein \citeref{6389711}. Another validation research approach, considered in 46 primary studies, consists of an \emph{experiment}, that is a formal, rigorous and controlled empirical investigation, where one factor or variable of the studied setting is manipulated, while all the other parameters are regulated at fixed levels \citeref{wohlin2012experimentation}. 
Most of these experiments are performed in simulation: the \emph{experimental testbeds} are employed only in 7 of these 46 works. 
As shown in Figure~\ref{fig:experimental-testbeds}, the \emph{quadruple-tank process} \citeref{845876}, that is a multivariable laboratory process consisting of four interconnected water tanks, is used in 3 primary studies [S081, S083, S107]. \emph{LandShark}\footnote{\url{http://www.blackirobotics.com/LandShark_UGV_UC0M.html}} robot, i.e. a fully electric unmanned ground vehicle developed by Black I Robotics, is used in other 3 works [S097, S099, S106]. Finally, micro grid experimental testbed consisting of three Siemens SENTRON PAC4200 smart meters connected into the network with YanHua Industry control machine, which is used to monitor all traffic of lab network and read the data from all meters, is used only in one primary study [S054]. The remaining 39 works that use experiments as a validation method are employing different simulation test systems, described in Section~\ref{subsec:simulation-test-system}. Notably, simulation experiments follow a good line of argumentation of the rest of the paper in 21 primary studies, while in the remaining 18 works the experiments are coupled with formal mathematical proofs.

Then, in 30 (i.e. 25.42\% of all) studies \emph{solution proposals} for specific problems are given, where the potential benefits and the applicability of a solution is simply shown through a small example or a line of argumentation; those solutions are either novel or a significant extension of existing ones. We want to point out that often this category corresponds to the results of theoretical research. There are 2 primary studies that use only a good line of argumentation [S014, S069], while sound argument is followed by an illustrative numerical example in 6 primary studies ([S050, S092, S096, S100, S104, S109]), or by an example applied to simulation test system in 22 works. The different simulation test systems found in our primary studies are described in Section~\ref{subsec:simulation-test-system}.

Finally, \emph{evaluation research}, where the techniques are implemented in practice with identification of problems in industry, is done only in one study [S072], in which the Gignac irrigation canal network is used to demonstrate the feasibility of stealthy deception attacks on water SCADA systems.

\subsection{Simulation model}\label{subsec:simulation-model}
\begin{figure}[!htbp]
	\centering
	\includegraphics[width=\columnwidth]{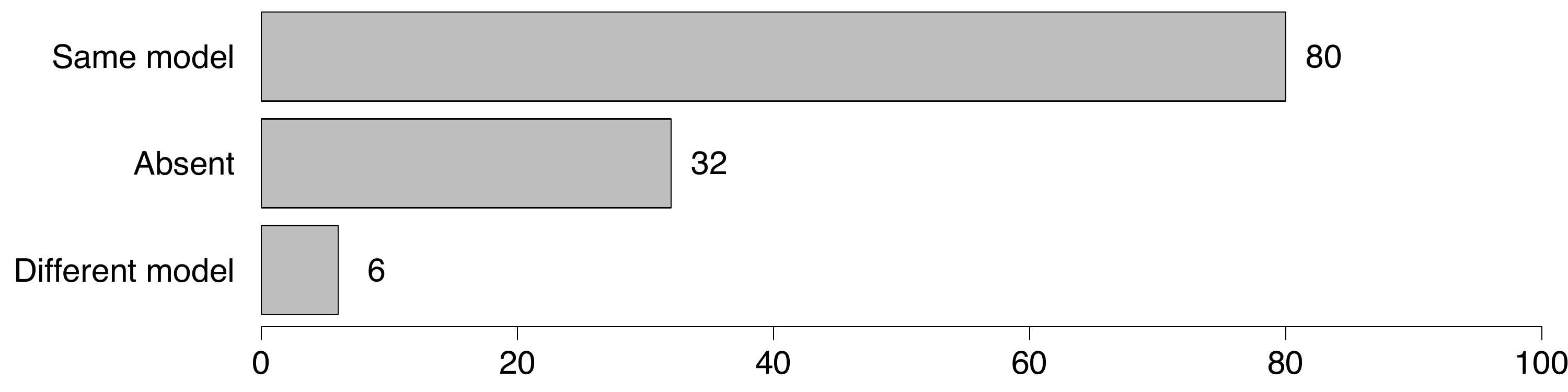}
	\caption{Distribution of primary studies by simulation model}
	\label{fig:plant_model_simulation}
\end{figure}

As in the case of plant models used by attackers,
also the plant models adopted for simulation purposes can be different from the plant models used in the analysis. As we can see from Figure~\ref{fig:plant_model_simulation}, an overwhelming majority of primary studies uses the same model of plant for both the analysis and simulation, while only in 6 studies (5.08\%) these models are different [S028, S030, S051, S057, S062, S067]. Those six studies are within the power transmission or electricity market application domains and use nonlinear AC model for simulation, while consider a DC model (sometimes together with AC model) for analysis purposes. It is worth to mention that in 32 primary studies there are no simulations. Those works account for those solution proposals and validation research papers already described in Subsection \ref{subsec:research-types-and-methods} that use only good line of argumentation, formal mathematical proofs and illustrative numerical examples as the research methods. The only exception is Tiwari et al. [S097], which uses LandShark robot as the experimental testbed, without relying on simulations.

\subsection{Simulation test system}\label{subsec:simulation-test-system}
\begin{figure}[!htbp]
	\centering
	\includegraphics[width=\columnwidth]{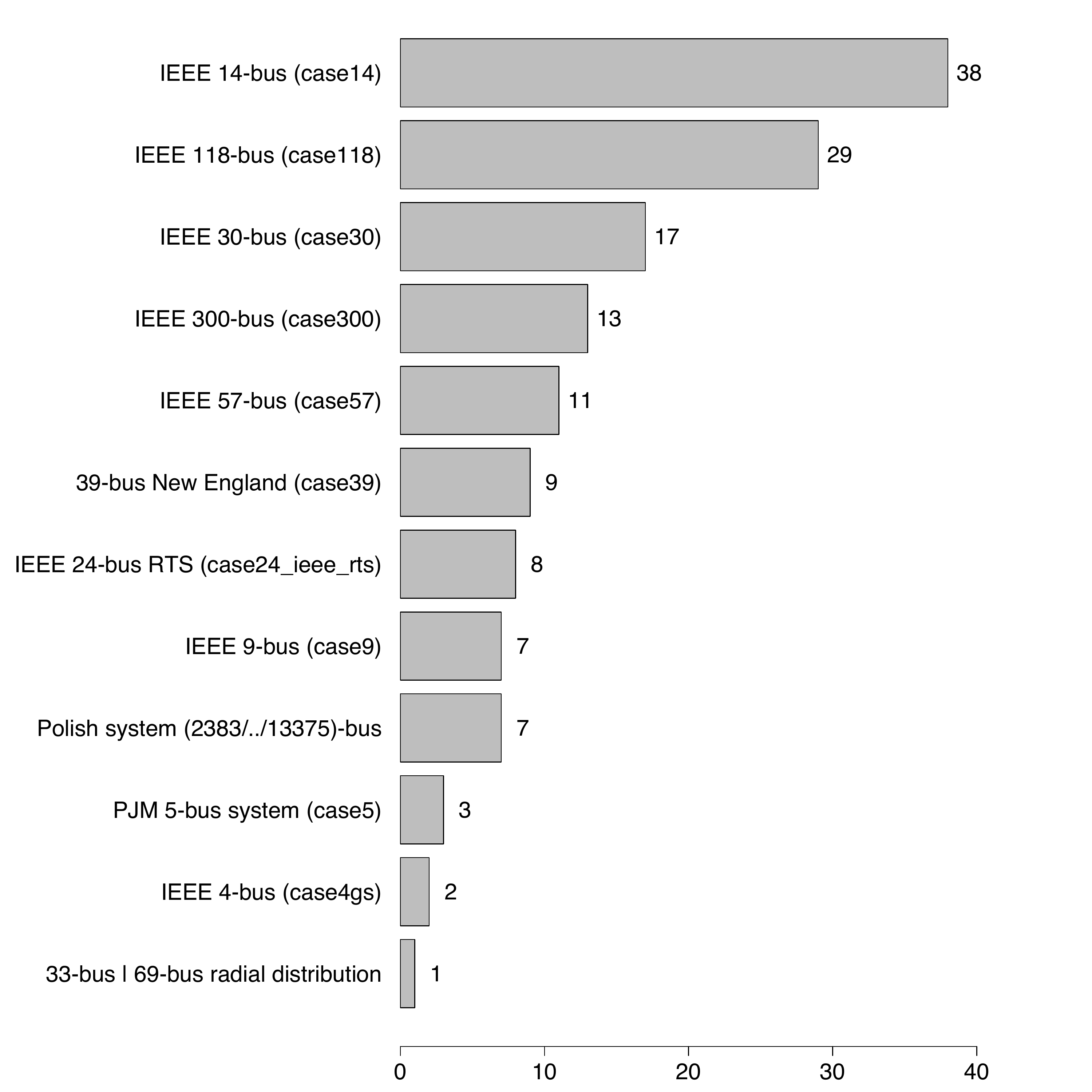}
	\caption{Distribution of power grid test cases}
	\label{fig:matpower-test-cases}
\end{figure}

As it was anticipated in the previous section, 85 primary studies (72,03\%)  use simulation test systems to validate the presented results. Within the power systems application domains, the simulation tool used in all but one primary study is \emph{MatPower} \citeref{5491276}. The distribution of its test cases is shown in Figure~\ref{fig:matpower-test-cases}.

The works studying applications to electricity market use a modified 5-bus PJM example (MatPower case5) \citeref{5589973}, which is employed in 3 primary studies, and IEEE 14-bus (case14), IEEE 30-bus (case30), IEEE 118-bus (case118) test systems. Generally speaking, IEEE 14-bus test system is the most used one, found in 38 works, treating mostly power transmission (in 34 studies), but also power generation (in 2 studies) and electricity market (in 8 studies). IEEE 30-bus test system is used in 17 primary studies, 16 of which are focused on power transmission only, and the remaining one on electricity market. IEEE 118-bus test system is second most adopted one, found in 29 primary studies, dealing power transmission (in 27 studies), power generation (in 2 studies), and electricity market (in 2 studies). 

Studies on power distribution use 33-bus \citeref{1266577} and 69-bus \citeref{Chakravorty2001129} radial distribution test systems in one primary study [S056], and IEEE 24-bus reliability test system (MatPower case24\_ieee\_rts) in another one [S018]. We recall that IEEE 24-bus RTS is based on IEEE RTS-79 \citeref{780914, 651632} and is used in 8 primary studies, all 8 focused on power transmission, 2 of which are dealing also with power generation.

39-bus New England test system (MatPower case39), obtained from Bills et al. \citeref{Bills-1970}, is used in 9 studies, 3 of which are about power generation and 8 are about power transmission.

The remaining test systems are all about power transmission. IEEE 4-bus test system (MatPower case4gs) is used in 2 studies; IEEE 9-bus (case9) is found in 9 studies; IEEE 57-bus (case57) is adopted by 11 and IEEE 300-bus (case300) by 13 studies, while MatPower cases representing the Polish 400, 220 and 110 kV networks during either peak or off-peak conditions are used in 7 studies. 

Power generation is also studied on two-area Kundur system test case \citeref{kundur1994power}, which parameters can be found in the Matlab Power System Toolbox \citeref{207380}, in two studies ([S005, S059]); and on multi-area load frequency control schemes installed with proportional-integral controllers, as described by Jiang et al. \citeref{6080746}, in one study [S118].

\begin{figure}[!htbp]
	\centering
	\includegraphics[width=\columnwidth]{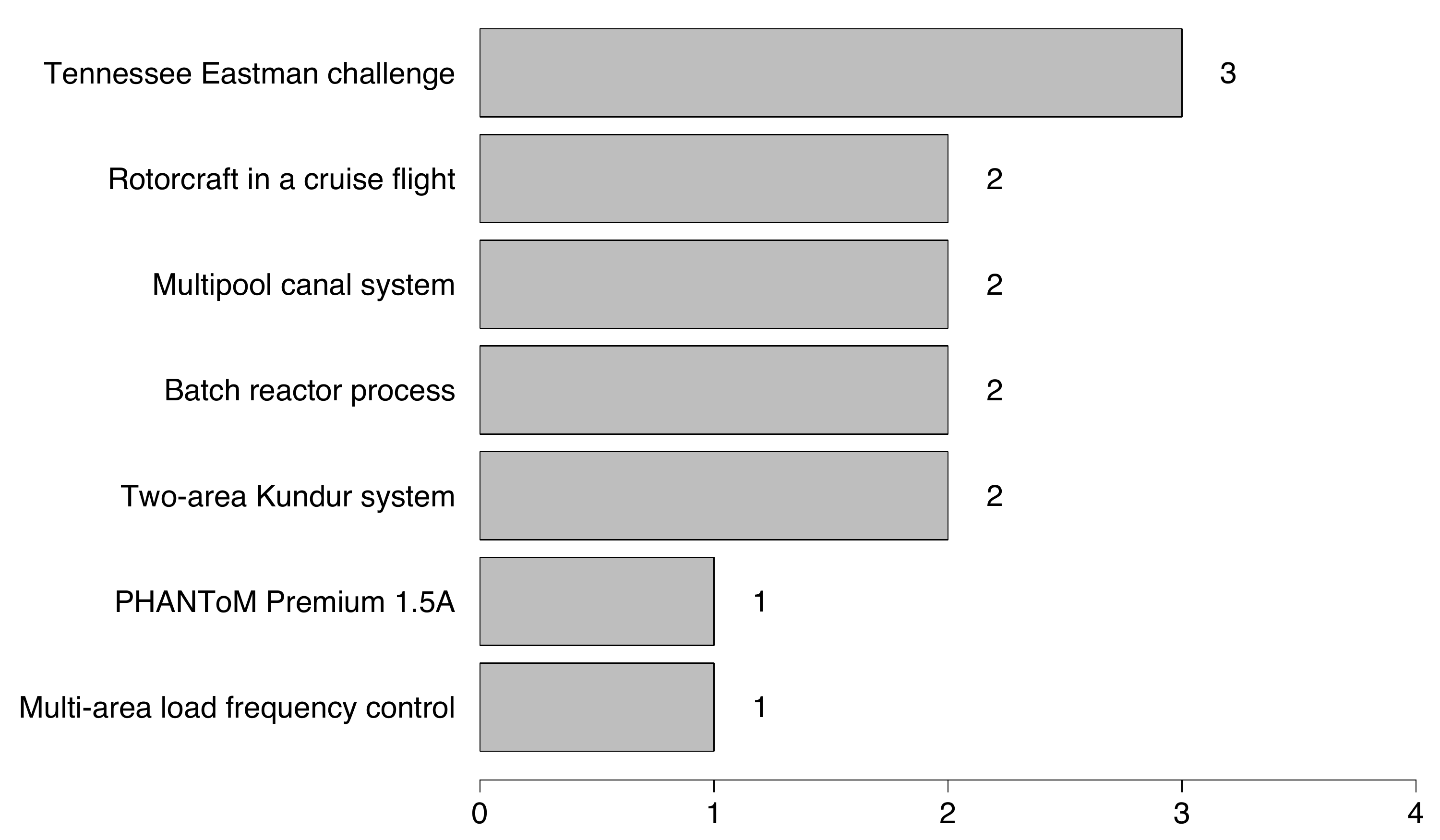}
	\caption{Distribution of other used test cases}
	\label{fig:other-test-cases}
\end{figure}

The other used test cases are summarized in Figure~\ref{fig:other-test-cases}. Irrigation system consisting of a cascade of a number of canal pools, as presented in Amin et al. \citeref{6307833}, is used in two primary studies [S072, S078]. Also an unstable batch reactor system presented by Walsh et al. \citeref{998034}, which is a fourth order unstable linear system with two inputs, is employed in two works [S087, S094]. Tennessee Eastman process control system model and associated multi-loop proportional-integral control law, as proposed by Ricker \citeref{ricker1993109}, is adopted in three studies [S070, S075, S094]. PHANToM Premium 1.5A \citeref{taati2008experimental}, that is a haptic device from SensAble Technologies, is used once in a simulation setup [S105]. Finally, a rotorcraft in a cruise flight \citeref{narendra1973identification} is simulated in two studies [S090, S093]. 

There are also 8 primary studies, which use ad hoc simulation test systems to validate their results. Specifically, Kwon et al. [S082] use Monte Carlo simulation with 1000 runs on an unmanned aerial system navigation system integrating the inertial navigation system and the global positioning system implemented in Matlab. D'Innocenzo et al. [S088] perform Matlab/Simulink simulations on the multi-hop wireless network deployed in a room to connect the temperature sensor to the variable-air-volume box, which is positioned nearby the room. Also Eyisi and Koutsoukos [S098] perform Matlab/Simulink simulations on a single-input single-output (SISO) system; it deals with a velocity control of a single joint robotic arm over a communication network. Bezzo et al. [S106] use robot operating system\footnote{\url{http://www.ros.org}} (ROS) based simulator emulating electromechanical and dynamical behavior of the real robot. In Park et al. [S108] simulations are carried out using a simple model of air traffic operations. Shoukry and Tabuada [S111] use an UGV model implemented in Matlab. Jones et al. [S113] simulate a train, which uses an electronically-controlled pneumatic braking system modeled as a classical hybrid automaton. Finally, Shoukry et al. developed a 
\textit{
``theory solver in Matlab and interfaced it with the pseudo-Boolean SAT solver SAT4J''
} [S117], where
the simulations are performed on linear dynamical systems with a variable number of sensors and system states.

It is not surprising that most advanced and realistic validation methods have been exploited in the power networks application domain. Despite research on CPS Security in this domain appears quite mature, a benchmark is still missing.

\subsection{Repeatability and availability of replication package}\label{subsec:repeatability}
The possibility of reproducing the evaluation or validation results provided by the authors is called repeatability, while the possibility of exploring changes to experiment parameters is known as workability. The repeatability process is a good scientific practice \citeref{bonnet2011repeatability}. The so called Artifact Evaluation Process\footnote{\url{http://www.artifact-eval.org}} is used in a number of conferences in computer science, and a similar concept of repeatability evaluation of computational elements has been introduced in cyber-physical systems domain in 2014 ACM Hybrid Systems Computation and Control (HSCC) conference\footnote{\url{http://www.cs.ox.ac.uk/conferences/hscc2016/re.html}}. However, such practice is rather new to several research communities working on CPS: we found no primary study with a replication package. Thus, we have isolated the information concerning the availability of a replication package and extended the simple dimension provided in Yuan et al. \citeref{Yuan2014SSS} in a way that \emph{repeatability} is considered \emph{high} when the authors provide enough details about
\begin{itemize}
\item the steps performed for evaluating or validating the study,
\item the developed or used software,
\item the used or simulated testbed, if any, and
\item any other additional resource,
\end{itemize}
in a way that interested third parties can be able to repeat the evaluation or validation of the study. Otherwise, we have \emph{low} repeatability. 

\begin{figure}[!htbp]
	\centering
	\includegraphics[width=\columnwidth]{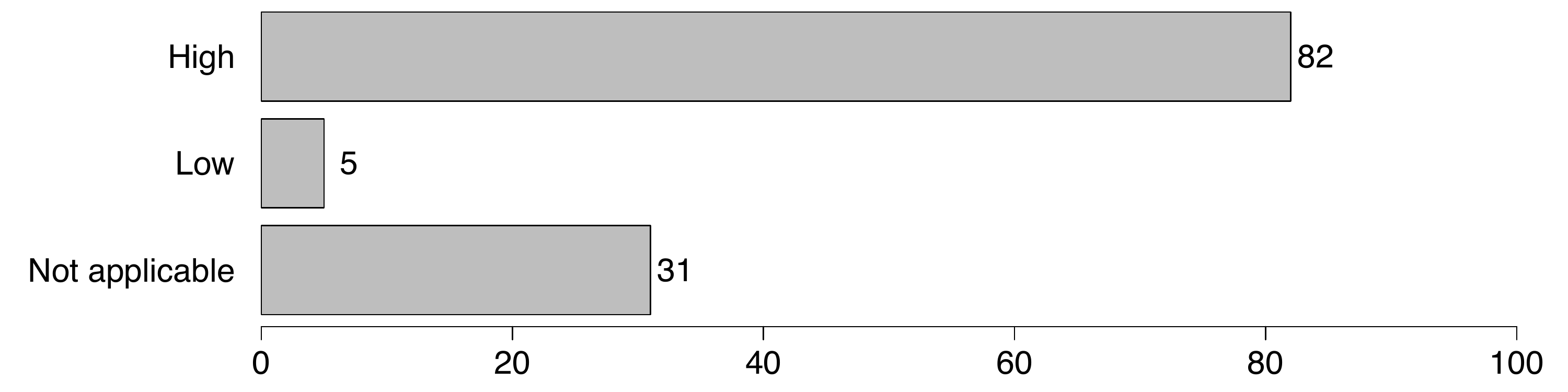}
	\caption{Distribution of primary studies by repeatability}
	\label{fig:repeatability}
\end{figure}

Such high-level definition of repeatability values has ensured that the primary studies using standard test systems from Section~\ref{subsec:simulation-test-system} and well known experimental testbeds have received high values of repeatability, where steps performed in their experiments, case studies and/or simulation examples have been described with enough details. On the other hand, the usage of some ad hoc simulation test system has caused some low values of repeatability assigned. As shown in Figure~\ref{fig:repeatability}, 82 studies (69.49\%) have a high repeatability value, and 5 studies (4.24\%) have a low repeatability score. As a note, we did not have the possibility to evaluate the repeatability of 31 studies (26.27\%) since they do not present any experiment, case study or simulation example. 

Overall, we advocate the improving of repeatability and workability of computational results of the papers by adopting the best practices of repeatability process and creating related replication packages, because we strongly believe in the usefulness of repeatability to empower others to build on top of the contributions of a paper\footnote{\url{http://evaluate.inf.usi.ch/artifacts/aea}} and thus accelerate scientific and technological progress.
\section{Implications for Future Research}\label{sec:future}

We discussed potential future research trends and challenges for CPS security throughout this paper in the context of the various discussions of obtained results (Sections \ref{sec:trends}, \ref{sec:characteristics}, and \ref{sec:validation}); in the following we discuss more general observations about implications for future research on CPS security.

CPS security is a relatively young research domain that is experiencing a strong \textbf{academic and industrial interest} in the very few years, and both European Commission and NSF are very oriented in financing research in this area. From the data obtained in this systematic mapping study it can be inferred that the potential of the developed results and methodologies in addressing realistic emerging problems in several application domains (first of all, power systems) is very promising. As a consequence it is predictable that CPS security will be a ``hot topic'' for the forthcoming years. Our investigation, based on the current state of the art, sheds some light on challenges that will possibly represent the next steps of research in CPS security.

From a \textbf{modeling} point of view this study shows that, as usual in the control theory community, most of the research is based on the model-based paradigm. However, as experience demonstrates, e.g. in the context of energy efficient control of building automation systems, in many CPS application domains the cost of modeling is much larger than the improvement margin in terms of efficiency/cost/performance. As a consequence, we expect that part of future research will be based on the data-based paradigm. This approach, based on ``learning'' techniques and thus strictly connected with the computer science research community, together with the recent large availability of (big!) data deriving from CPS infrastructures, can also be of help towards a more realistic and systematic modeling/mapping of attack/defense models/strategies/architectures.

From a \textbf{validation} point of view, selected papers, as illustrated in the previous sections, address a wide range of application domains, system architectures, problem formulations and theoretical foundations: this makes it very difficult to compare different solutions to similar problems, and we believe that time is mature for the development of academic or industrial benchmarks, test-beds and demonstrators. This could also help in disseminating how research on CPS security can make the difference in each application domain. 

From the point of view of the \textbf{societal and industrial impact}, it is easy to infer from the selected papers that, even thought realistic applications are almost always the main motivation for research, a strong synergy between real industrial/societal problems and theoretical investigation and results is still not apparent from the scientific literature. It is also true that our research questions did not include analysis of relevant projects related to CPS security, however most of the selected papers do not directly relate to or derive from direct collaboration between industry and academia: we hope and expect that this will happen in the near future. Also, we were unable to find research devoted to formal certification with reference to international standards, whose satisfaction is often the biggest barrier for testing and applying novel methods and technologies. Finally, we observe the lack of workshops or symposia with the explicit target of catalysing collaboration between industry and academia on specific applications.

\section{Threats to Validity}\label{sec:threats}

We assessed the level of quality of our study by applying the quality checklist proposed by Petersen et al. in 2015. The goal of Petersen's quality checklist is to assess an objective quality rating for systematic mapping studies.
According to the metrics defined in Petersen's quality checklist, we achieve an outstanding score of 54\%, defined as the ratio of the number of actions taken in comparison to the total number of actions reported in the quality checklist. \textit{The quality score of our study is far beyond the scores obtained by existing systematic mapping studies in the literature}, which have a distribution with a median of 33\% and 48\% as absolute maximum value.

Overall, the high quality of our study has being ensured by producing a detailed research protocol document in which all of its steps have been subject to three external reviews by independent researchers (see Section~\ref{sec:method}) and by conducting our study by following the well-accepted and updated guidelines of systematic review/mapping study~\citeref{kitchenham2007guidelines,petersen2015guidelines}. In the following we detail the main threats to validity of our study and how we alleviated them.

\noindent \textbf{Conclusion validity}.
Conclusion validity refers to the relationship between the extracted data, the produced map, and the resulting findings~\citeref{wohlin2012experimentation}.

In order to mitigate possible conclusion validities, first of all we defined the search terms systematically and we document procedures in our research protocol, so that our research can be replicated by other researchers interested in the topic.
Moreover, we documented and used a rigorously defined data extraction form, so that we could reduce possible biases that may happen during the data extraction process; also, in so doing we had the guarantee that the data extraction process has been consistent to our research questions. 

On the same line, the classification scheme could have been another source of threats to the conclusion validity of our study; indeed, other researchers may identify classification schemes with different facets and attributes. In this context, we mitigated this bias by (i) performing an external evaluation by independent researchers who were not involved in our research, and (ii) having the data extraction process conducted by the principle researcher and validated by the secondary researcher.

\noindent \textbf{Internal validity}.
Internal validity is concerned with the degree of control of our study design with respect to potential extraneous variables influencing the study itself.

In this case, having a rigorously defined protocol with a rigorous data extraction form has surely helped in mitigating biases related to the internal validity of our research. Also, for what concerns the data analysis validity, the threats have been minimal since we employed well-assessed descriptive statistics when dealing with quantitative data. When considering qualitative data, the sensitivity analysis performed on all extracted data has helped in having good internal validity.

\noindent \textbf{Construct validity}.
It concerns the validity of extracted data with respect to our research questions.
Construct validity concerns the selection of the primary studies with respect to how they really represent the population in light of what is investigated. 

Firstly, as described in Section~\ref{sec:search}, the automatic search has been performed on multiple electronic databases to get relevant studies independently of publishers' policies and business concerns.
Moreover, we are reasonably confident about the construction of the search string used in our automatic search since the used terms have been identified by rigorously applying a systematic procedure (i.e., the quasi-gold standard systematic procedure as defined in \citeref{zhang2011identifying}). Moreover, the automatic search is complemented by the snowballing activity performed during the search and selection activity of our review process (see Figure~\ref{fig:search}), thus making us reasonably confident about our search strategy. 
Since our automated search strategy actually relies on search engines quality and on how researchers write their abstracts, the set of primary selected studies have been extended by means of the backward and forward snowballing procedure.

After having collected all relevant studies from the automatic search, we rigorously screened them according to well-documented inclusion and exclusion criteria (see Section~\ref{sec:selection}); this selection stage has been performed by the principle researcher, under the supervision of the secondary researcher.
Also, in order to assess the quality of the selection process, both principle and secondary researchers assessed a random sample of studies, and inter-researcher agreement has been statistically measured with very good results (i.e., we obtained a Cohen-Kappa coefficient of inter-rater agreement of more than 0.80).


\noindent \textbf{External validity}.
It concerns the generizability of the produced map and of the discovered findings~\citeref{wohlin2012experimentation}.

In our research, the most severe threat related to external validity consists in having a set of primary studies that is not representative of the whole research on security for cyber-physical systems. 
In order to mitigate this possible threat, we employed a search strategy consisting of both automatic search and backward-forward snowballing of selected studies. Using these two search strategies in combination empowered us in mitigating this threat to validity. Also, having a set of well-defined inclusion and exclusion criteria contributed to reinforcing the external validity of our study.

A potential source of issues regarding the external validity of our study can be the fact that only studies published in the English language have been selected in our search. This decision may result in a possible threat to validity because potentially important primary studies published in other languages may have not been selected in our research. However, the English language is the most widely used language for scientific papers, so this bias can be reasonably considered as minimal.

Similarly, grey literature (e.g., white papers, not-peer-reviewed scientific publications, etc.) is not included in our research; this potential bias is intrinsic to our study design, since we want to focus exclusively on the state of the art presented in high-quality scientific papers, and thus undergoing a rigorous peer-reviewed publication process is a well-established requirement for this kind of scientific works.

\section{Conclusions and Future Work}\label{sec:conclusions}
The main goal of this research is to analyse the publication trends, characteristics, and validation strategies of existing methods and techniques for CPS security from a researcher's point of view.
In order to achieve this goal we designed and conducted an empirical study that provides a detailed overview of publication trends, venues, and research groups active on CPS security, and a
thorough classification providing an empirically validated foundation for evaluating existing solutions for cyber-physical systems security. The main contribution of this research is to provide a systematic map of research on CPS security; the map has been carried out methodologically in order to warrant the quality of the analysis and results. Additionally, another main contribution of our research is the definition of a sound and complete comparison framework for both existing and future research on CPS security. These contributions will benefit researchers proposing new approaches for CPS security, or willing to better understand or refine existing ones.

We selected a total of 118 primary studies as a result of the systematic mapping process, each of them belonging to different research areas, such as automatic control, networked systems, smart grid, security for information systems.
The main findings emerging from our study are summarized in Section \ref{sec:intro} and explained in details in Sections \ref{sec:trends}, \ref{sec:characteristics}, and \ref{sec:validation}. The resulting implications for the future research are presented in Section \ref{sec:future}.

As future work we are planning to extend this study in order to enlarge its scope to (1) papers weakly related to CPS security but not included (such as typical distributed problems of reaching consensus in the presence of malicious agents, as discussed in Section \ref{subsec:defense-scheme}) and (2) papers/technical reports that derive from relevant academic and industrial projects focused on CPS security.

Also, based on the learning of this work, our future scientific research will be oriented to address CPS security problems providing non-trivial mathematical models of the interaction between physical systems and non-idealities due to communication protocols, in particular regarding wireless sensor and actuator networks. 
\section{Acknowledgements}\label{sec:acks}
Our thanks to Paolo Tell for his valuable comments, suggestions, and feedback on an early version of this work. We are also thankful to Fabio Pasqualetti and Chung-Wei Lin
for their useful and constructive comments. 
\appendices
\section{Research Team}\label{app:team}

Four researchers worked on this study, each of them with a specific role within the research team:
\begin{itemize}
\item[-] \textit{Principal researcher}: PhD student with knowledge about cyber-physical systems, and security for software and control systems; he performed the majority of activities from planning the study to reporting;
\item[-] \textit{Secondary researcher}: assistant professor with background on cyber-physical systems, control theory, networked control systems; he has been mainly involved in the conducting of the study, specially in supporting the primary researcher during the activities of \textit{comparison framework definition} and \textit{data synthesis};
\item[-] \textit{Research methodologist}: post-doctoral researcher with expertise in empirical methods applied to software systems and systematic literature reviews; he has been mainly involved in (i) the planning phase of the study, and (ii) supporting the principle researcher during the whole study, e.g., by reviewing the data extraction form, selected primary studies, extracted data, produced reports, etc.;
\item[-] \textit{Advisor}: senior researcher with many-years expertise in analysis and control of nonlinear and hybrid systems, embedded control systems, networked control systems. She made final decisions on conflicts and options to ``avoid endless discussions''~\citeref{tertiaryAli}, and supported other researchers during the data synthesis and findings synthesis activities.
\end{itemize}

From a geographical point of view, the research team has been locally distributed in Italy, thus having a very low communication overhead and lower chances of misunderstandings.
\section{Selected Primary Studies}\label{app:primary}
\footnotesize
\begin{flushleft}

\begin{tabular}{@{}l p{7.65cm}@{}}
\textbf{[S001]} & \vspace{-2.2mm} \bibentry{Liu:2011:FDI:1952982.1952995} \\
\textbf{[S002]} & \vspace{-2.2mm} \bibentry{6032057} \\
\textbf{[S003]} & \vspace{-2.2mm} \bibentry{bobba2010detecting} \\
\textbf{[S004]} & \vspace{-2.2mm} \bibentry{6882830} \\
\textbf{[S005]} & \vspace{-2.2mm} \bibentry{Vrakopoulou:2015:1} \\
\textbf{[S006]} & \vspace{-2.2mm} \bibentry{5717318} \\
\textbf{[S007]} & \vspace{-2.2mm} \bibentry{5766111} \\
\textbf{[S008]} & \vspace{-2.2mm} \bibentry{6148224} \\
\textbf{[S009]} & \vspace{-2.2mm} \bibentry{5751206} \\
\textbf{[S010]} & \vspace{-2.2mm} \bibentry{6545301} \\
\textbf{[S011]} & \vspace{-2.2mm} \bibentry{6102368} \\
\textbf{[S012]} & \vspace{-2.2mm} \bibentry{6102326} \\
\textbf{[S013]} & \vspace{-2.2mm} \bibentry{6787098} \\
\textbf{[S014]} & \vspace{-2.2mm} \bibentry{6102319} \\
\textbf{[S015]} & \vspace{-2.2mm} \bibentry{6504815} \\
\textbf{[S016]} & \vspace{-2.2mm} \bibentry{6490324} \\
\end{tabular}

\begin{tabular}{@{}l p{7.65cm}@{}} \\[-1.5mm]
\textbf{[S017]} & \vspace{-2.2mm} \bibentry{6840294} \\
\textbf{[S018]} & \vspace{-2.2mm} \bibentry{5976424} \\
\textbf{[S019]} & \vspace{-2.2mm} \bibentry{6486007} \\
\textbf{[S020]} & \vspace{-2.2mm} \bibentry{6693798} \\
\textbf{[S021]} & \vspace{-2.2mm} \bibentry{talebi2012secure} \\
\textbf{[S022]} & \vspace{-2.2mm} \bibentry{6194239} \\
\textbf{[S023]} & \vspace{-2.2mm} \bibentry{6275516} \\
\textbf{[S024]} & \vspace{-2.2mm} \bibentry{Wei:2015:1} \\
\textbf{[S025]} & \vspace{-2.2mm} \bibentry{6547838} \\
\textbf{[S026]} & \vspace{-2.2mm} \bibentry{6376274} \\
\textbf{[S027]} & \vspace{-2.2mm} \bibentry{6503599} \\
\textbf{[S028]} & \vspace{-2.2mm} \bibentry{6547837} \\
\textbf{[S029]} & \vspace{-2.2mm} \bibentry{6840318} \\
\textbf{[S030]} & \vspace{-2.2mm} \bibentry{en7031517} \\
\textbf{[S031]} & \vspace{-2.2mm} \bibentry{6362259} \\
\textbf{[S032]} & \vspace{-2.2mm} \bibentry{6563147} \\
\textbf{[S033]} & \vspace{-2.2mm} \bibentry{6740901} \\
\textbf{[S034]} & \vspace{-2.2mm} \bibentry{7058419} \\
\textbf{[S035]} & \vspace{-2.2mm} \bibentry{SEC:SEC835} \\
\textbf{[S036]} & \vspace{-2.2mm} \bibentry{6305450} \\
\end{tabular}

\begin{tabular}{@{}l p{7.65cm}@{}}   \\[-1.5mm]
\textbf{[S037]} & \vspace{-2.2mm} \bibentry{6840319} \\
\textbf{[S038]} & \vspace{-2.2mm} \bibentry{6785301} \\
\textbf{[S039]} & \vspace{-2.2mm} \bibentry{6849307} \\
\textbf{[S040]} & \vspace{-2.2mm} \bibentry{Li2014156} \\
\textbf{[S041]} & \vspace{-2.2mm} \bibentry{6859403} \\
\textbf{[S042]} & \vspace{-2.2mm} \bibentry{7039384} \\
\textbf{[S043]} & \vspace{-2.2mm} \bibentry{7031948} \\
\textbf{[S044]} & \vspace{-2.2mm} \bibentry{6939486} \\
\textbf{[S045]} & \vspace{-2.2mm} \bibentry{7007750} \\
\textbf{[S046]} & \vspace{-2.2mm} \bibentry{7007752} \\
\textbf{[S047]} & \vspace{-2.2mm} \bibentry{Rahman:2014:MTD:2663474} \\
\textbf{[S048]} & \vspace{-2.2mm} \bibentry{6897944} \\
\textbf{[S049]} & \vspace{-2.2mm} \bibentry{7036880} \\
\textbf{[S050]} & \vspace{-2.2mm} \bibentry{aminidynamic} \\
\textbf{[S051]} & \vspace{-2.2mm} \bibentry{6996007} \\
\textbf{[S052]} & \vspace{-2.2mm} \bibentry{7001709} \\
\textbf{[S053]} & \vspace{-2.2mm} \bibentry{soltanjoint} \\
\textbf{[S054]} & \vspace{-2.2mm} \bibentry{Liu2014} \\
\textbf{[S055]} & \vspace{-2.2mm} \bibentry{7084114} \\
\textbf{[S056]} & \vspace{-2.2mm} \bibentry{Anwar2014} \\
\end{tabular}

\begin{tabular}{@{}l p{7.65cm}@{}}   \\[-1.5mm]
\textbf{[S057]} & \vspace{-2.2mm} \bibentry{6922163} \\
\textbf{[S058]} & \vspace{-2.2mm} \bibentry{7035067} \\
\textbf{[S059]} & \vspace{-2.2mm} \bibentry{7057677} \\
\textbf{[S060]} & \vspace{-2.2mm} \bibentry{6982207} \\
\textbf{[S061]} & \vspace{-2.2mm} \bibentry{6074981} \\
\textbf{[S062]} & \vspace{-2.2mm} \bibentry{6657769} \\
\textbf{[S063]} & \vspace{-2.2mm} \bibentry{6214211} \\
\textbf{[S064]} & \vspace{-2.2mm} \bibentry{6522551} \\
\textbf{[S065]} & \vspace{-2.2mm} \bibentry{6410468} \\
\textbf{[S066]} & \vspace{-2.2mm} \bibentry{6831166} \\
\textbf{[S067]} & \vspace{-2.2mm} \bibentry{kim2014dynamic} \\
\textbf{[S068]} & \vspace{-2.2mm} \bibentry{7050307} \\
\textbf{[S069]} & \vspace{-2.2mm} \bibentry{Amin:2009:1} \\
\textbf{[S070]} & \vspace{-2.2mm} \bibentry{7011170} \\
\textbf{[S071]} & \vspace{-2.2mm} \bibentry{Mo:2012:IAC:2185505.2185514} \\
\textbf{[S072]} & \vspace{-2.2mm} \bibentry{Amin:2010:SDA:1755952.1755976} \\
\textbf{[S073]} & \vspace{-2.2mm} \bibentry{5717544} \\
\textbf{[S074]} & \vspace{-2.2mm} \bibentry{5717166} \\
\textbf{[S075]} & \vspace{-2.2mm} \bibentry{Cardenas:2011:AAP:1966913.1966959} \\
\textbf{[S076]} & \vspace{-2.2mm} \bibentry{7070734} \\
\end{tabular}

\begin{tabular}{@{}l p{7.65cm}@{}}   \\[-1.5mm]
\textbf{[S077]} & \vspace{-2.2mm} \bibentry{6587520} \\
\textbf{[S078]} & \vspace{-2.2mm} \bibentry{7011176} \\
\textbf{[S079]} & \vspace{-2.2mm} \bibentry{6727407} \\
\textbf{[S080]} & \vspace{-2.2mm} \bibentry{7011006} \\
\textbf{[S081]} & \vspace{-2.2mm} \bibentry{Teixeira2015135} \\
\textbf{[S082]} & \vspace{-2.2mm} \bibentry{kwon2014analysis} \\
\textbf{[S083]} & \vspace{-2.2mm} \bibentry{6483441} \\
\textbf{[S084]} & \vspace{-2.2mm} \bibentry{Xue2014852} \\
\textbf{[S085]} & \vspace{-2.2mm} \bibentry{foroush2013multi} \\
\textbf{[S086]} & \vspace{-2.2mm} \bibentry{Zhu:2013:1} \\
\textbf{[S087]} & \vspace{-2.2mm} \bibentry{Shoukry:2013:MCC:2461446.2461460} \\
\textbf{[S088]} & \vspace{-2.2mm} \bibentry{D'Innocenzo2015ecc} \\
\textbf{[S089]} & \vspace{-2.2mm} \bibentry{6623749} \\
\textbf{[S090]} & \vspace{-2.2mm} \bibentry{kwon2013analytical} \\
\textbf{[S091]} & \vspace{-2.2mm} \bibentry{keller2014input} \\
\textbf{[S092]} & \vspace{-2.2mm} \bibentry{barreto2013ds} \\
\textbf{[S093]} & \vspace{-2.2mm} \bibentry{6759880} \\
\textbf{[S094]} & \vspace{-2.2mm} \bibentry{7039433} \\
\textbf{[S095]} & \vspace{-2.2mm} \bibentry{7054460} \\
\textbf{[S096]} & \vspace{-2.2mm} \bibentry{6881627} \\
\end{tabular}

\begin{tabular}{@{}l p{7.65cm}@{}}   \\[-1.5mm]
\textbf{[S097]} & \vspace{-2.2mm} \bibentry{Tiwari:2014:SES:2566468.2566483} \\
\textbf{[S098]} & \vspace{-2.2mm} \bibentry{Eyisi:2014:EAD:2566468.2566472} \\
\textbf{[S099]} & \vspace{-2.2mm} \bibentry{6843720} \\
\textbf{[S100]} & \vspace{-2.2mm} \bibentry{6859001} \\
\textbf{[S101]} & \vspace{-2.2mm} \bibentry{baisecurity} \\
\textbf{[S102]} & \vspace{-2.2mm} \bibentry{6859478} \\
\textbf{[S103]} & \vspace{-2.2mm} \bibentry{7070677} \\
\textbf{[S104]} & \vspace{-2.2mm} \bibentry{zhang2014online} \\
\textbf{[S105]} & \vspace{-2.2mm} \bibentry{Liu20144570} \\
\textbf{[S106]} & \vspace{-2.2mm} \bibentry{6943080} \\
\textbf{[S107]} & \vspace{-2.2mm} \bibentry{li2014stochastic} \\
\textbf{[S108]} & \vspace{-2.2mm} \bibentry{6882825} \\
\textbf{[S109]} & \vspace{-2.2mm} \bibentry{7040293} \\
\textbf{[S110]} & \vspace{-2.2mm} \bibentry{7039631} \\
\textbf{[S111]} & \vspace{-2.2mm} \bibentry{7039940} \\
\textbf{[S112]} & \vspace{-2.2mm} \bibentry{7039974} \\
\textbf{[S113]} & \vspace{-2.2mm} \bibentry{7039487} \\
\textbf{[S114]} & \vspace{-2.2mm} \bibentry{bezzomarkovian} \\
\textbf{[S115]} & \vspace{-2.2mm} \bibentry{Dadras:2015:VPA:2714576} \\
\end{tabular}

\begin{tabular}{@{}l p{7.65cm}@{}}   \\[-1.5mm]
\textbf{[S116]} & \vspace{-2.2mm} \bibentry{mishrasecure} \\
\textbf{[S117]} & \vspace{-2.2mm} \bibentry{shoukry2015sound} \\
\textbf{[S118]} & \vspace{-2.2mm} \bibentry{yuan2015resilient} \\
\end{tabular}

\end{flushleft} 


\bibliographystyleref{IEEEtran} 
\bibliographyref{references}


\end{document}